\documentclass[aps,preprint,floats,nofootinbib]{revtex4}
\usepackage{amsmath}
\usepackage{amssymb}
\usepackage{latexsym}
\usepackage{epsf}
\usepackage{graphicx}

\newlength{\defbaselineskip}
\newcommand{\setlinespacing}[1]%
           {\setlength{\baselineskip}{#1 \defbaselineskip}}

\def\lsim{\mathrel{\raise.3ex\hbox{$<$\kern-.75em\lower1ex\hbox{$\sim$}}}} 
\def\gsim{\mathrel{\raise.3ex\hbox{$>$\kern-.75em\lower1ex\hbox{$\sim$}}}} 
\begin{document}

\preprint{
\hfill
\begin{minipage}[t]{3in}
\begin{flushright}
\vspace{0.0in}
FERMILAB--PUB--06--198--A\\
\end{flushright}
\end{minipage}
}

\hfill$\vcenter{\hbox{}}$

\vskip 0.5cm

\title {Determining Supersymmetric Parameters With Dark Matter Experiments}
\author{Dan Hooper$^1$ and Andrew M.~Taylor$^2$}
\address{$^1$Theoretical Astrophysics Group, Fermilab, Batavia, IL  60510, USA \\ $^2$ Department of Astrophysics, University of Oxford, Oxford OX1 3RQ, UK}
\date{\today}

\bigskip

\begin{abstract}
In this article, we explore the ability of direct and indirect dark matter experiments to not only detect neutralino dark matter, but to constrain and measure the parameters of supersymmetry. In particular, we explore the relationship between the phenomenological quantities relevant to dark matter experiments, such as the neutralino annihilation and elastic scattering cross sections, and the underlying characteristics of the supersymmetric model, such as the values of $\mu$ (and the composition of the lightest neutralino), $m_A$ and $\tan \beta$. We explore a broad range of supersymmetric models and then focus on a smaller set of benchmark models. We find that by combining astrophysical observations with collider measurements, $\mu$ can often be constrained far more tightly than it can be from LHC data alone. In models in the $A$-funnel region of parameter space, we find that dark matter experiments can potentially determine $m_A$ to roughly $\pm 100$ GeV, even when heavy neutral MSSM Higgs bosons ($A$, $H_1$) cannot be observed at the LHC. The information provided by astrophysical experiments is often highly complementary to the information most easily ascertained at colliders.
\end{abstract}

\maketitle

\tableofcontents
\newpage

\section{Introduction}

A great deal of effort has been directed to developing methods of detecting particle dark matter. Over the past years and decades, numerous studies have been conducted to assess the prospects for these various techniques \cite{review}. If dark matter consists of neutralinos, or another weakly interacting particle with a TeV-scale mass, it is likely that one or more of these techniques will make the first detection of dark matter particles within the next several years.

Of all of the candidates for dark matter that have been proposed, none has received as much attention as the lightest neutralino in models of supersymmetry. Supersymmetry is theoretically attractive for a variety of reasons. Among the most compelling is its ability to provide a natural solution to the hierarchy problem~\cite{susyreview}, and a common scale for the unification of the forces of the Standard Model~\cite{gut}. From the standpoint of providing a dark matter candidate, the lightest neutralino is naturally stable by virtue of R-parity conservation \cite{neutralinodm}, and in many models is produced in the early universe in a quantity similar to the measured density of cold dark matter~\cite{wmap}. 

Astrophysical techniques for detecting neutralinos include direct and indirect detection experiments. Direct detection experiments attempt to observe neutralinos scattering elastically off of target nuclei. Indirect detection experiments, in contrast, attempt to detect the annihilation products of neutralinos, including gamma-rays, neutrinos, positrons, anti-protons and anti-deuterons.

Direct and indirect detection measurements are of critical importance in determining the identity of dark matter. Even if collider experiments were to observe a long-lived, weakly interacting, massive particle that appears to be a suitable dark matter candidate, such experiments will never be able to determine whether a particle is stable over cosmological timescales. To determine whether the dark matter of our universe is made up of such a particle (either entirely, or in part), direct and indirect detection experiments will be needed.

But looking beyond the mere detection of dark matter, what will these astrophysical observations reveal to us about the nature of particle dark matter? In particular, is it possible to determine the properties of the lightest neutralino and the corresponding supersymmetry model by direct or indirect dark matter detection experiments? In this article, we attempt to address these questions. While extracting such information from direct and indirect dark matter detection experiments is challenging, it may be possible in many scenarios. We have found that direct detection measurements of the neutralino's spin-independent elastic scattering cross section, rates in neutrino telescopes, and the brightness of gamma-ray lines (from $\chi^0_1 \chi^0_1 \rightarrow \gamma \gamma$ and $\chi^0_1 \chi^0_1 \rightarrow \gamma Z$) are among the most useful astrophysical probes for determining the properties of supersymmetry. In most cases, the value of $\mu$ (or alternatively, the composition of the lightest neutralino) can be constrained far more tightly if astrophysical data is included than it can be by LHC data alone. Furthermore, in some models (those in the $A$-funnel region of supersymmetric parameter space) the mass of the CP-odd Higgs boson can be determined by astrophysical experiments to within roughly $\pm 100$ GeV, even if it cannot be observed at the LHC.

\section{The Composition of the Lightest Neutralino}

\begin{figure}[t]
\includegraphics[width=2.8in,angle=0]{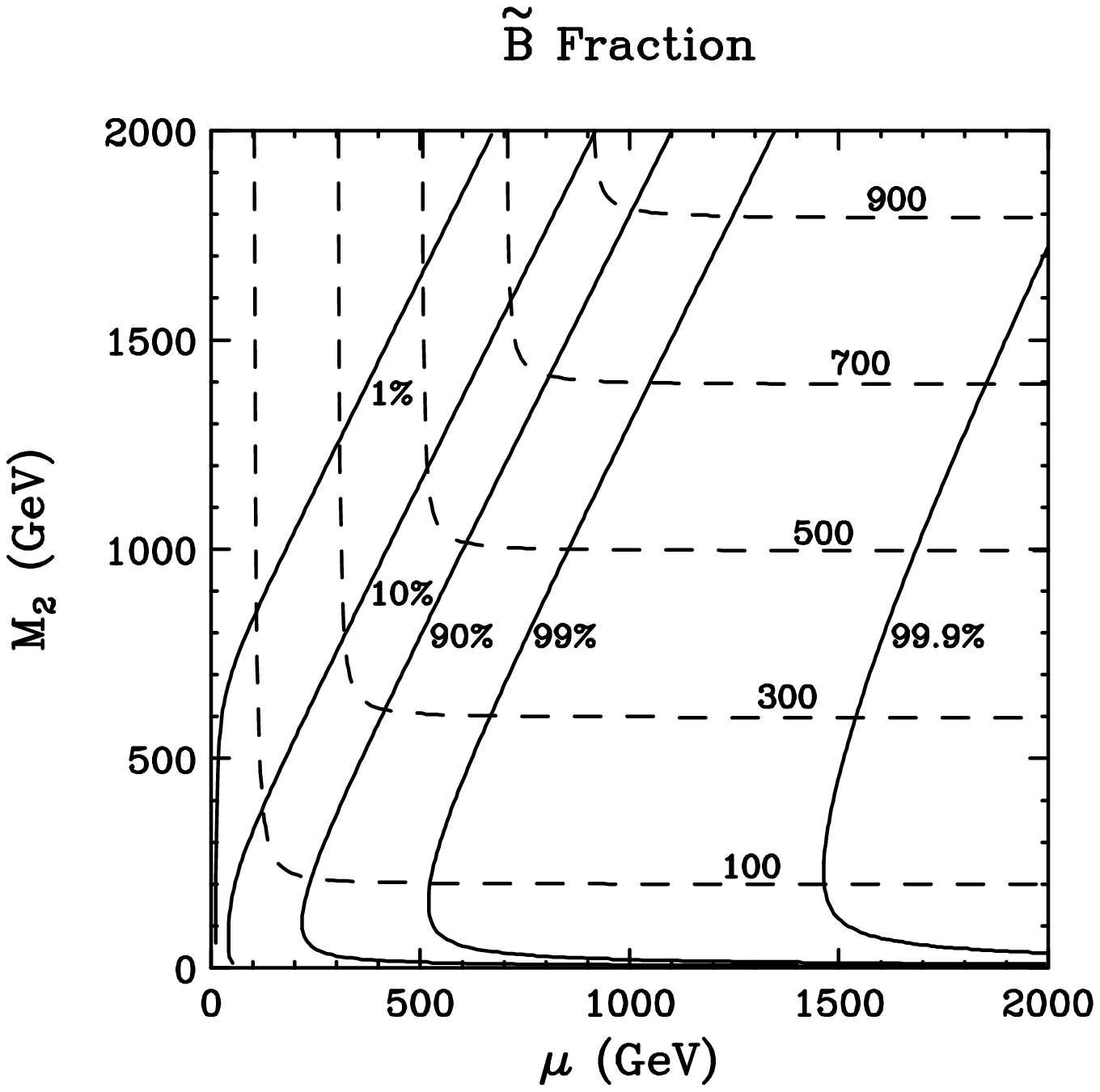}
\includegraphics[width=2.8in,angle=0]{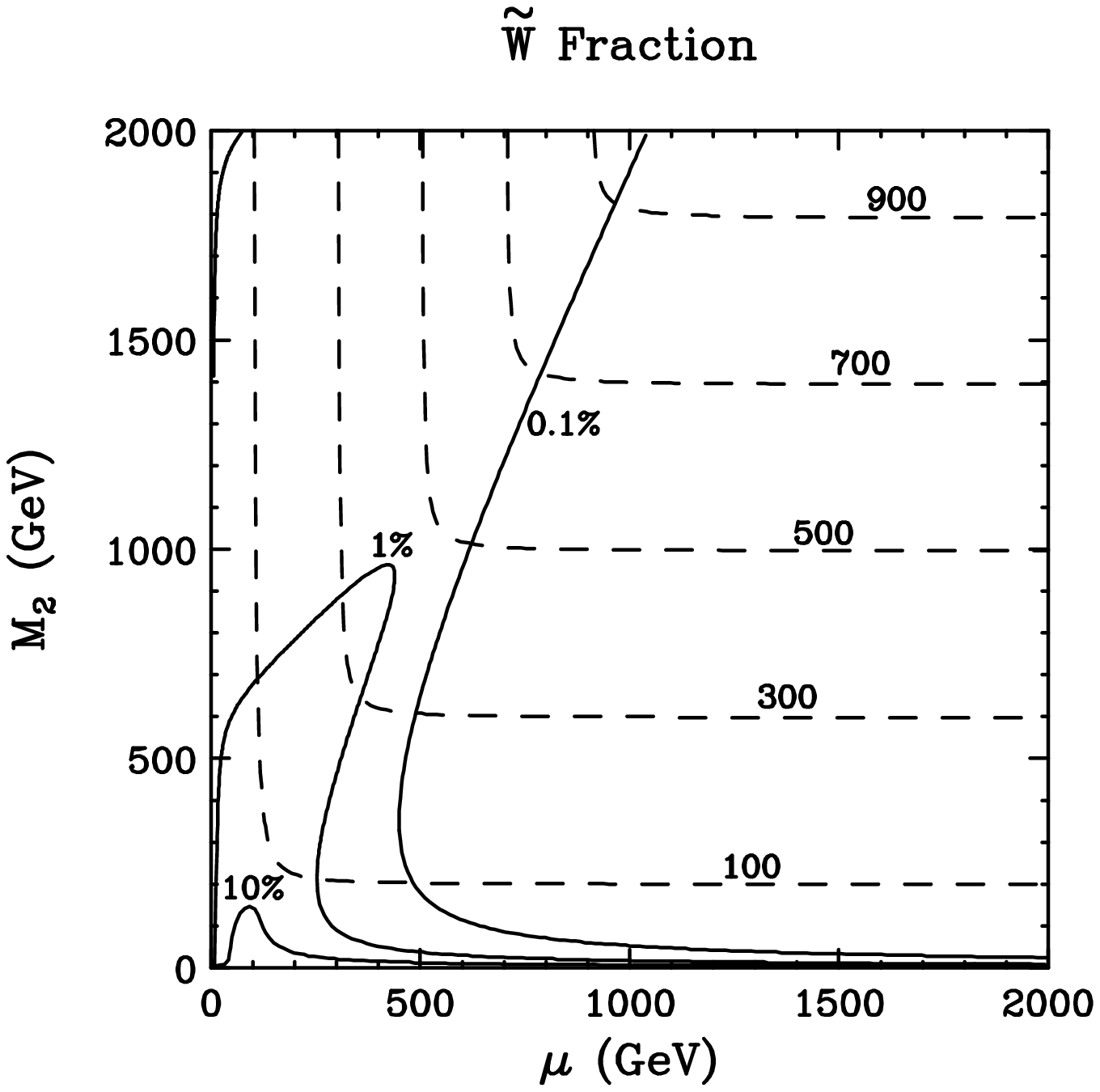}
\\
\includegraphics[width=2.8in,angle=0]{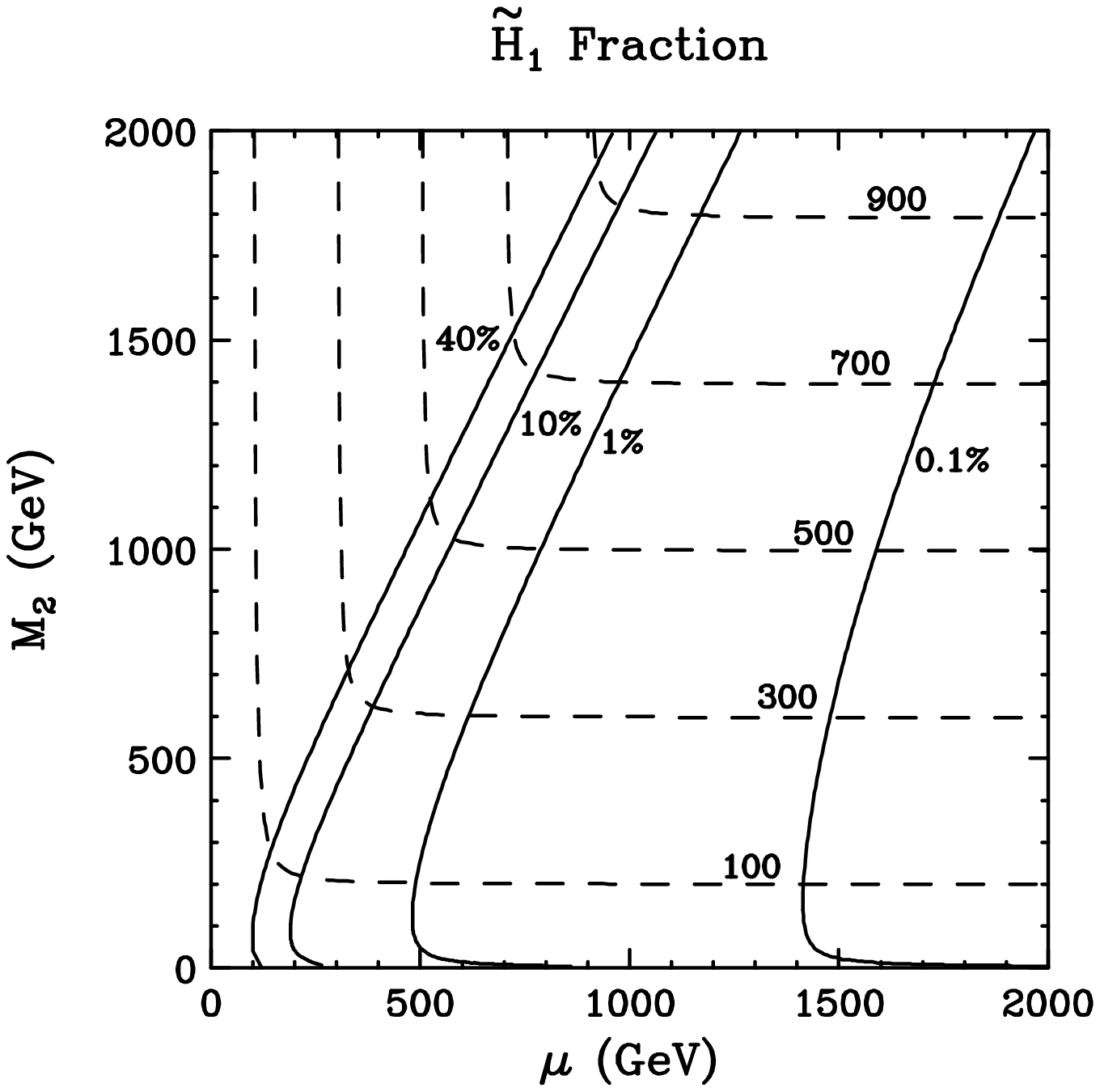}
\includegraphics[width=2.8in,angle=0]{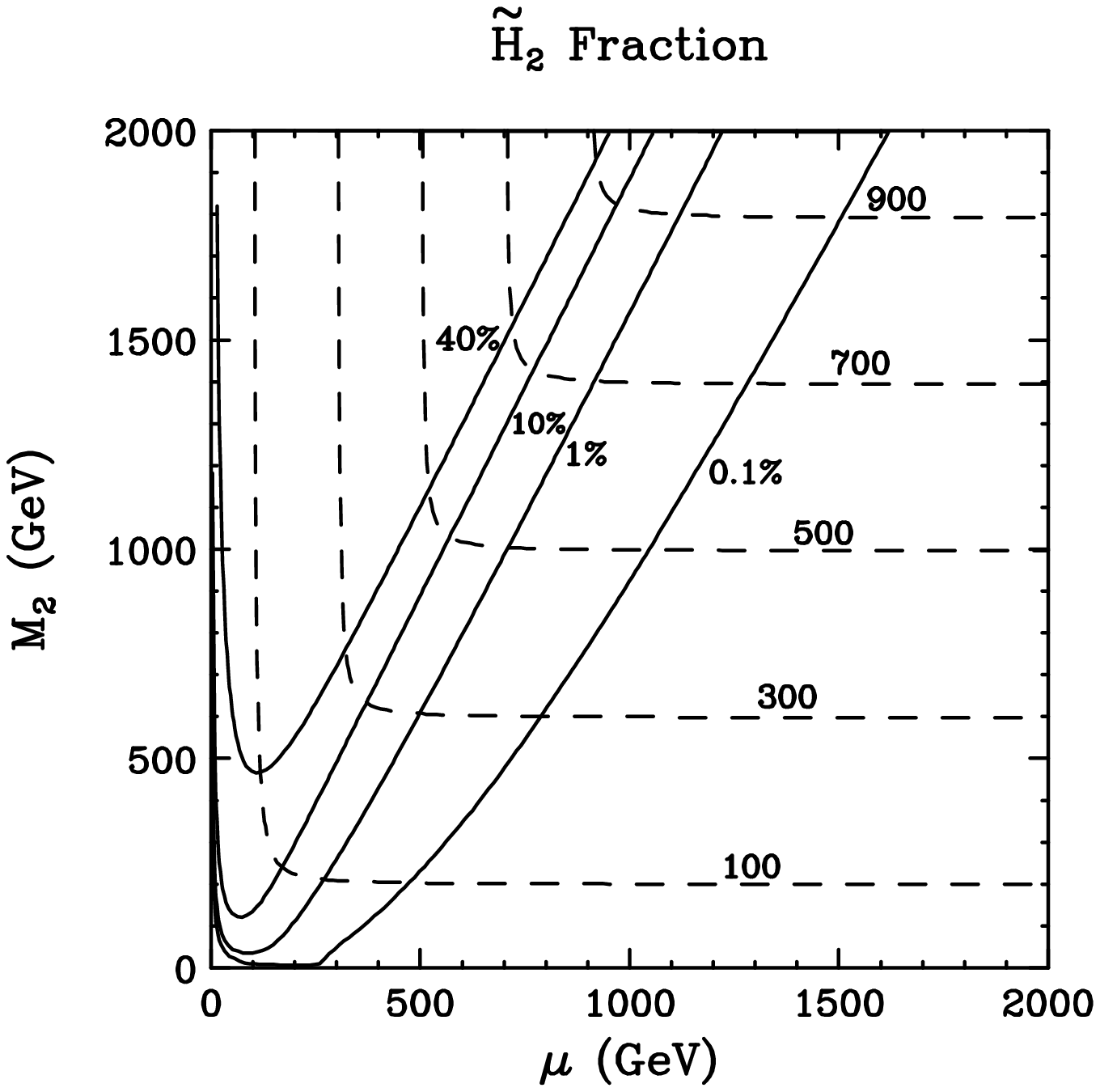}
\caption{Contours of constant bino, wino and higgsino fractions of the lightest neutralino in the $M_2$-$\mu$ plane. The GUT relation, $M_1 = \frac{5}{3} \tan^2\theta_W M_2$, has been adopted, and $\tan \beta=10$ has been used. Also shown as dashed lines are contours of constant neutralino mass, in GeV.}
\label{comp10}
\end{figure}

In the Minimal Supersymmetric Standard Model (MSSM), the neutral electroweak gauginos ($\tilde{B}, \tilde{W}$) and higgsinos ($\tilde{H_1}, \tilde{H_2}$) have the same quantum numbers and, therefore, mix into four mass eigenstates called neutralinos. The neutralino mass matrix in the $\widetilde{B}$-$\widetilde{W}$-$\widetilde{H}_1$-$\widetilde{H}_2$ basis is given by
\begin{equation}
\arraycolsep=0.01in
{\cal M}_N=\left( \begin{array}{cccc}
M_1 & 0 & -M_Z\cos \beta \sin \theta_W^{} & M_Z\sin \beta \sin \theta_W^{}
\\
0 & M_2 & M_Z\cos \beta \cos \theta_W^{} & -M_Z\sin \beta \cos \theta_W^{}
\\
-M_Z\cos \beta \sin \theta_W^{} & M_Z\cos \beta \cos \theta_W^{} & 0 & -\mu
\\
M_Z\sin \beta \sin \theta_W^{} & -M_Z\sin \beta \cos \theta_W^{} & -\mu & 0
\end{array} \right)\;,
\end{equation}
where $M_1$, $M_2$ and $\mu$ are the bino and wino masses, and the higgsino mass parameter, respectively. $\theta_W$ is the Weinberg angle, and $\tan \beta$ is the ratio of the vacuum expectation values of the up and down Higgs doublets. This matrix can be diagonalized into mass eigenstates by
\begin{equation}
M_{\chi^0}^{\rm{diag}} = N^{\dagger}  M_{\chi^0} N.
\end{equation}
In terms of the elements of the matrix, $N$, the lightest neutralino ($\chi^0_1$, or simply $\chi^0$) is the following mixture of gauginos and higgsinos:
\begin{equation}
\chi^0_1 = N_{11}\tilde{B}     +N_{12} \tilde{W}^3
          +N_{13}\tilde{H}_1 +N_{14} \tilde{H}_2.
\end{equation}
The absolute square of these coefficients are known as the bino, wino and higgsino fractions of the lightest neutralino, respectively. The composition of the lightest neutralino is a function of four supersymmetric parameters, $M_1$, $M_2$, $\mu$ and $\tan \beta$. This becomes further simplified if the gaugino masses are assumed to evolve to a single value at the scale of Grand Unification, as they often are. Requiring that $M_1$, $M_2$ and $M_3$ ($M_3$ being the gluino mass) evolve to the same value at the GUT scale yields the following ratios at the electroweak scale:
\begin{eqnarray}
M_1 &=& \frac{5}{3} \tan^2\theta_W M_2 \approx 0.5 M_2, \label{m1m2}\\
M_3 &=&  \frac{\alpha_S}{\alpha_{EM}} \sin^2 \theta_W  M_2 \approx 3.7 M_2. 
\end{eqnarray}
Adopting this relationship between $M_1$ and $M_2$, we plot the composition of the lightest neutralino as a function of $M_2$ and $\mu$ in figure~\ref{comp10} for the case of $\tan \beta =10$. In each frame, the solid lines represent contours of constant bino, wino or higgsino fraction, as labeled. The dashed lines represent contours of constant neutralino mass, in GeV. For $M_2/2 \ll \mu$ the lightest neutralino is almost entirely bino-like, whereas for $M_2/2 \gg \mu$ it is a nearly pure-higgsino, containing large quantities of both $\tilde{H_1}$ and $\tilde{H_2}$. As a result of $M_1$ always being significantly smaller than $M_2$, the lightest neutralino is never primarily wino-like. This, of course, is not always true if the gaugino masses do not obey the GUT relationship of Eq.~\ref{m1m2}.

\section{Neutralino Annihilations}

The annihilation cross section of the lightest neutralino is important for determining its thermal abundance in the universe, and for determining the fluxes of neutralino annihilation products that may be observed in indirect detection experiments. Furthermore, by studying future signals from indirect detection experiments, such as the spectrum of gamma-rays generated in the annihilations of neutralino dark matter, the annihilation modes of the lightest neutralino could (in principle) be identified. Neutralino annihilations which generate these signals take place in situations in which the neutralinos are traveling at velocities far below the speed of light. Therefore, we can focus here on the annihilation modes preferred by neutralinos in the low velocity limit. 

Neutralino annihilations can produce a large variety of final states, including pairs of fermions and pairs of gauge or Higgs bosons. We will briefly summarize the most relevant aspects of these processes here. For a more complete discussion of neutralino annihilations, see Ref.\cite{jungmanreview} and references therein.

\subsection{Neutralino Annihilations to Fermions}

Neutralinos can annihilate to fermion pairs through three classes of tree level Feynman diagrams: s-channel Higgs boson ($H_1$, $H_2$ or $A$) exchange, t-channel sfermion exchange, or s-channel $Z$ boson exchange. In the low velocity limit, the exchange of $A$, sfermions and $Z$ each contribute. The cross section resulting from Higgs exchange is given by:
\begin{equation}
\sigma_{\chi^0_1 \chi^0_1 \rightarrow A \rightarrow f\bar{f}} \, v \, (v \rightarrow 0) = \sum_{f} \frac{4 G^2_F c_f m^2_{\chi^0_1} m^2_W m^2_f  \tan^2 \beta \, C^2_{\chi^0_1 \chi^0_1 A} (1-m^2_f/m^2_{\chi^0_1})^{1/2}}{\pi [(4 m^2_{\chi^0_1}- m^2_A)^2 + m^2_A \Gamma^2_A]}, 
\end{equation}
where the sum is over fermion species, $c_f$ is a color factor (3 for quarks, 1 for leptons), $m_A$ is the CP-odd Higgs mass, and $\Gamma_A$ is its width. The expression shown is valid in the case of down-type fermions. For up-type fermions in the final state, the $\tan^2 \beta$ should be replaced by $\cot^2 \beta$. The neutralino-neutralino-Higgs coupling is given by, $C_{\chi^0_1 \chi^0_1 A} \approx (g_2 N_{12} - g_1 N_{11}) (N_{14} \cos \beta - N_{13} \sin \beta)$. This contribution to the annihilation cross section is most large in the case of a mixed gaugino-higgsino with large $\tan \beta$ and with a mass near the $A$-pole ($m_A \sim 2 m_{\chi^0_1}$), known as the funnel region.

The contribution from sfermion exchange is given by:
\begin{equation}
\sigma_{\chi^0_1 \chi^0_1 \rightarrow \tilde{f} \rightarrow f\bar{f}} \, v \, (v \rightarrow 0) = \sum_{f} \frac{4 G^2_F c_f m^2_{\chi^0_1} m^4_W \, C^2_{\tilde{f}} (1-m^2_f/m^2_{\chi^0_1})^{1/2}}{\pi [m^2_{\chi^0_1}(1-m^2_f/m^2_{\chi^0_1}) + m^2_{\tilde{f}}]^2}, 
\end{equation}
where, in the small $N_{12}$ limit, $C_{\tilde{f}} \approx (m_f N_{11} N_{13}^* \tan \theta_W/2 m_W \cos\beta) + (m^3_f |N_{13}|^2/2 m^2_W m_{\chi^0_1} \cos^2 \beta) + m_f \tan^2 \theta_W |N_{11}|^2 (0.25+2 e^2_f + e_f)$ for down-type fermions, and $C_{\tilde{f}} \approx (-m_f N_{11} N_{14}^* \tan \theta_W/2 m_W \sin\beta) + (m^3_f |N_{14}|^2/2 m^2_W m_{\chi^0_1} \sin^2 \beta) + m_f \tan^2 \theta_W |N_{11}|^2 (0.25+2 e^2_f - e_f)$ for up-type fermions. $e_f$ is the charge of the fermion. Unlike in the case of $A$ exchange, this process can contribute significantly for either a mixed bino-higgsino or a pure bino.

The contribution from $Z$ boson exchange is given by:
\begin{equation}
\sigma_{\chi^0_1 \chi^0_1 \rightarrow Z \rightarrow f\bar{f}} \, v \, (v \rightarrow 0) =  \sum_{f} \frac{G^2_F c_f m^2_f \, (|N_{13}|^2-|N_{14}|^2)^2 (1-m^2_f/m^2_{\chi^0_1})^{1/2}}{4 \pi},  
\end{equation}
which scales simply as the square of the difference of the two higgsino fractions, $(|N_{13}|^2-|N_{14}|^2)^2$.

Before moving on to neutralino annihilations to bosonic final states, a few comments are in order. Firstly, all annihilation channels to fermions strongly prefer heavy fermions in the low velocity limit ($\sigma v \propto m^2_f$). Secondly, interference terms between each of these three channels can be important and, in some cases, dominate (see Ref.~\cite{griest} for the complete cross sections, including interference terms).

\subsection{Neutralino Annihilations to Gauge Bosons}

Neutralinos can annihilate to $W^+ W^-$ though t-channel chargino exchange, as well as s-channel exchange of a $Z$ boson or CP-even Higgs boson. The diagrams involving $Z$ and Higgs bosons do not contribute in the low velocity limit, however. The chargino exchange diagrams result in an annihilation cross section given by:
\begin{eqnarray}
&&\sigma_{\chi^0_1 \chi^0_1 \rightarrow W^+ W^-} \, v \, (v \rightarrow 0) =  \frac{4 G^2_F m^7_W ((m^2_{\chi^0_1}/m^2_{W})-1)^{3/2}}{\pi m_{\chi^0_1}} \nonumber  \\
&\times& \bigg[\frac{(N_{13} \sin \phi_-/\sqrt{2} +N_{12} \cos \phi_-)^2 + (-N_{14} \sin \phi_+/\sqrt{2} +N_{12} \cos \phi_+)^2}{m^2_W - m^2_{\chi^0_1} - m^2_{\chi^+_1}}  \nonumber \\
&+& \frac{(N_{13} \cos \phi_-/\sqrt{2} -N_{12} \sin \phi_-)^2 + (-N_{14} \cos \phi_+/\sqrt{2} -N_{12} \sin \phi_+)^2}{m^2_W - m^2_{\chi^0_1} - m^2_{\chi^+_2}}\bigg]^2,  
\end{eqnarray}
where $\phi_+$ and $\phi_-$ are the angles appearing in the chargino mixing matrices. 

Neutralino annihilations to $ZZ$ can take place through t-channel neutralino exchange and s-channel Higgs exchange. Again, the s-channel processes do not contribute in the low velocity limit. The neutralino exchange diagrams yield:
\begin{eqnarray}
\sigma_{\chi^0_1 \chi^0_1 \rightarrow ZZ} \, v \, (v \rightarrow 0) =  \frac{8 G^2_F m^7_Z ((m^2_{\chi^0_1}/m^2_{Z})-1)^{3/2}}{\pi m_{\chi^0_1} \cos^4 \theta_W} \times \bigg[ \sum_{i=1,4}  \frac{N_{i3} N_{13}^* -N_{i4} N_{14}^*}{m^2_Z - m^2_{\chi^0_1} - m^2_{\chi^0_i}}  \bigg]^2. 
\end{eqnarray}
Note that neither of these expressions contain the bino-content of the neutralino. Both channels to $W^+ W^-$ and $ZZ$ are most important for neutralinos with a large higgsino component (or a large wino component in the case of $W^+ W^-$).

\subsection{Neutralino Annihilations to Final States Including Higgs Bosons}

Neutralinos can also annihilate to pairs of Higgs bosons, or to a Higgs boson along with a $W^{\pm}$ or $Z$ boson. Possible final states include $H_1 H_1$, $H_1 H_2$, $H_2 H_2$, $AA$, $A H_1$, $A H_2$, $H^+ H^-$, $Z H_1$, $Z H_2$, $Z A$ and $W^{\pm} H^{\mp}$, although processes to $H_1 H_1$, $H_1 H_2$, $H_2 H_2$, $AA$, $Z A$ and  $H^+ H^-$ do not contribute in the low velocity limit. Of the contributing final states, those including a Higgs and a gauge boson ($Z H_1$, $Z H_2$ and $W^{\pm} H^{\mp}$) are generally more important than Higgs-only final states, so we focus on those here.

Neutralino annihilations to $Z H_1$ and $Z H_2$ can proceed through the t-channel exchange of a neutralino, or the s-channel exchange of $Z$ or $A$, all of which contribute in the low velocity limit. Similarly, neutralinos can annihilate to $W^{\pm} H^{\mp}$ through t-channel chargino exchange, or the s-channel exchange of $H_2$, $H_1$ or $A$, of which chargino and $A$ diagrams contribute in the low velocity limit.

The cross sections to these final states are larger for higgsino-like neutralinos, but decrease more slowly with increasing bino-fraction than annihilations to $ZZ$ or $W^+ W^-$. In particular, annihilations through an $A$ to $Z H_1$ and $Z H_2$ are proportional to $|N_{13}|^2$ rather than some combination of $N_{13}$ and $N_{14}$ to the fourth power, as is the case for annihilations to $ZZ$ or $W^+ W^-$. For this reason, annihilations to $Z H_1$, $Z H_2$ and $W^{\pm} H^{\mp}$ dominate over those to $ZZ$ or $W^+ W^-$ for neutralinos with a substantial bino fraction.

As the expressions for neutralino annihilations to final states including Higgs bosons are very lengthy, we do not reproduce them here. For further details, we refer the reader to Ref.~\cite{jungmanreview} and references therein.

\subsection{Neutralino Annihilations to $\gamma \gamma$ and $\gamma Z$}

Although neutralinos do not annihilate at tree level to $\gamma \gamma$ or $\gamma Z$, a large number of one-loop diagrams contribute to these processes~\cite{lines}. Although the cross sections for annihilations to these final states are considerably smaller than to heavy fermions, gauge or Higgs bosons, they are important because they produce mono-energetic gamma-ray lines, potentially providing a clear and distinctive signature in indirect detection experiments. Neutralino annihilations to $\gamma \gamma$ and $\gamma Z$ result in gamma-ray lines with energies of $E_{\gamma} = m_{\chi^0_1}$ and $E_{\gamma} = m_{\chi^0_1}-(m^2_Z/4 m_{\chi^0_1})$, respectively.

Although the expressions for the cross sections to these final states are too long to reproduce here, we will briefly point out a few common features. In most cases, the largest contributions come from diagrams with a chargino-$W^{\pm}$ loop. Each term in the amplitude from these processes is proportional to one of the higgsino fractions ($|N_{13}|^2$, $|N_{14}|^2$), to the wino-fraction ($|N_{12}|^2$), or to some combination of these three ($N_{13} N_{14}^*$, etc.). The bino content of the lightest neutralino does not contribute to this class of diagrams. Roughly speaking, we expect the annihilation cross section to gamma-ray lines to scale with the square of the higgsino fraction (or possibly the wino fraction) of the lightest neutralino. Even a pure-higgsino or pure-wino, however, has an annihilation cross section to $\gamma \gamma$ and $\gamma Z$ of around or less than $10^{-28}$ cm$^3$/s, which is less than 1\% of the value needed to generate the observed thermal abundance. Lines are, therefore, always a small fraction of neutralino annihilation final states.

In addition to diagrams with a chargino-$W^{\pm}$ loop, diagrams involving a sfermion-fermion loop can generate the leading contribution to gamma-ray lines, in particular in the case of light sfermions.

\begin{figure}[tbp]
\includegraphics[width=2.3in,angle=-90]{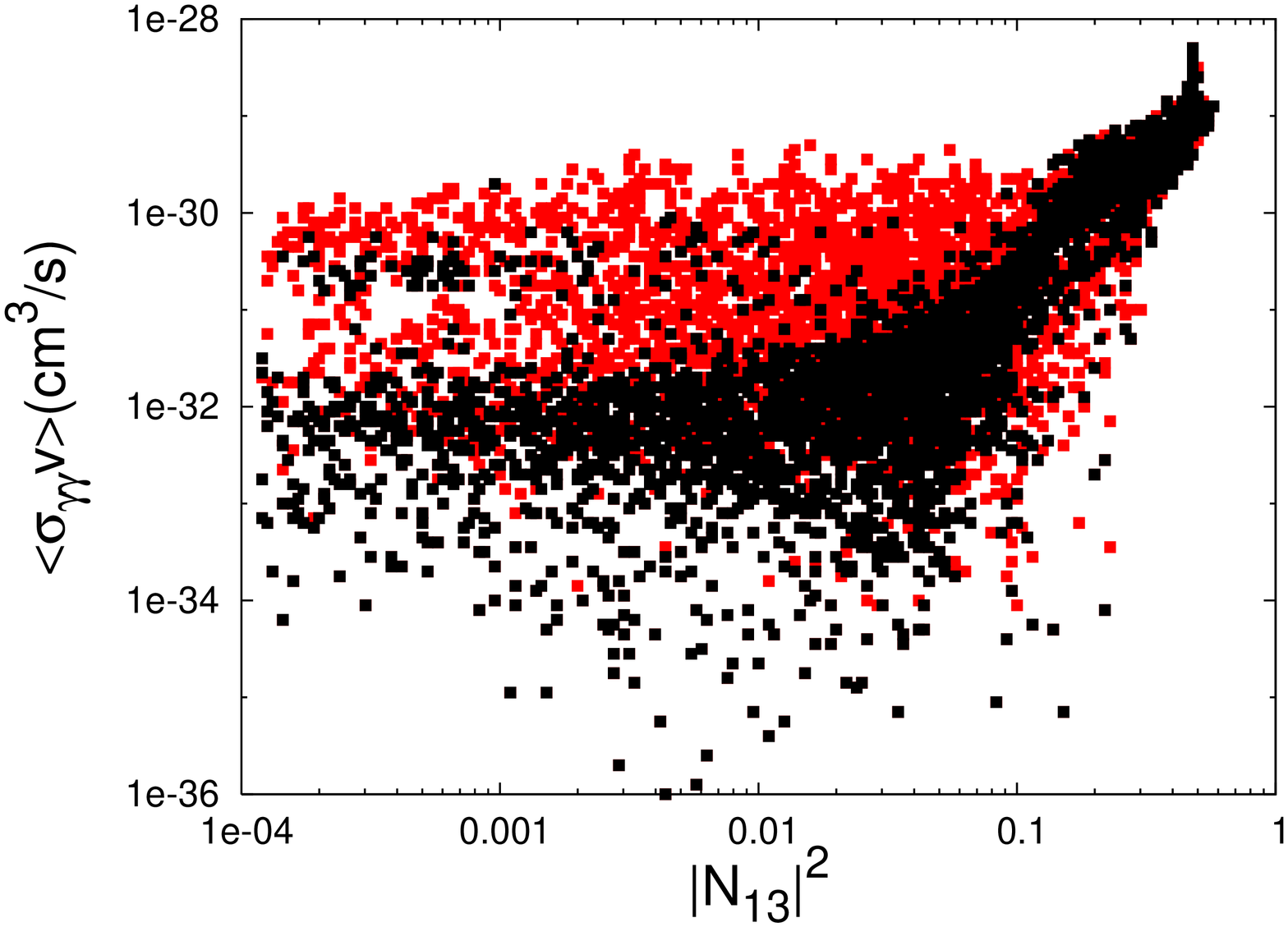}
\includegraphics[width=2.3in,angle=-90]{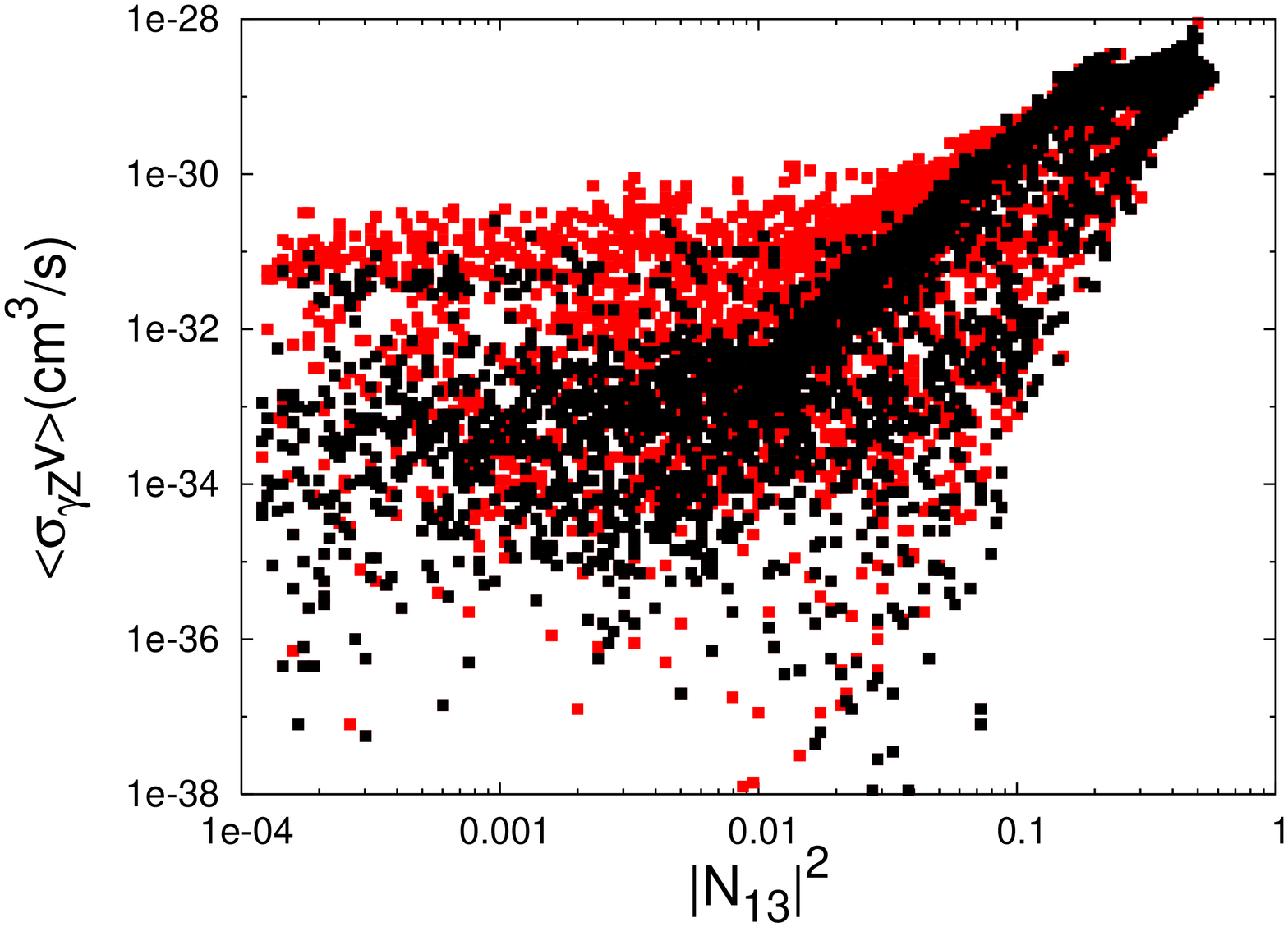}
\\
\includegraphics[width=2.3in,angle=-90]{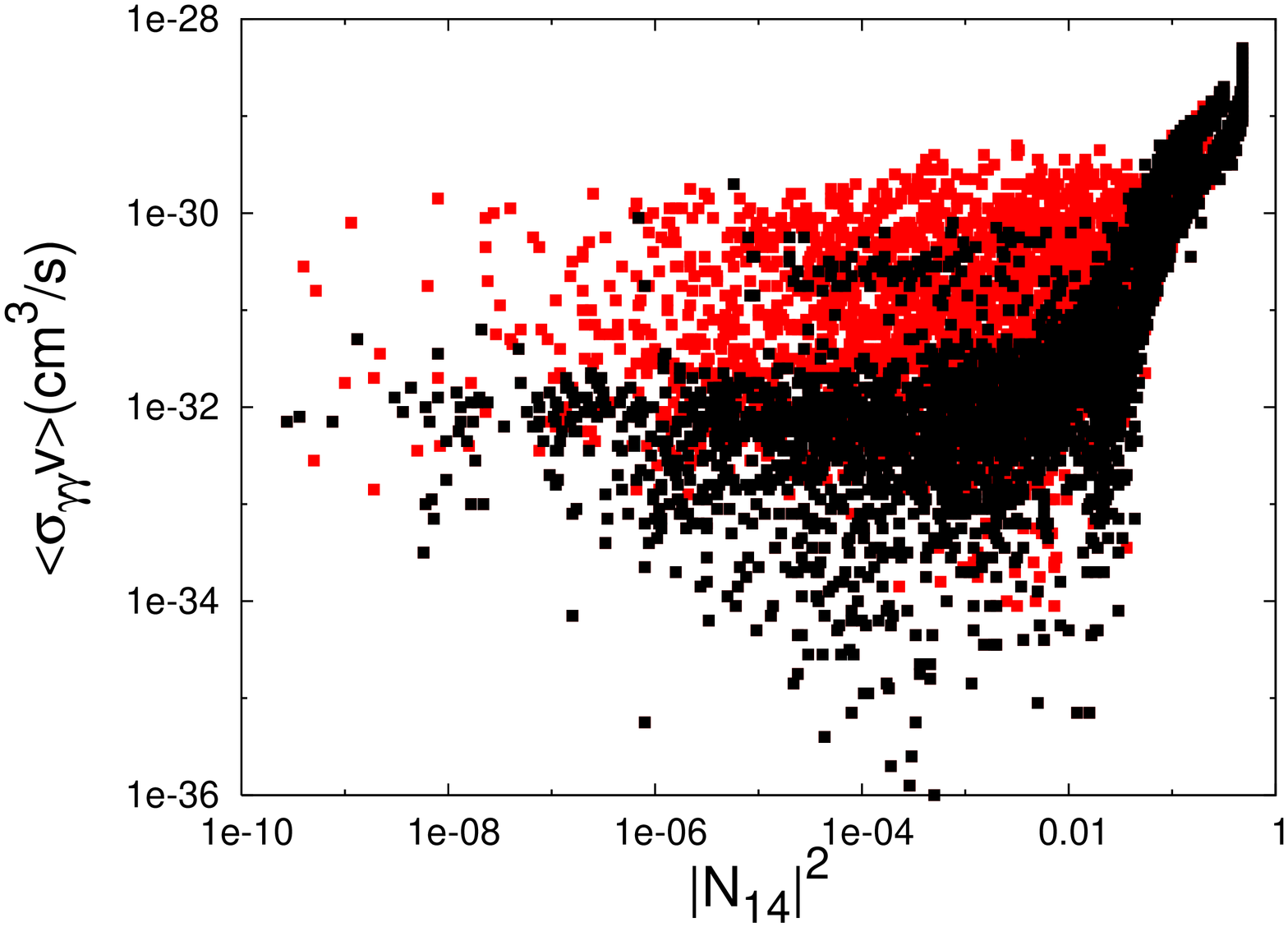}
\includegraphics[width=2.3in,angle=-90]{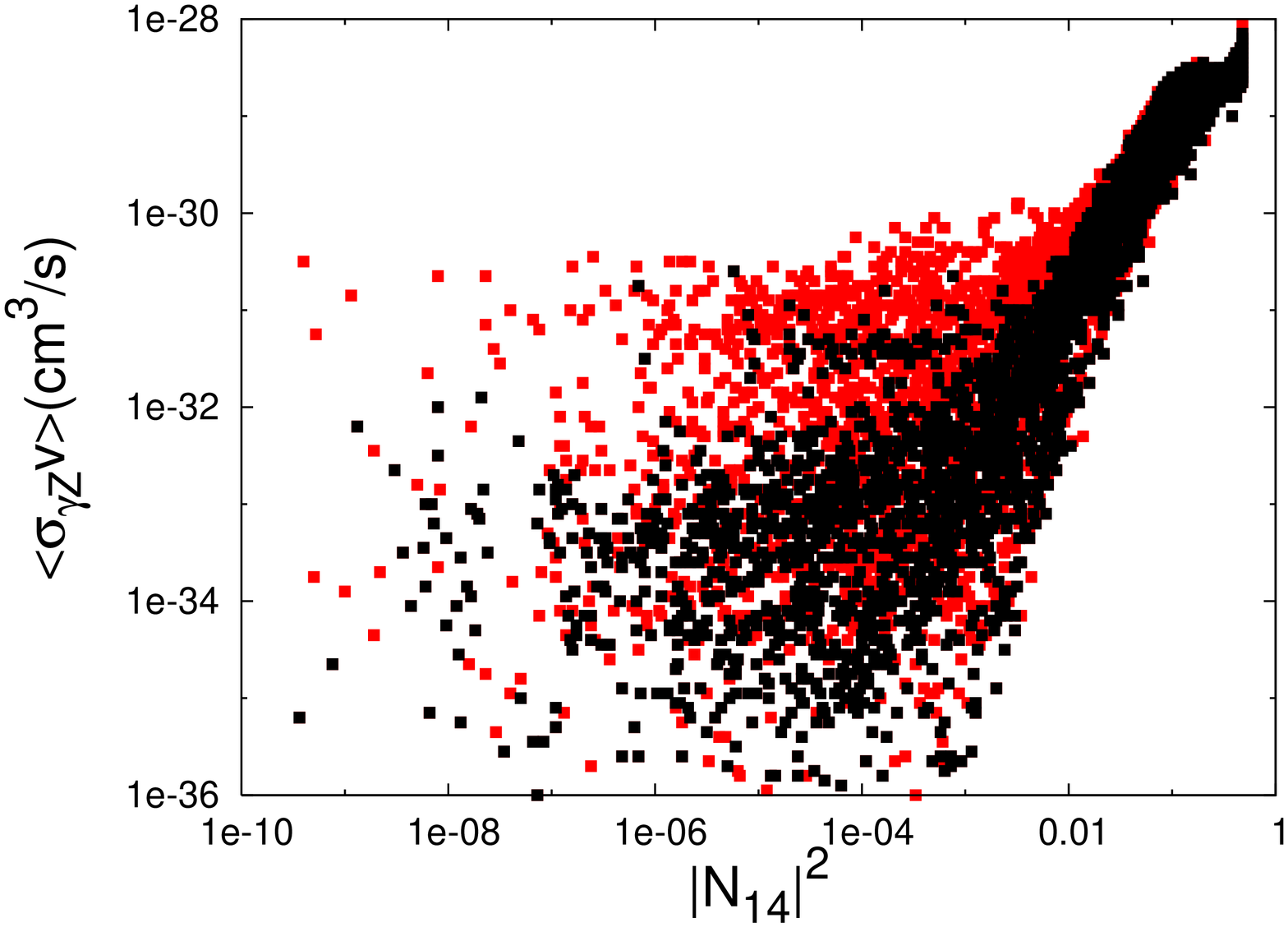}
\\
\includegraphics[width=2.3in,angle=-90]{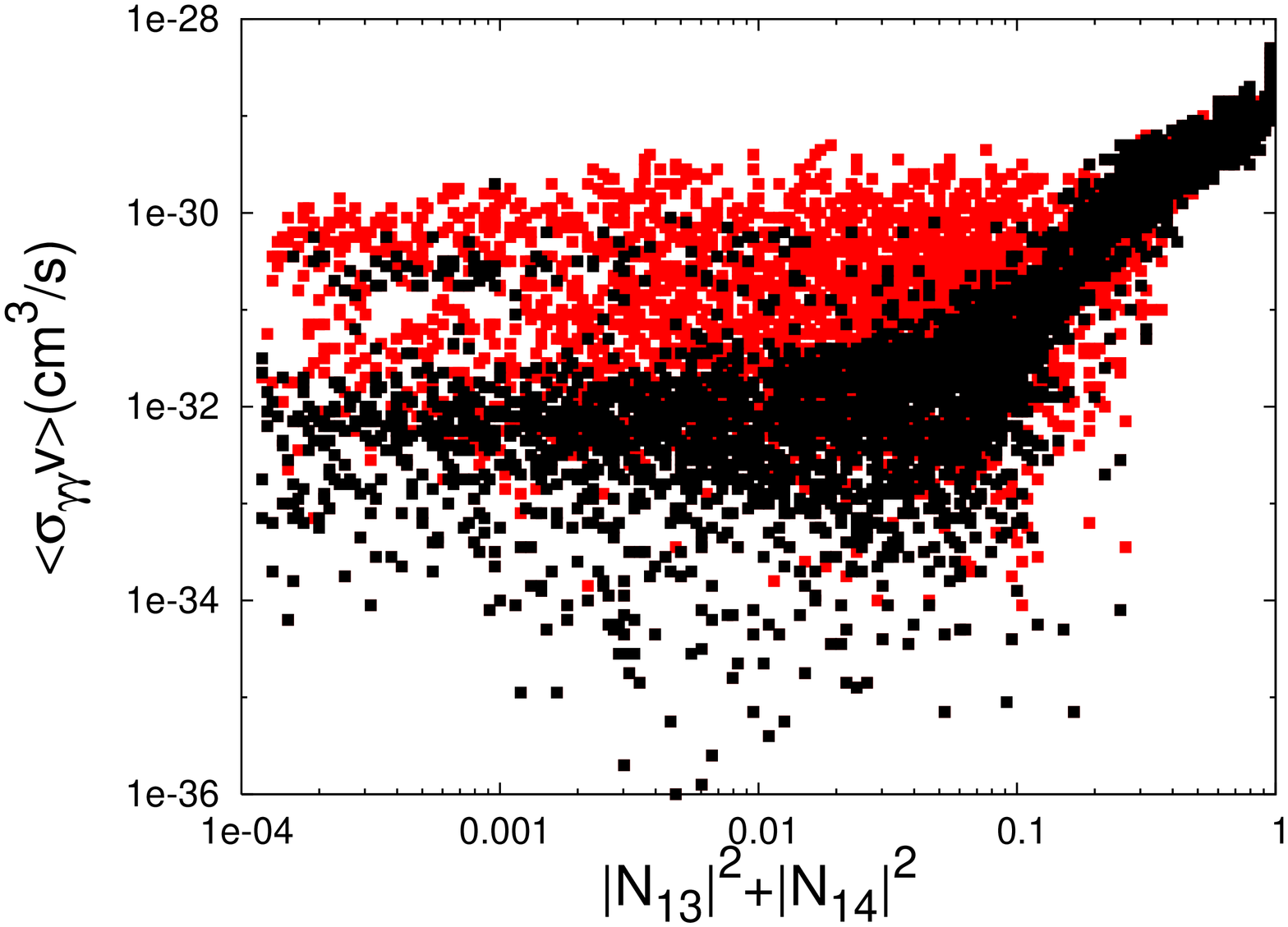}
\includegraphics[width=2.3in,angle=-90]{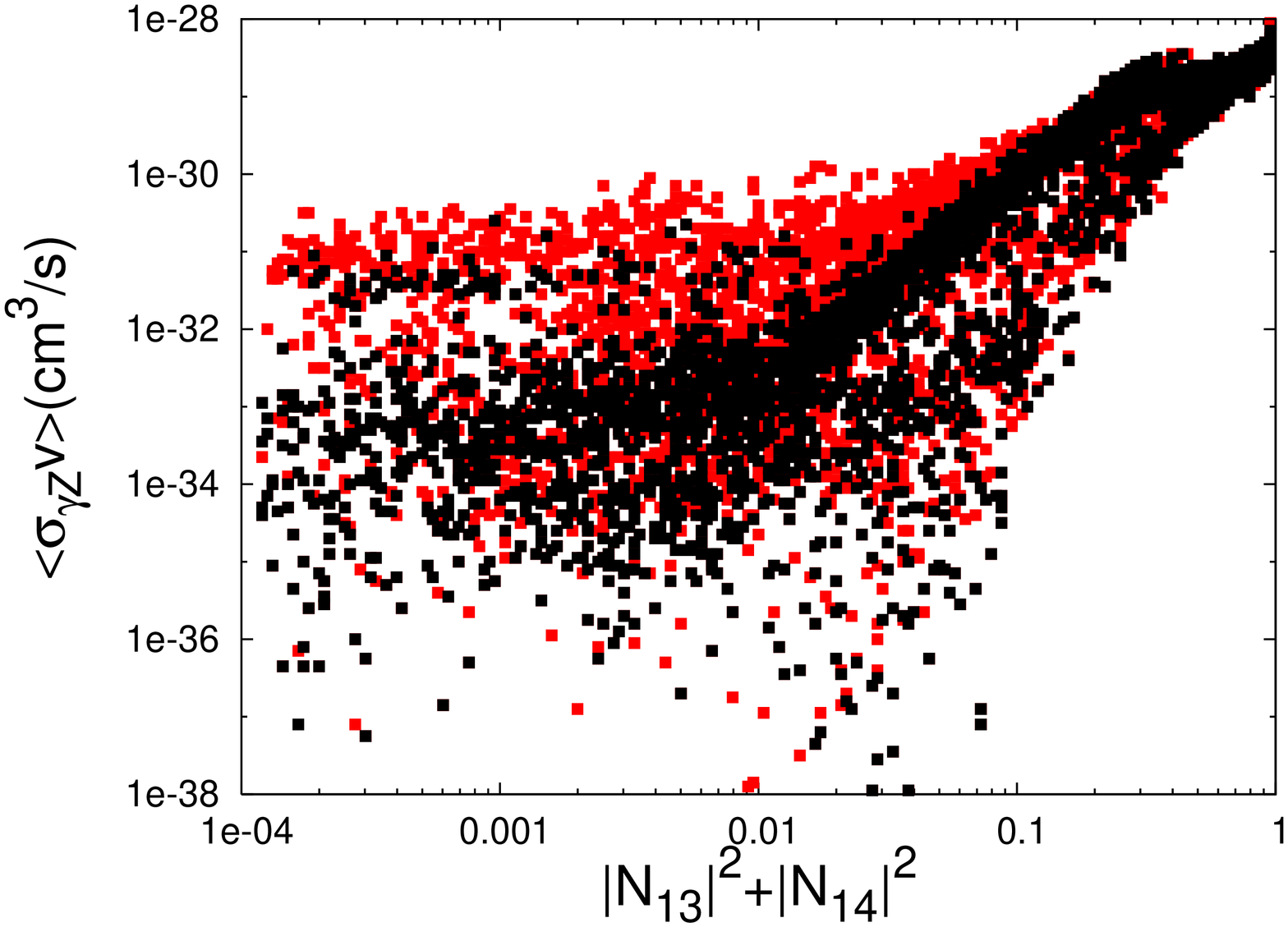}
\\
\caption{The neutralino annihilation cross section to $\gamma \gamma$ and $\gamma Z$ final states as a function of its higgsino content, $|N_{13}|^2$, $|N_{14}|^2$ and  $|N_{13}|^2+|N_{14}|^2$. Dark (black) points represent models with heavy squarks ($m_{\tilde{q}}> 1$ TeV), while lighter (red) points have no such constraint. Each point shown represents a set of parameters within the MSSM that is not in violation of direct collider constraints and that generates a thermal relic density within the $2\sigma$ range measured by WMAP ($0.119 > \Omega_{\chi^0} > 0.0787$)\cite{wmap}. The GUT relationship between the gaugino masses has been adopted.}
\label{linesplot}
\end{figure}

In Fig.~\ref{linesplot}, we plot the neutralino's annihilation cross section to gamma-ray lines (in the low velocity limit) as a function of higgsino composition. In models with heavy squarks (black points each represent models with $m_{\tilde{q}}> 1$ TeV), these cross sections are almost always dominated by $W^{\pm}$-chargino loop diagrams, and therefore a strong correlation emerges between the higgsino fractions and the cross sections to $\gamma \gamma$ and $\gamma Z$. This correlation is less strong among models with light squarks. These results, and the results shown in the other scatter plots included in this paper, were calculated using the DarkSUSY package \cite{darksusy}.

\subsection{Neutralino Annihilation Summary}

\begin{figure}[tbp]
\includegraphics[width=3.2in,angle=0]{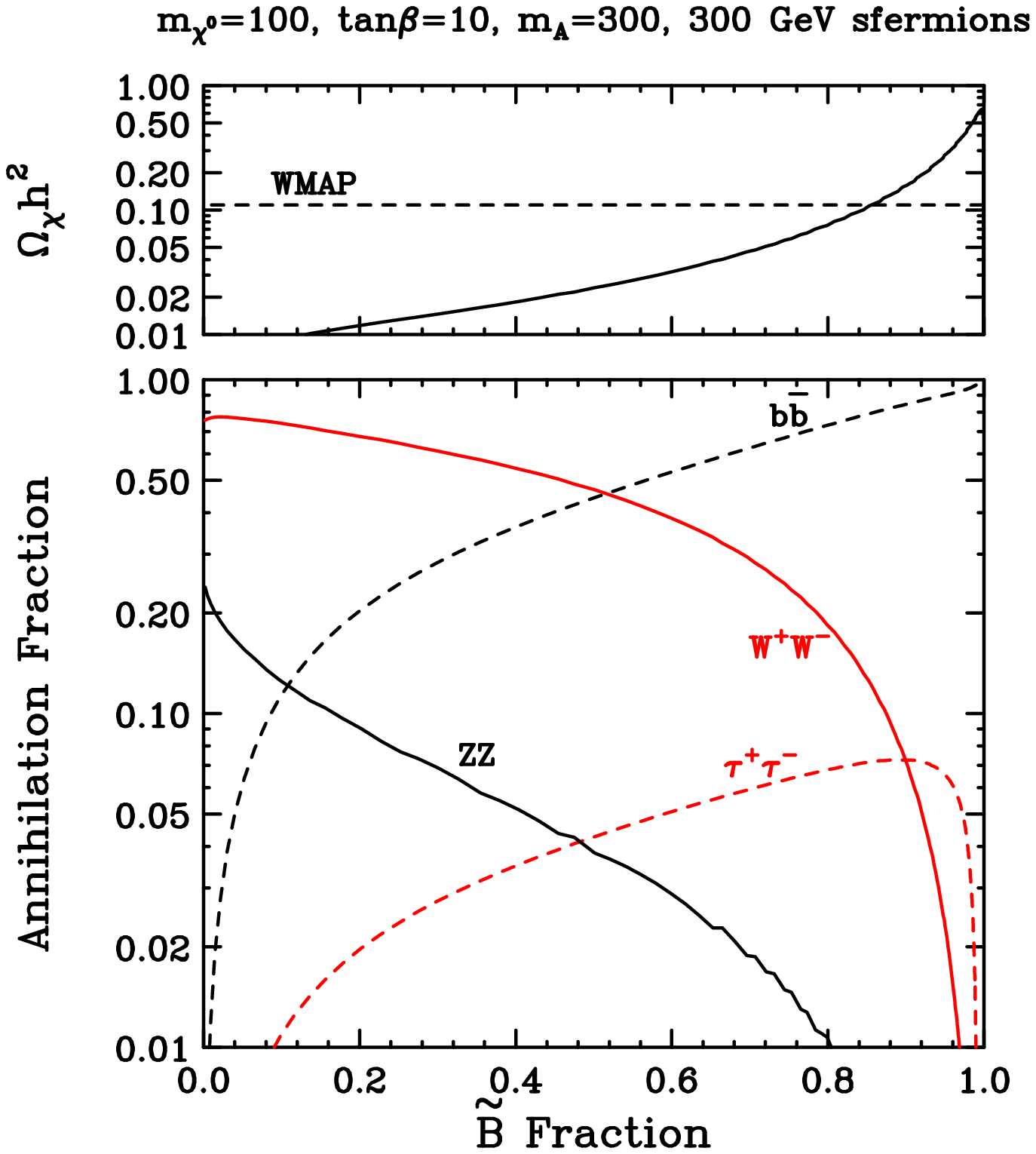}
\includegraphics[width=3.2in,angle=0]{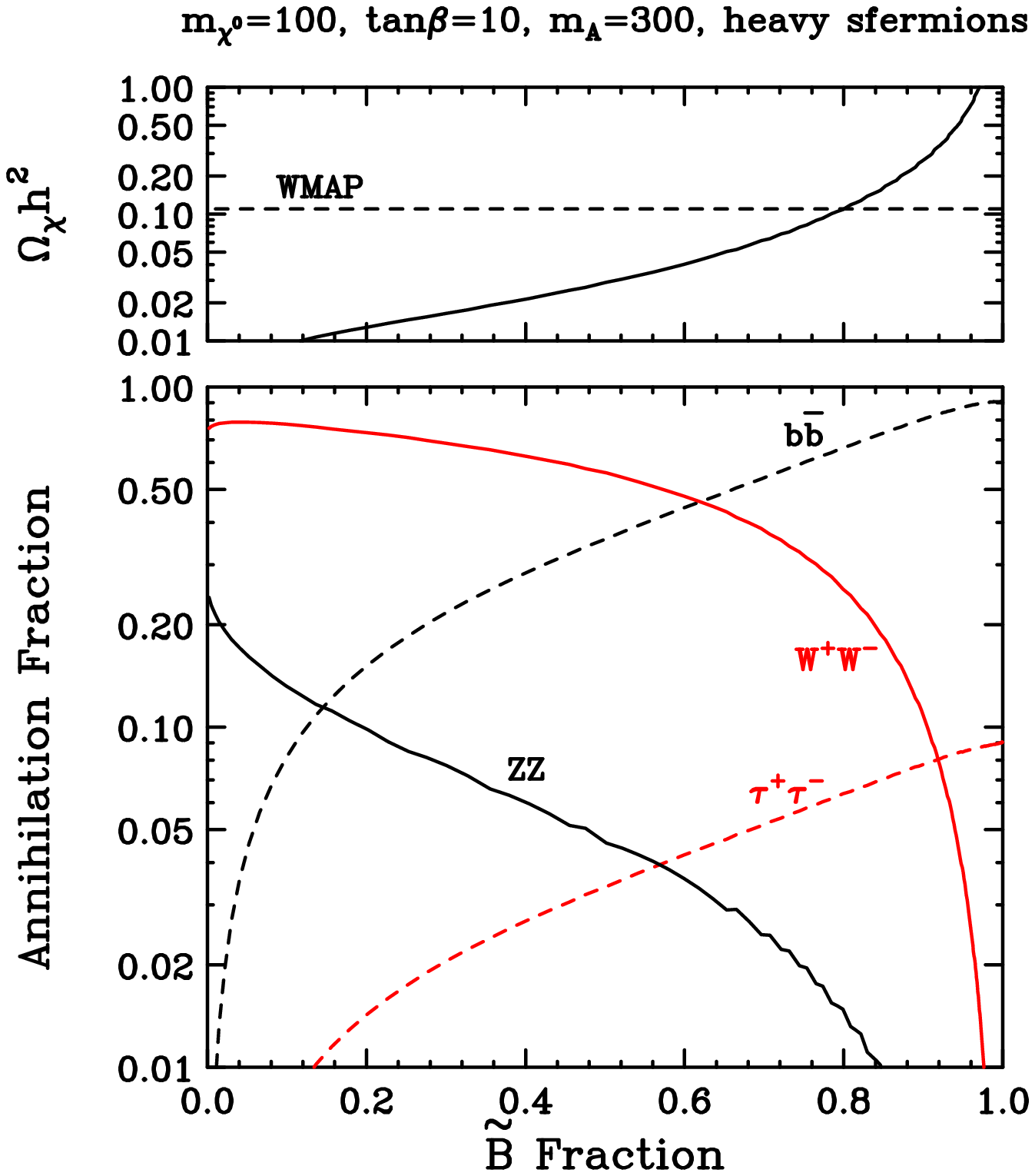}
\\
\includegraphics[width=3.2in,angle=0]{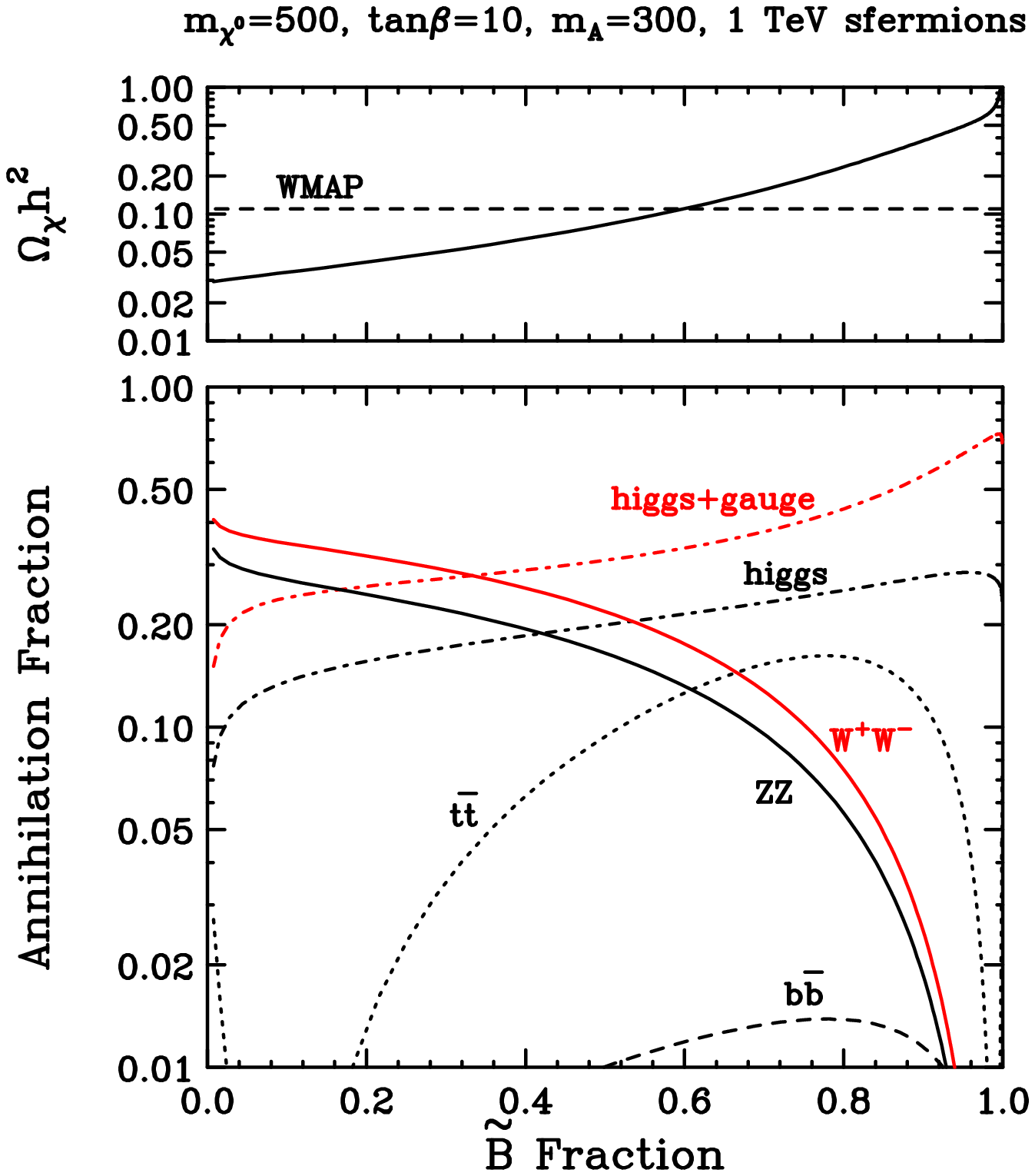}
\includegraphics[width=3.2in,angle=0]{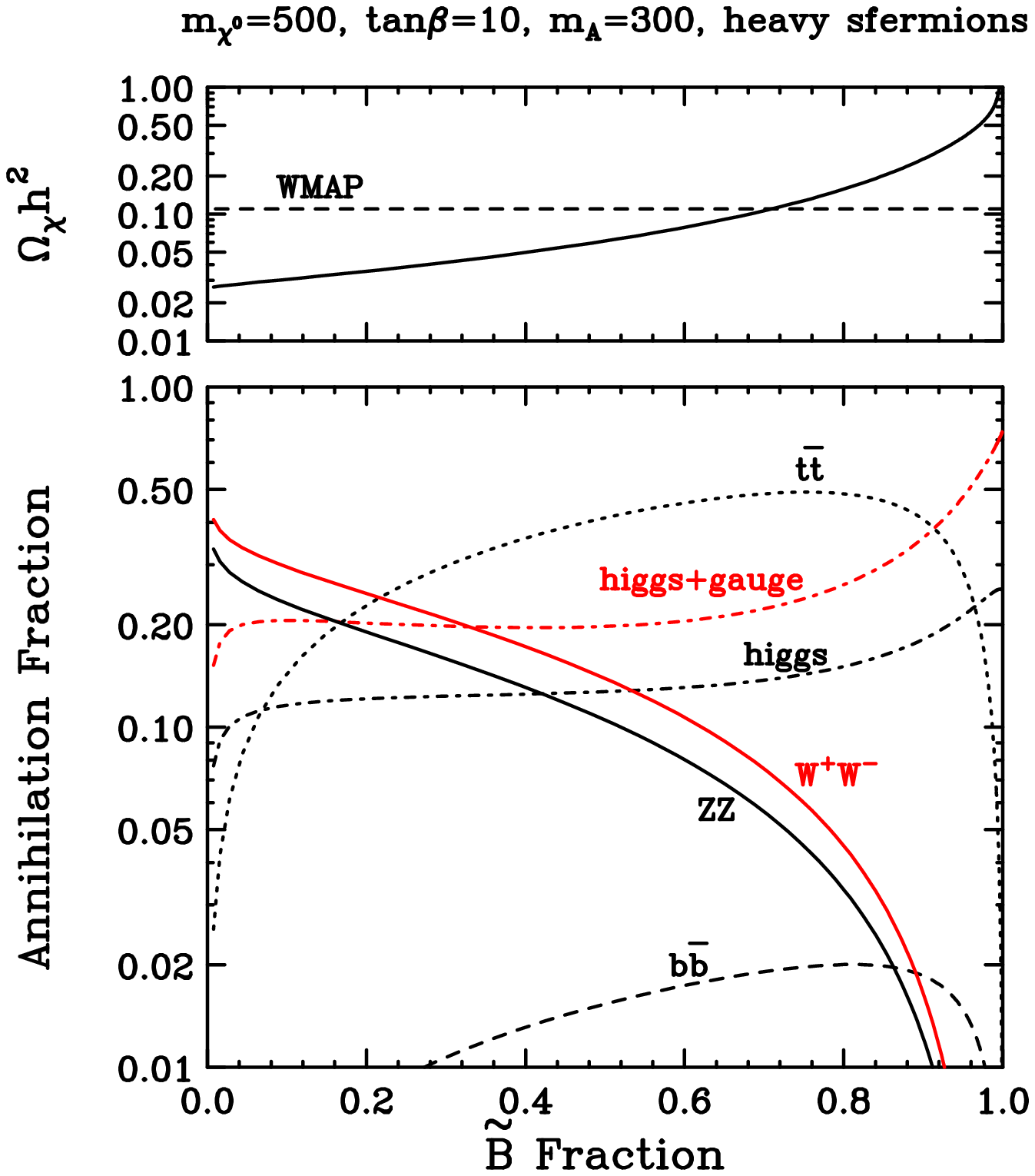}
\caption{The dominant neutralino annihilation modes for $m_{\chi^0_1}=$100 and 500 GeV, as a function of its bino content. $\tan \beta$ has been set to 10. The GUT relationship between $M_1$ and $M_2$ has been used, and in each frame $\tan\beta=10$ and $m_A$=300 GeV. In the left frames, 300 GeV and 1 TeV sleptons and squarks have been used, while in the right frame they have been made very heavy for illustration. Also shown above each frame is the relic abundance of the lightest neutralino compared to the quantity measured by WMAP. See the text for more details.}
\label{ann10}
\end{figure}

\begin{figure}[t]
\includegraphics[width=3.2in,angle=0]{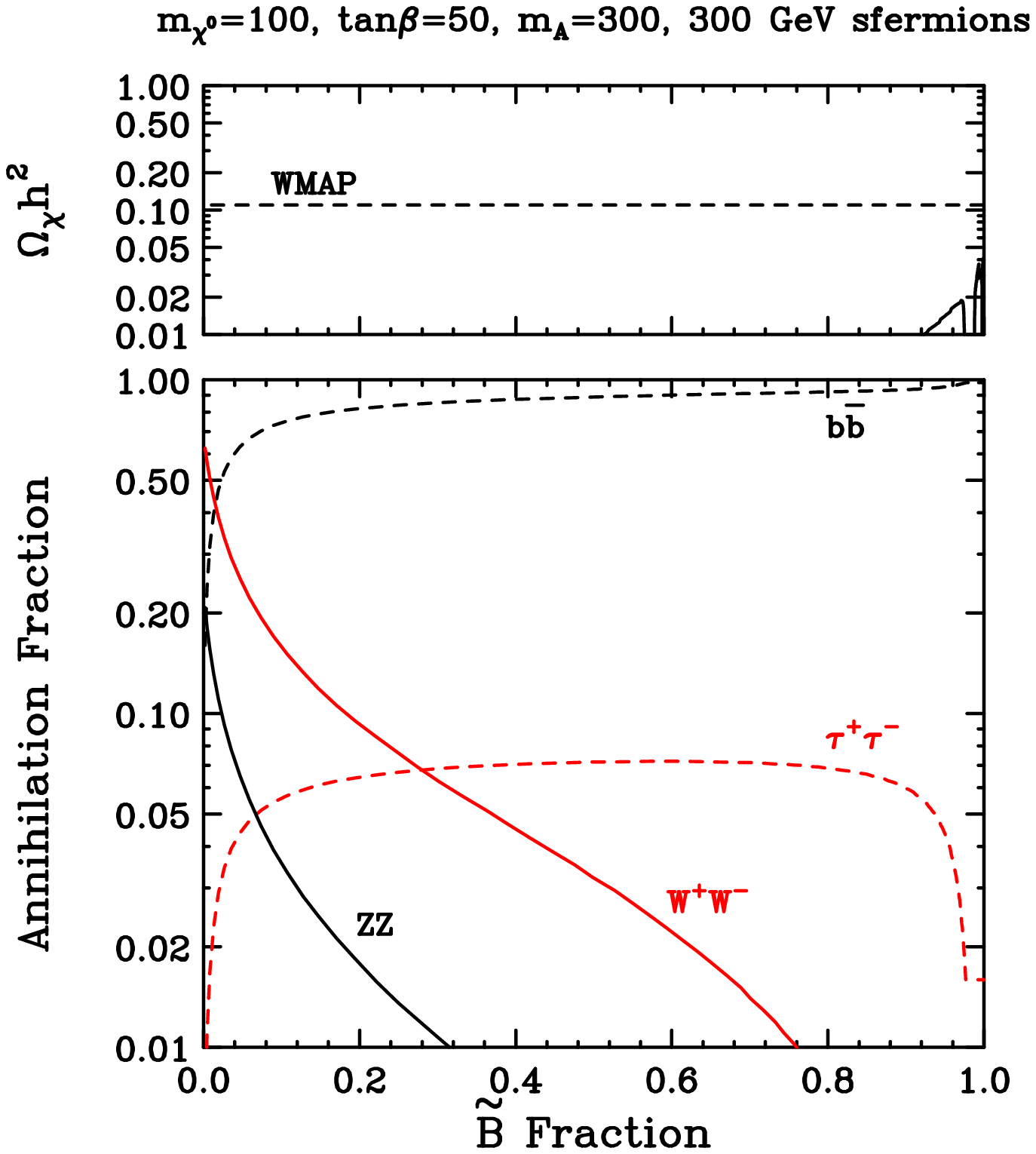}
\includegraphics[width=3.2in,angle=0]{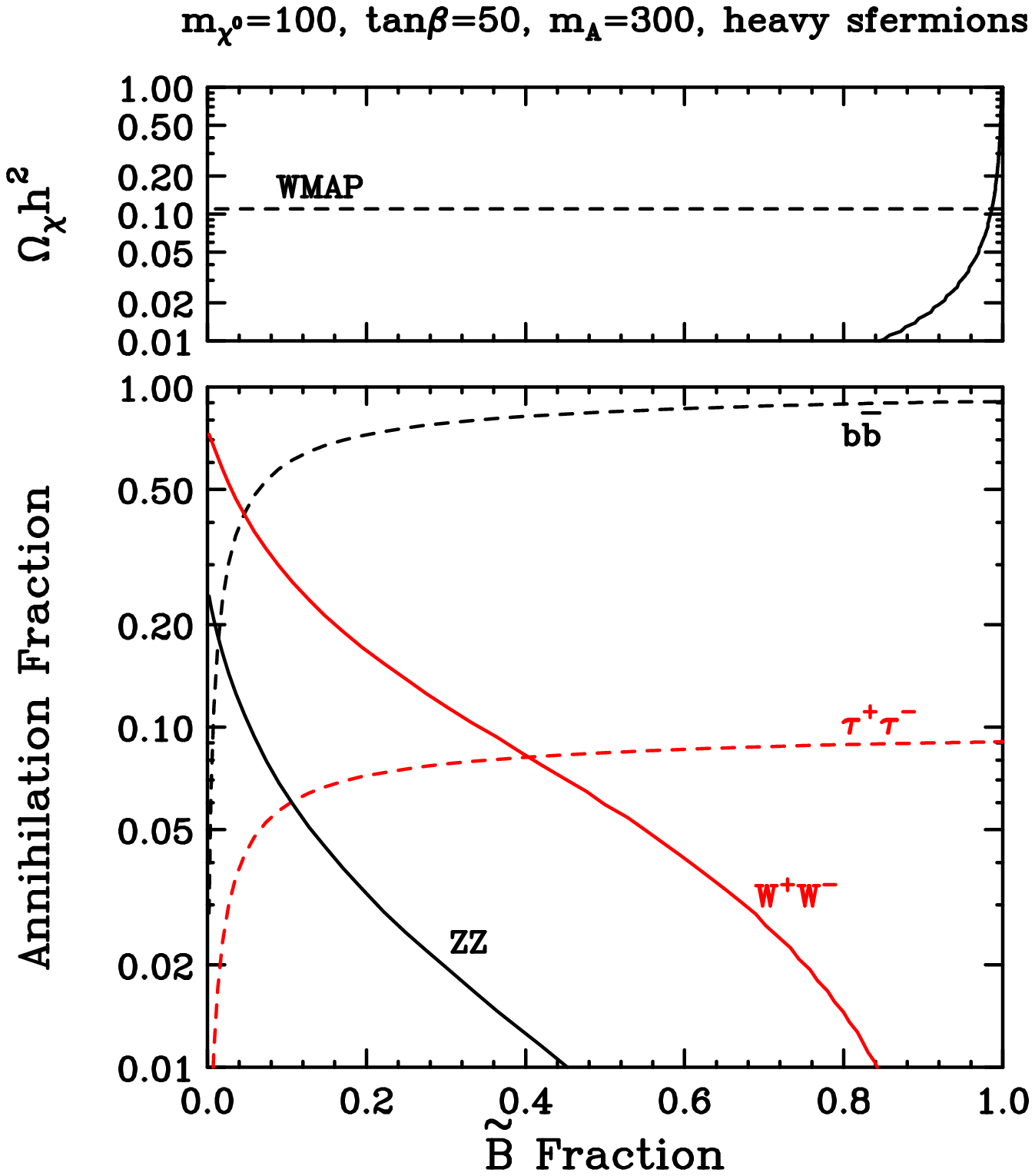}
\\
\includegraphics[width=3.2in,angle=0]{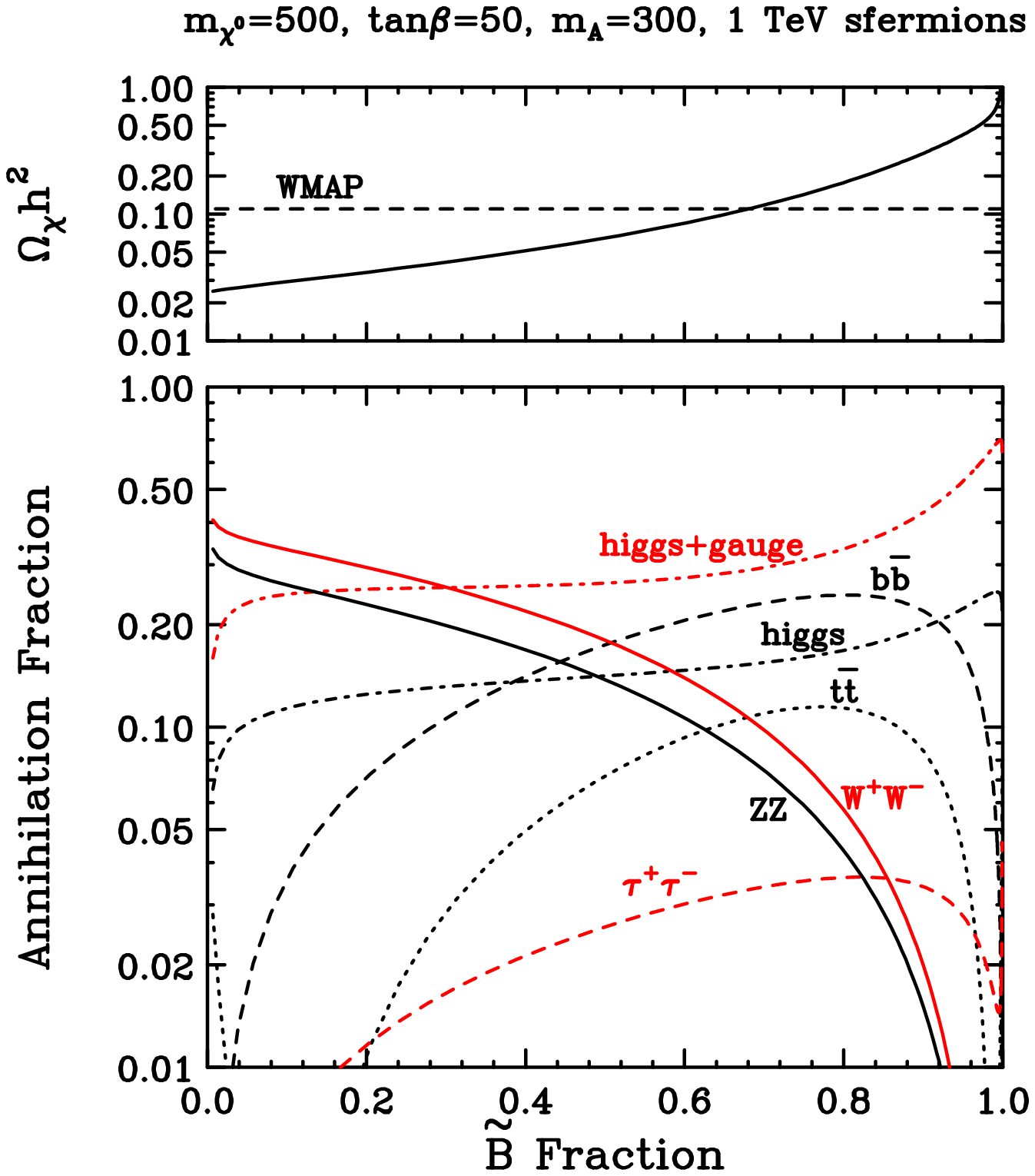}
\includegraphics[width=3.2in,angle=0]{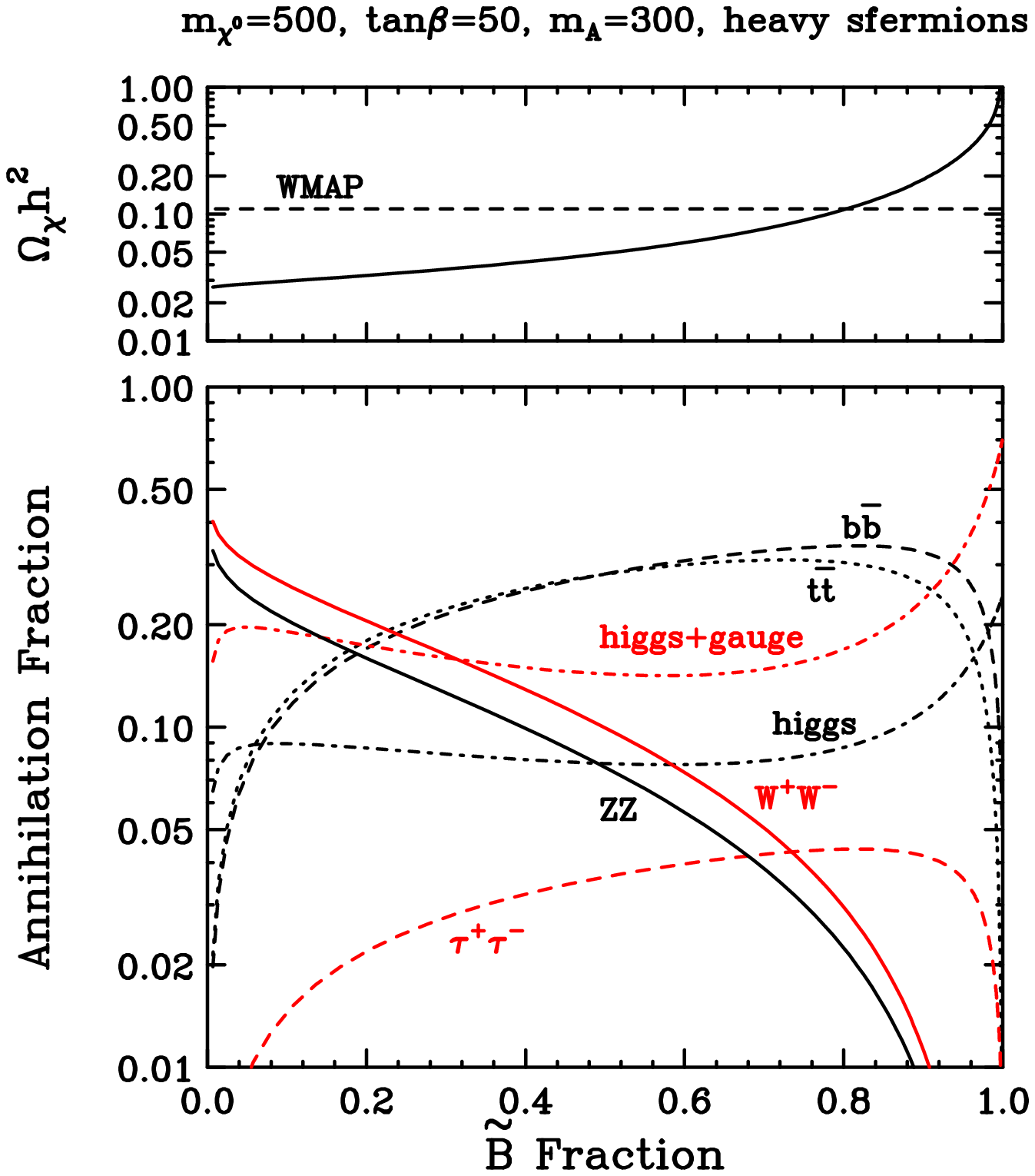}
\caption{The same as in figure~\ref{ann10}, but with $\tan \beta =50$. See the text for more details.}
\label{ann50}
\end{figure}

To summarize the content of this section, we will here briefly discuss under what circumstances the various above mentioned neutralino annihilation modes are likely to dominate. To study this more concretely, we have plotted in figures~\ref{ann10} and \ref{ann50} the fraction of neutralino annihilations which go to various final states as a function of the neutralino's bino-content. Each of the frames in these figures follow a contour of constant mass, such as those shown in figure~\ref{comp10}. 

As a first case, consider a somewhat light neutralino, such that Higgs bosons, gauge bosons and top quarks are not kinematically accessible final states. Such a neutralino will annihilate almost completely to bottom quarks and tau leptons. If the sbottoms and staus are sufficiently heavy to not contribute, the leading annihilation channel will be through $A$ exchange. The ratio of bottoms to taus produced through this channel is simply the ratio of the masses squared and the color factor, $3 m^2_b/m^2_{\tau} \sim 10$. Although interference terms from $Z$ and $A$ exchange diagrams can modify this ratio somewhat, annihilations to bottom quarks with a tau admixture of a few to ten percent is robustly predicted in the case of a light neutralino with much heavier sfermions. 

If the sfermions are not so heavy, however, and sfermion exchange diagrams to fermion pairs are substantial, then the ratio of annihilations producing bottoms and taus can be much larger or smaller. This depends on the sbottom and stau masses and on competition between sfermion exchange diagram and various interference terms. For example, in figures~\ref{ann10} and \ref{ann50}, changing the sfermion masses from very heavy (10 TeV) to 300 GeV \footnote{Not all sfermion masses are set to 300 GeV and 1000 GeV in figures~\ref{ann10} and \ref{ann50} once splitting has been taken into consideration. Although we have minimized the mass splitting by adopting $A_{t,b,\tau}$ soft terms much smaller than the sfermion masses, off diagonal terms in the sfermion mass matrices are proportional to $\mu m_f/\tan\beta$ for up-type fermions and $\mu m_f \tan\beta$ for down type fermions, and can be large for bino-like neutralinos.} has the effect of slightly increasing the fraction of annihilations which produce taus. 

We next consider a neutralino heavier than the gauge bosons. In this case, since annihilations to $W^+W^-$ and $ZZ$ have cross sections which approximately scale with $(|N_{13}|^2 - |N_{14}|^2)^2$, neutralinos annihilate primarily to fermions unless their higgsino fraction is rather large. 

If the lightest neutralino is heavy enough to annihilate to Higgs boson pairs, or to a gauge boson along with a Higgs boson, these modes are often important. As said before, annihilations to $Z H_1$ and $Z H_2$ are in cases proportional to $N^2_{13}$ rather than to the forth power, as is the case for annihilations to $ZZ$ or $W^+ W^-$. For bino-like or mixed bino-higgsino neutralinos, these annihilations (especially those to $Z H_1$, $Z H_2$ and $W^{\pm} H^{\mp}$) dominate over those to $ZZ$ or $W^+ W^-$. Whether these modes occur more often than those to fermion pairs depends on features such as the value of $\tan \beta$, the sfermion and $A$ masses, and the precise composition of the neutralino. For heavy neutralinos, annihilations to top quarks can also be significant, especially in models with low or moderate $\tan \beta$.

In addition to the annihilation fractions, in figures~\ref{ann10} and~\ref{ann50} we plot the thermal relic abundance of the lightest neutralino, and compare this with the quantity of cold dark matter measured by WMAP~\cite{wmap} ($0.119 > \Omega_{CDM} > 0.0787$ at the $2 \sigma$ confidence level). This quantity was calculated using the DarkSUSY package \cite{darksusy}. 

\clearpage
\clearpage

\newpage

\section{Neutralino-Nuclei Elastic Scattering}

Neutralinos scatter elastically with nuclei through both scalar and axial-vector couplings. Before moving on to direct and indirect detection methods, we review here the relevant elastic scattering cross sections for a neutralino.

\subsection{Scalar (Spin-Independent) Scattering}
\label{si}

A neutralino scattering elastically with a nucleus through scalar interactions does so with a cross section of:
\begin{equation}
\label{sig}
\sigma_{SI} \approx \frac{4 m^2_{\chi^0} m^2_{A}}{\pi (m_{\chi^0}+m_A)^2} [Z f_p + (A-Z) f_n]^2,
\end{equation}
where $m_A$, $Z$ and $A$ are the mass, atomic number and atomic mass of the target nucleus. $f_p$ and $f_n$ are the neutralino couplings to protons and neutrons, given by:
\begin{equation}
f_{p,n}=\sum_{q=u,d,s} f^{(p,n)}_{T_q} a_q \frac{m_{p,n}}{m_q} + \frac{2}{27} f^{(p,n)}_{TG} \sum_{q=c,b,t} a_q  \frac{m_{p,n}}{m_q},
\end{equation}
with $f^{(p)}_{T_u} \approx 0.020$,  $f^{(p)}_{T_d} \approx 0.026$,  $f^{(p)}_{T_s} \approx 0.118$,  $f^{(n)}_{T_u} \approx 0.014$,  $f^{(n)}_{T_d} \approx 0.036$ and $f^{(n)}_{T_s} \approx 0.118$. These quantities are subject to uncertainties in the relevant nuclear physics measurements (for more information, see Ref.~\cite{nuc}). $a_q$ are the neutralino-quark couplings. The first term in this expression corresponds to interactions with the quarks in the nucleus, either through t-channel CP-even Higgs exchange, or s-channel squark exchange. The second term corresponds to interactions with the gluons in the nucleus through a heavy quark/squark loop diagram. $f^{(p)}_{TG}$ is given by $1 -f^{(p)}_{T_u}-f^{(p)}_{T_d}-f^{(p)}_{T_s} \approx 0.84$, and analogously, $f^{(n)}_{TG} \approx 0.83$. To account for finite momentum transfer, the calculation should also include the appropriate form factor.

The neutralino-quark coupling, in which all of the SUSY model-dependent information is contained, is given by~\cite{scatteraq}:
\begin{eqnarray}
\label{aq}
a_q & = & - \frac{1}{2(m^{2}_{1i} - m^{2}_{\chi})} Re \left[
\left( X_{i} \right) \left( Y_{i} \right)^{\ast} \right] 
- \frac{1}{2(m^{2}_{2i} - m^{2}_{\chi})} Re \left[ 
\left( W_{i} \right) \left( V_{i} \right)^{\ast} \right] \nonumber \\
& & \mbox{} - \frac{g_2 m_{q}}{4 m_{W} B} \left[ Re \left( 
\delta_{1} [g_2 N_{12} - g_1 N_{11}] \right) D C \left( - \frac{1}{m^{2}_{H_1}} + 
\frac{1}{m^{2}_{H_2}} \right) \right. \nonumber \\
& & \mbox{} +  Re \left. \left( \delta_{2} [g_2 N_{12} - g_1 N_{11}] \right) \left( 
\frac{D^{2}}{m^{2}_{H_2}}+ \frac{C^{2}}{m^{2}_{H_1}} 
\right) \right],
\end{eqnarray}
where
\begin{eqnarray}
X_{i}& \equiv& \eta^{\ast}_{11} 
        \frac{g_2 m_{q}N_{1, 5-i}^{\ast}}{2 m_{W} B} - 
        \eta_{12}^{\ast} e_{i} g_1 N_{11}^{\ast}, \nonumber \\
Y_{i}& \equiv& \eta^{\ast}_{11} \left( \frac{y_{i}}{2} g_1 N_{11} + 
        g_2 T_{3i} N_{12} \right) + \eta^{\ast}_{12} 
        \frac{g_2 m_{q} N_{1, 5-i}}{2 m_{W} B}, \nonumber \\
W_{i}& \equiv& \eta_{21}^{\ast}
        \frac{g_2 m_{q}N_{1, 5-i}^{\ast}}{2 m_{W} B} -
        \eta_{22}^{\ast} e_{i} g_1 N_{11}^{\ast}, \nonumber \\
V_{i}& \equiv& \eta_{22}^{\ast} \frac{g_2 m_{q} N_{1, 5-i}}{2 m_{W} B}
        + \eta_{21}^{\ast}\left( \frac{y_{i}}{2} g_1 N_{11},
        + g_2 T_{3i} N_{12} \right)
\label{xywz}
\end{eqnarray}
where throughout $i=1$ for up-type quarks and $i=2$ for down type quarks. $m_{1i}, m_{2i}$ denote elements of the appropriate 2 x 2 squark mass matrix and $\eta$ is the matrix which diagonalizes that matrix. $y_i$, $T_{3i}$ and $e_i$ denote hypercharge, isospin and electric charge of the quarks. 

For scattering off of up-type quarks:
\begin{eqnarray}
\delta_{1} = N_{13},\,\,\,\, \delta_{2} = N_{14}, \,\,\,\, B = \sin{\beta},\,\,\,\, C = \sin{\alpha}, \,\,\,\, D = \cos{\alpha},
\end{eqnarray}
whereas for down-type quarks:
\begin{eqnarray}
\delta_{1} = N_{14},\,\,\,\, \delta_{2} = -N_{13}, \,\,\,\, B = \cos{\beta},\,\,\,\, C = \cos{\alpha}, \,\,\,\, D = -\sin{\alpha}.
\end{eqnarray}
Here, $\alpha$ is the Higgs mixing angle. 

The first two terms of Eq.~\ref{aq} correspond to interactions through the exchange of a squark, while the final term is generated through Higgs exchange. As in the treatment of the annihilation cross section to fermions, the cross section resulting from Higgs exchange terms is proportional to the higgsino fraction (or possibly the difference of the two higgsino fractions) and bino fraction, thus favoring a mixed neutralino. Squark exchange diagrams, on the other hand, contribute for either a mixed neutralino or a pure bino.

Taking a closer look at the contribution from Higgs exchange, note that if $m_{A,H_1} \gg m_{H_2}$, then $\cos \alpha \approx 1$ and $\sin \alpha \approx 0$. In this limit, the above expressions simplify considerably, yielding $a_q/m_q \approx g_1 g_2 N_{11} N_{14}^*/4 m_W \sin \beta m^2_{H_2}$ for up-type quarks and  $a_q/m_q \approx -g_1 g_2 N_{11} N_{13}^*/4 m_W \cos \beta m^2_{H_1}$ for down-type quarks. The cross section therefore scales with terms such as $|N_{11}|^2 |N_{14}|^2/m^4_{H_2}$ and $|N_{11}|^2 |N_{13}|^2 \tan^2 \beta/m^4_{H_1}$. The contributions from squark exchange can be similar in importance and, in the case of nearly diagonal squark mass matrices, scale as $|N_{11}|^2 |N_{14}|^2/m^4_{\tilde{q}}$ for up-type quarks and $|N_{11}|^2 |N_{13}|^2 \tan^2 \beta /m^2_{\tilde{q}}$ for down-type quarks~\cite{directsusypar}. If any of these contributions are similar in magnitude, interference terms can also be important.

\begin{figure}[tbp]
\includegraphics[width=3.2in,angle=0]{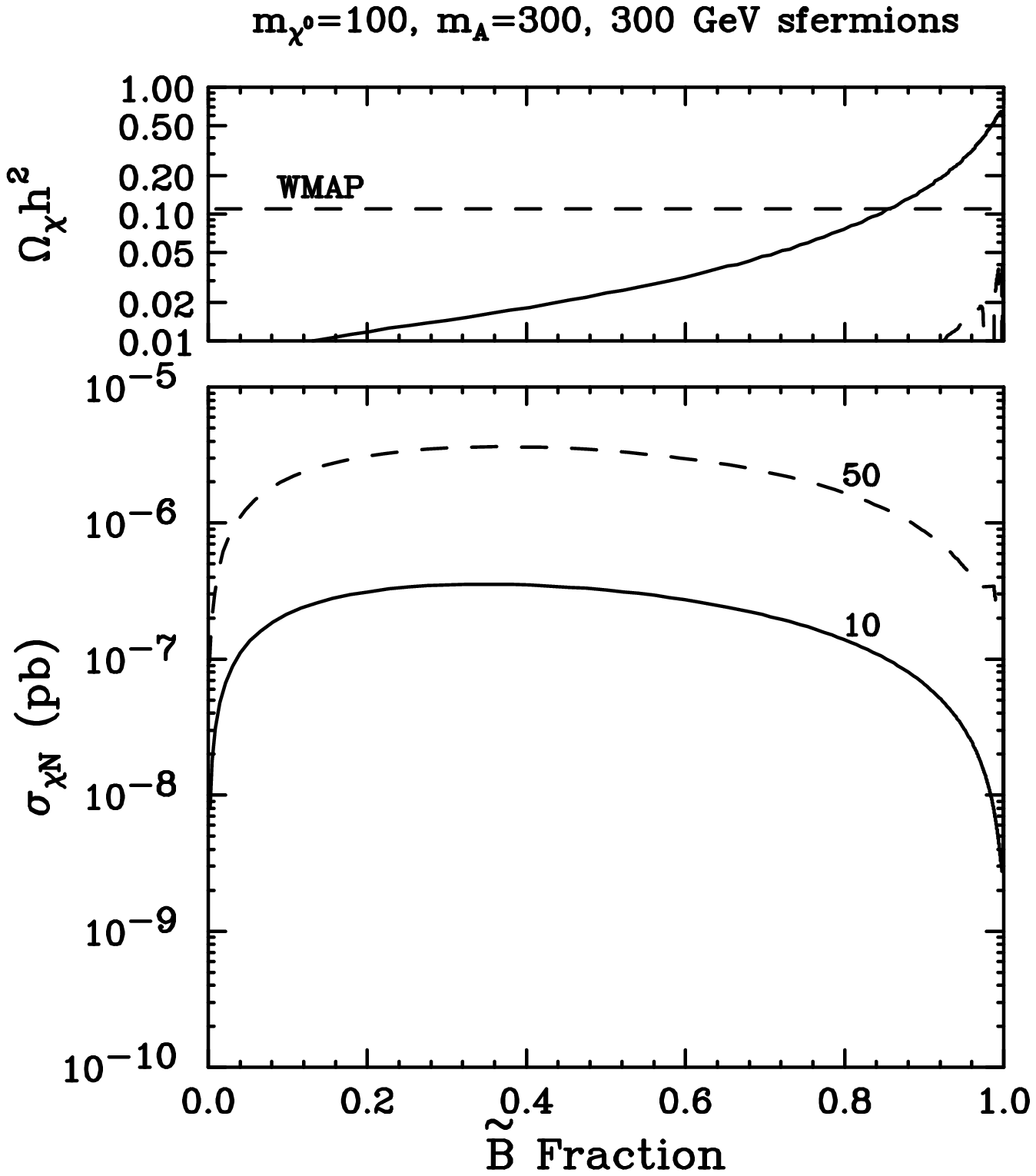}
\includegraphics[width=3.2in,angle=0]{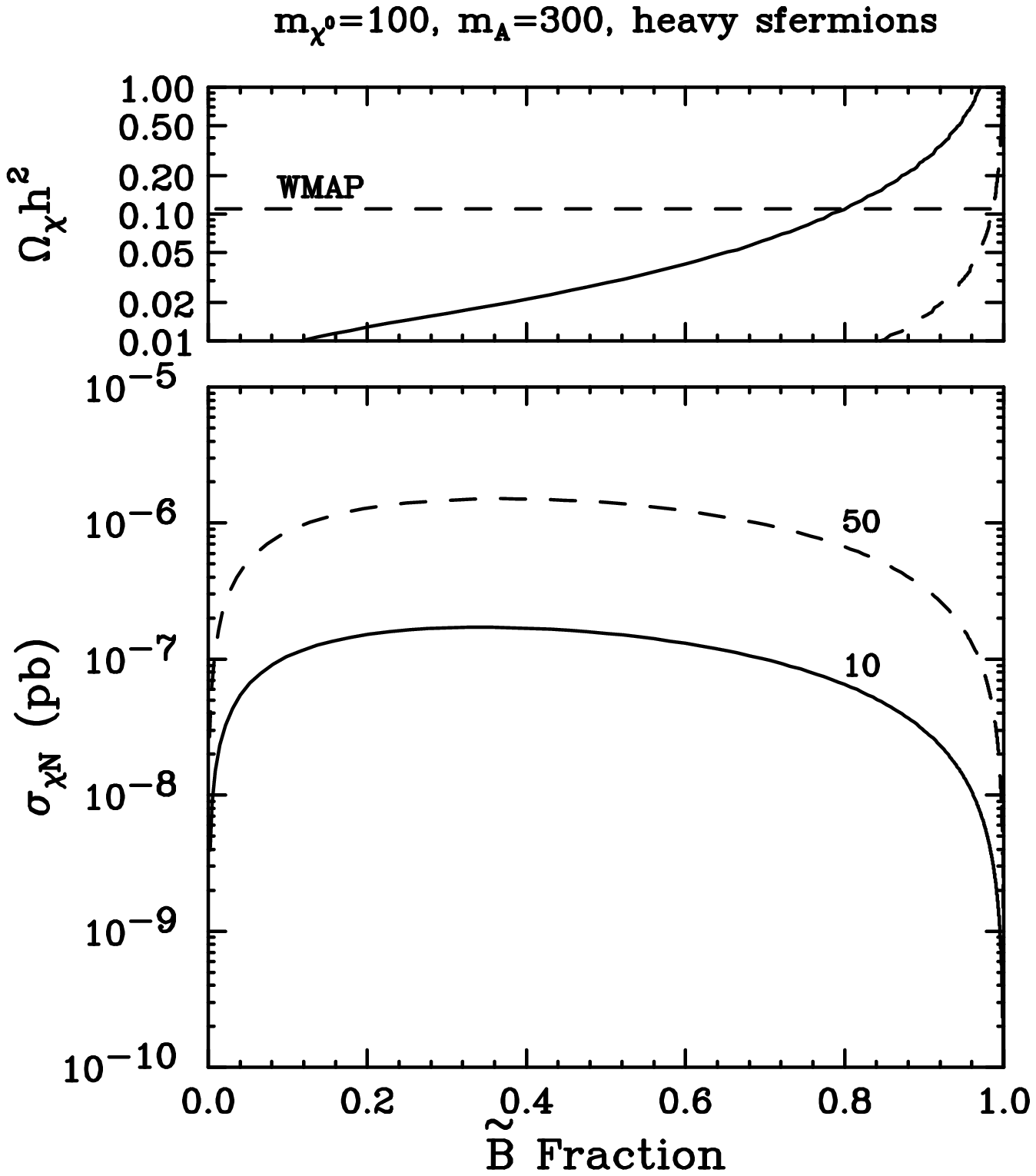}
\\
\includegraphics[width=3.2in,angle=0]{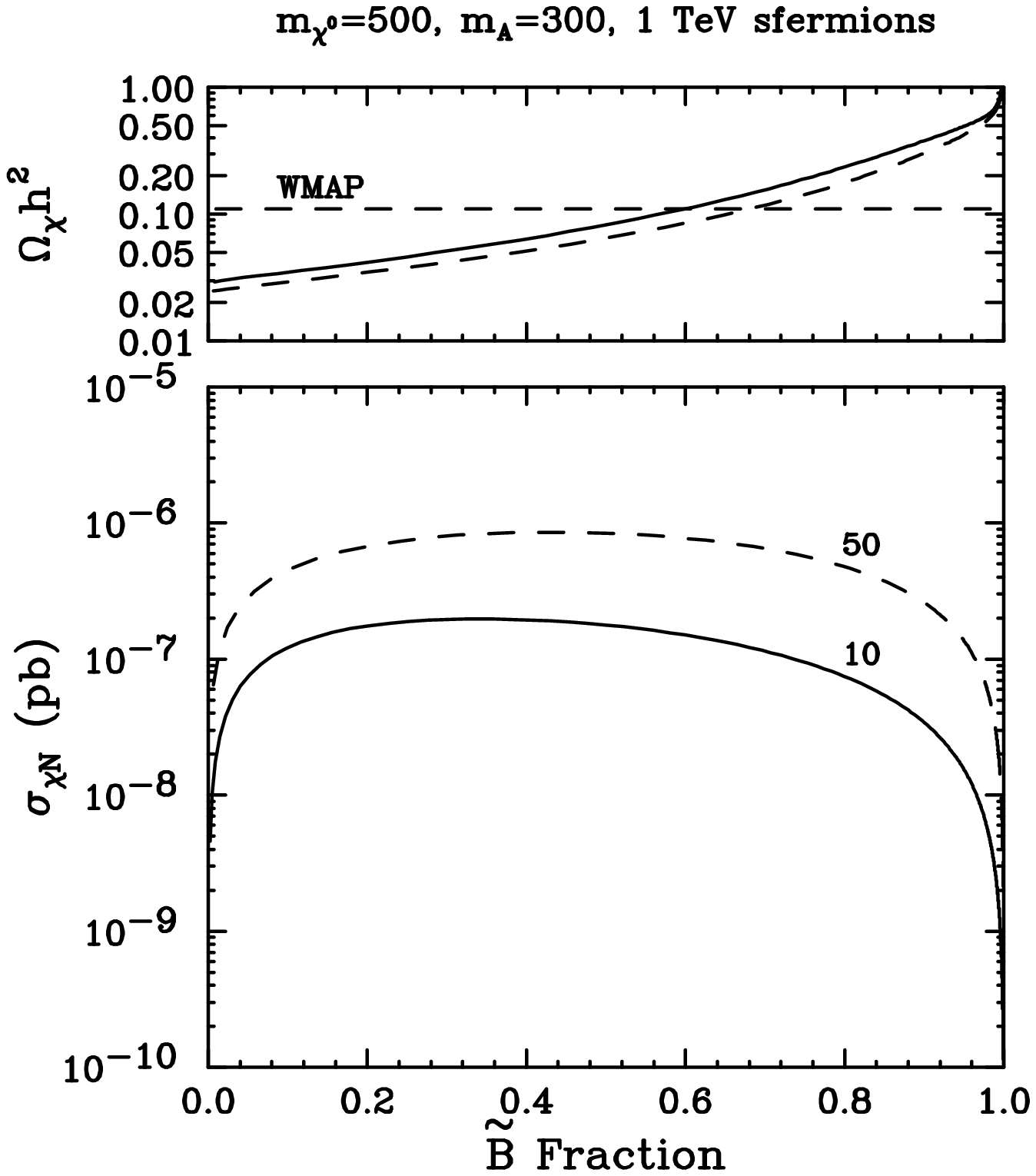}
\includegraphics[width=3.2in,angle=0]{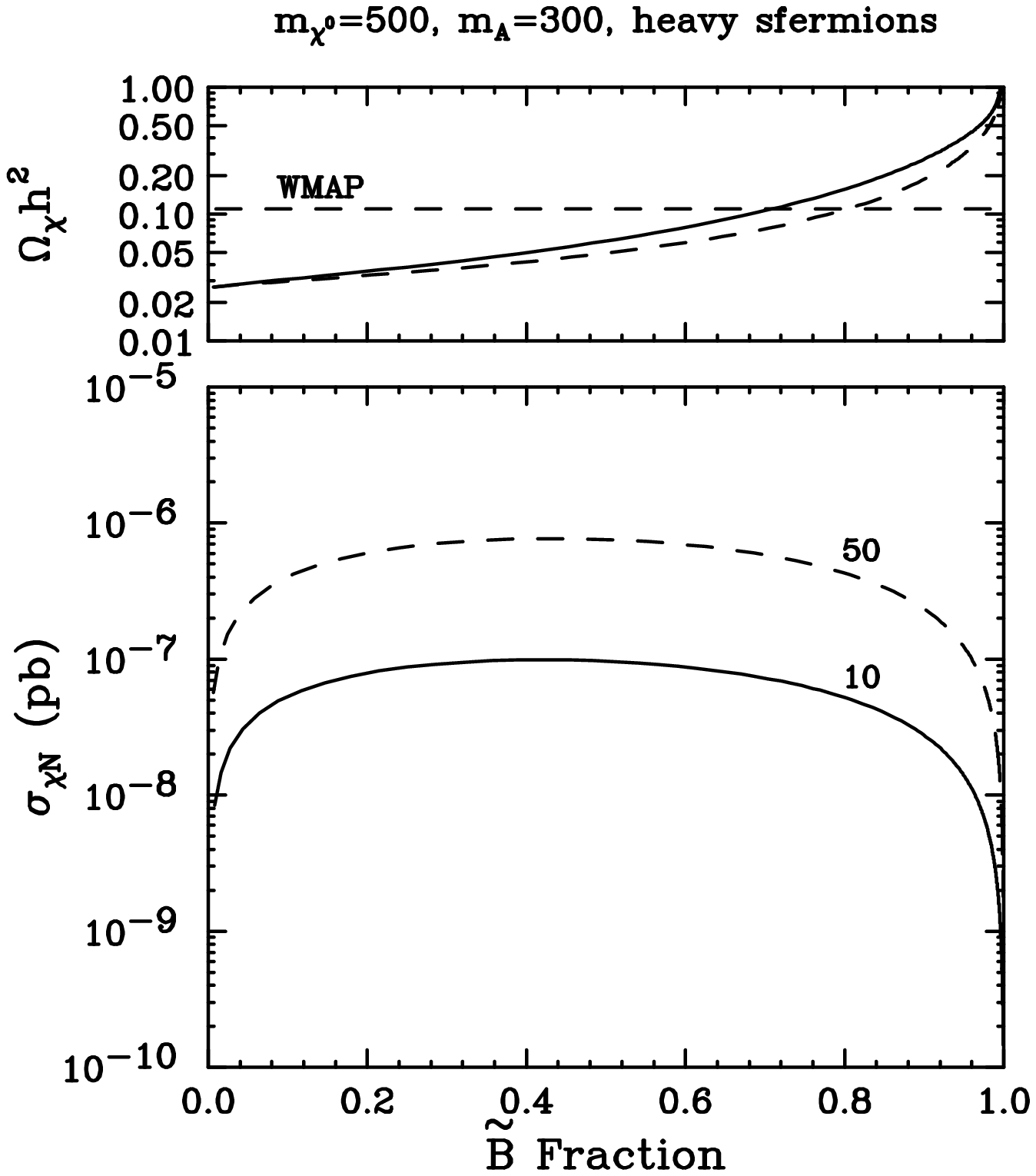}
\caption{The scalar (spin-independent) neutralino-nucleon elastic scattering cross section for $m_{\chi^0_1}=$100 and 500 GeV, as a function of its bino content. The GUT relationship between $M_1$ and $M_2$ has been used, and in each frame $\tan\beta=10,50$ (as labeled) and $m_A$=300 GeV. In the left frames, 300 GeV and 1 TeV sleptons and squarks have been used, while in the right frame they have been made very heavy for illustration (10 TeV). Current bound on this quantity from the CDMS experiment is $2 \times 10^{-7}$ pb or $7 \times 10^{-7}$ pb for a 100 GeV or 500 GeV neutralino, respectively \cite{cdms}.  Also shown above each frame is the relic abundance of the lightest neutralino compared to the quantity measured by WMAP. See the text for more details.}
\label{elsi}
\end{figure}

\begin{figure}[tbp]
\includegraphics[width=3.2in,angle=0]{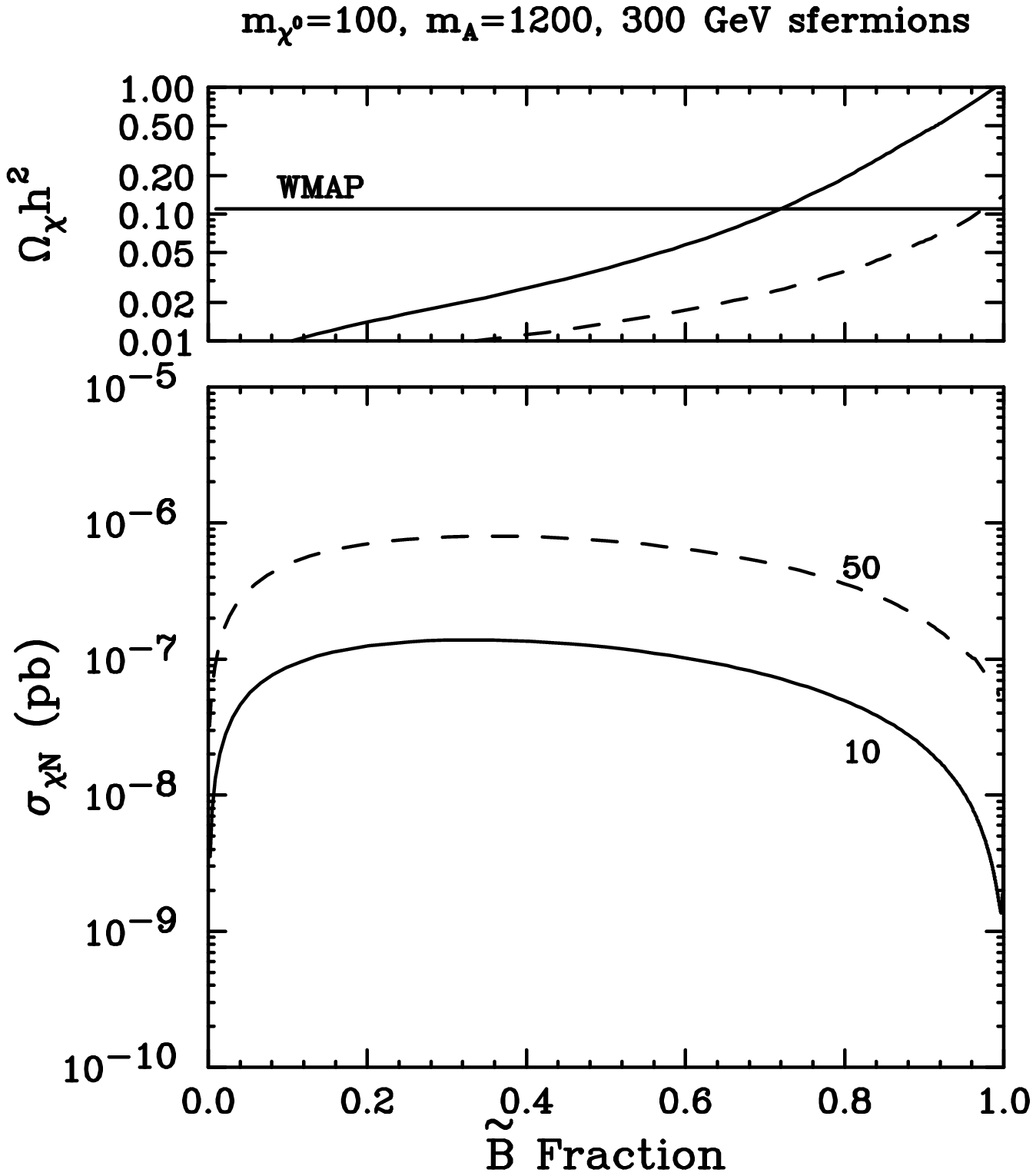}
\includegraphics[width=3.2in,angle=0]{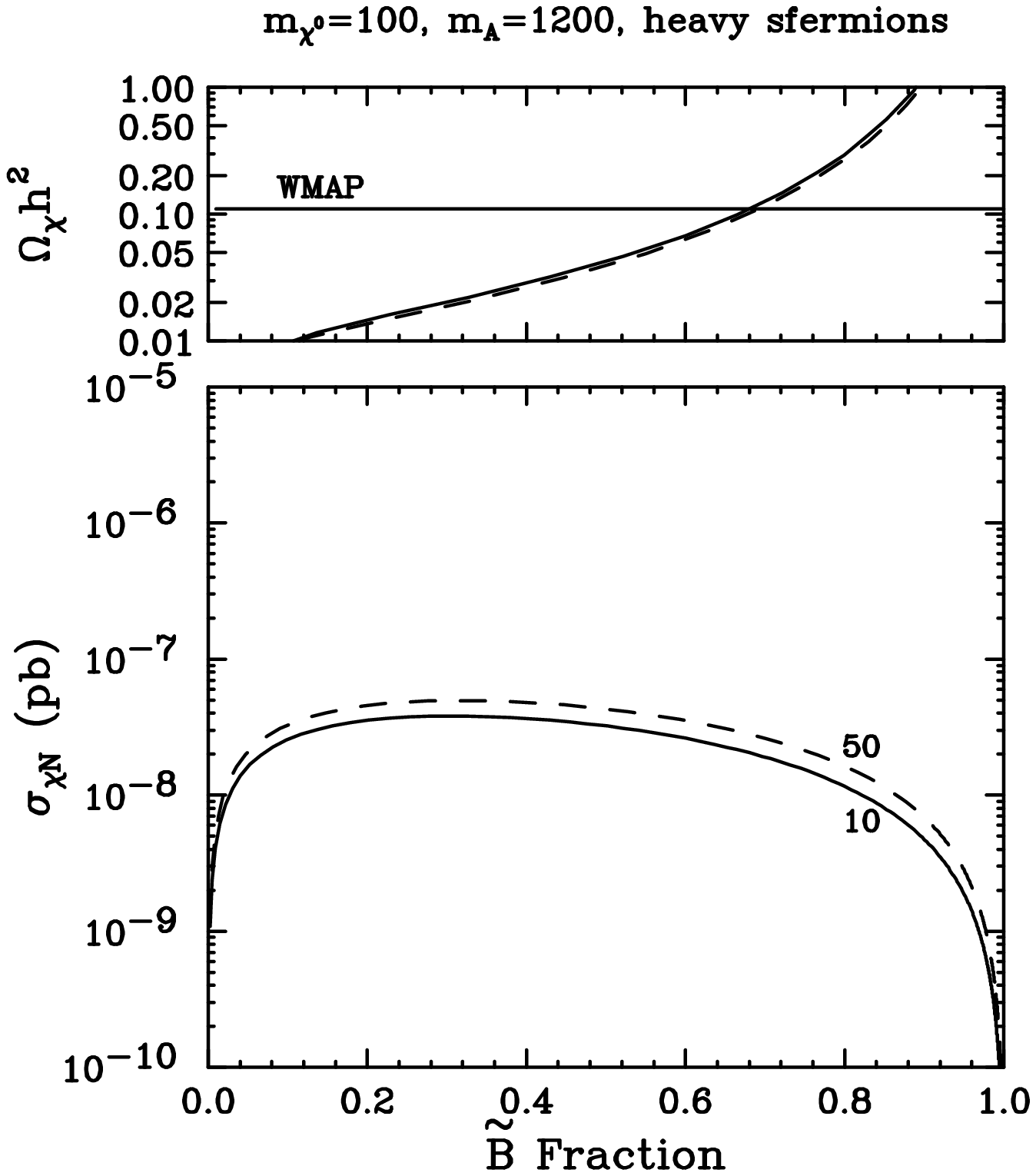}
\\
\includegraphics[width=3.2in,angle=0]{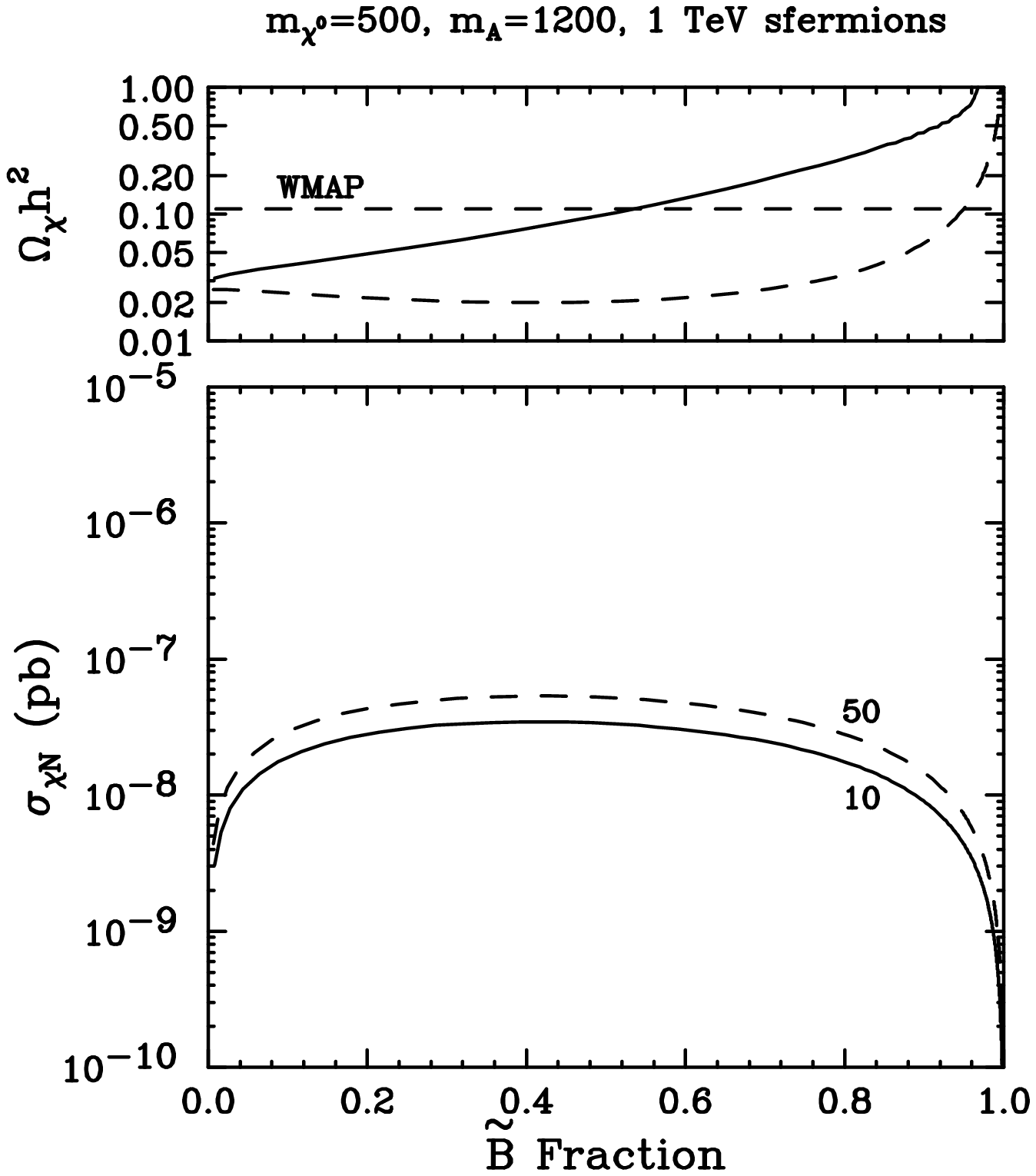}
\includegraphics[width=3.2in,angle=0]{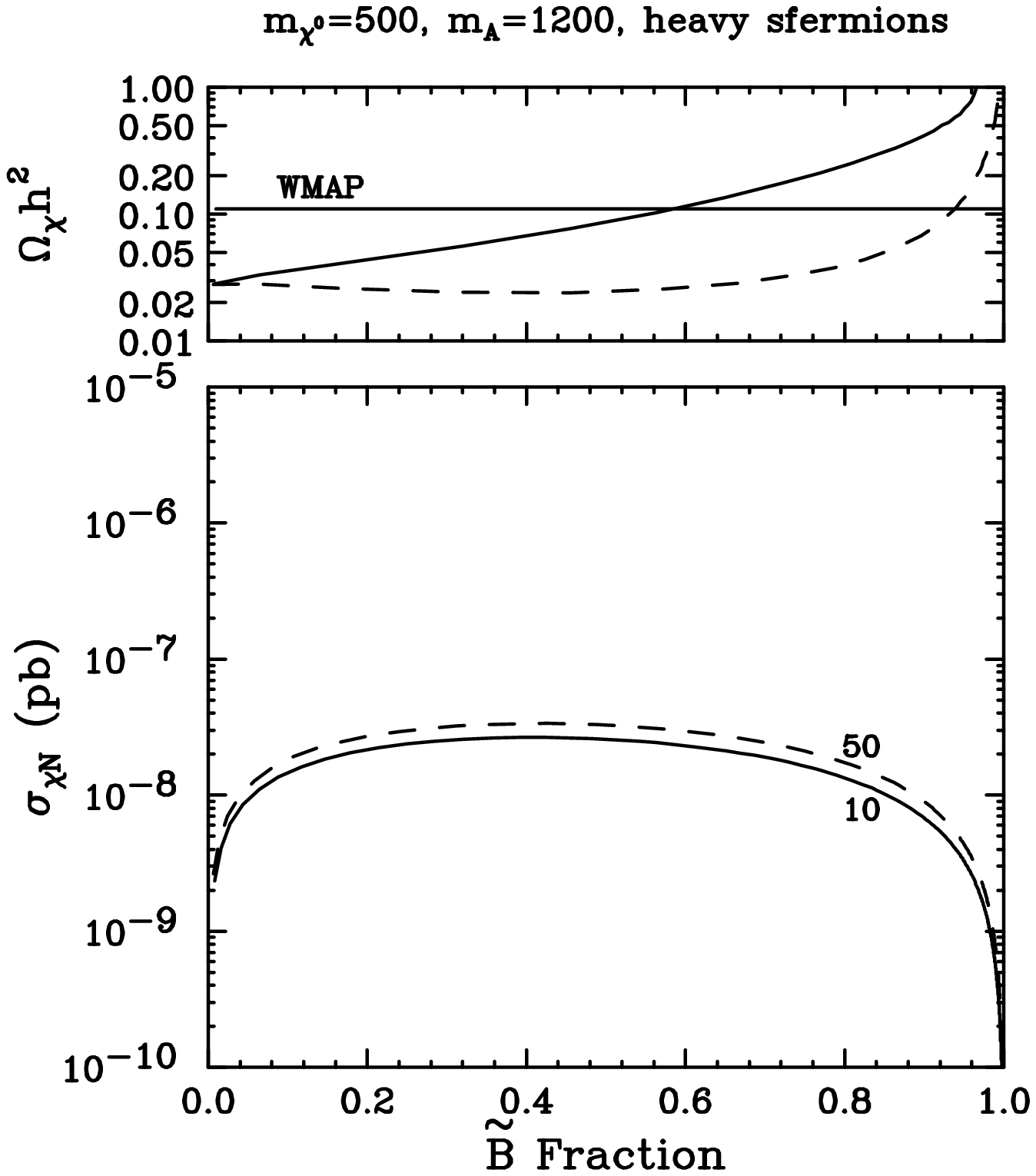}
\caption{The same as in figure~\ref{elsi}, but with $m_A=$1200 GeV. See the text for more details.}
\label{elsima}
\end{figure}

In figures~\ref{elsi} and~\ref{elsima}, we plot the scalar elastic scattering cross section of the lightest neutralino as a function of its bino-content for various choices of $\tan \beta$, $m_A$, $m_{\chi^0_1}$ and sfermion masses.

\subsection{Axial-Vector (Spin-Dependent) Scattering}

Although current direct detection experiments are primarily sensitive to spin-independent scattering, the elastic scattering of neutralinos with nuclei through axial-vector couplings is relevant to the capture rate of neutralinos in the Sun, and the corresponding indirect detection rate in neutrino telescopes. With this in mind, we will briefly summarize the most important aspects of this cross section here.

Axial-vector elastic scattering can be induced by the t-channel exchange of a $Z$ boson and by the s-channel exchange of a squark. The latter of these processes always generates a small cross section, ($\sigma_{SD} \sim 10^{-6}$ pb or less with a nucleon), however, which is well below the reach of planned direct detection experiments, and well below the magnitude needed to generate an observable flux of neutrinos from the Sun. Elastic scattering through $Z$ exchange, however, can be much more significant. 

The axial-vector elastic scattering cross section of a neutralino with a nucleus through a $Z$ boson simply follows from the $\chi^0_1-\chi^0_1-Z$ coupling and is proportional to $\sigma_{SD} \propto (|N_{13}|^2-|N_{14}|^2)^2$. This can lead to a cross section not far below $\sim 10^{-2}$ pb in extreme cases. As we will discuss in section~\ref{neutrino}, this cross section can potentially generate an observable flux of neutrinos from the Sun if $|N_{13}|^2-|N_{14}|^2$ is a few percent or larger.

\clearpage
\clearpage

\section{Deducing Dark Matter Properties From Astrophysical Observables}

The first step toward identifying the features of the supersymmetric model represented in nature (the lightest neutralino's composition, $\tan \beta$, etc.) is to determine the relationships between observable astrophysical quantities and the neutralino's phenomenological characteristics. In our study, we will consider the following astrophysical observables: 
\begin{itemize}
\item{The rate in direct dark matter detection experiments}
\item{The flux of high-energy neutrinos from the Sun, generated through neutralino annihilations}
\item{The shape of the gamma-ray spectrum from neutralino annihilations (including lines)}
\item{The annihilation rate of neutralinos in the local halo, as inferred by the cosmic positron flux}
\end{itemize}
These can, in turn, potentially be used to infer the following phenomenological characteristics of the lightest neutralino:
\begin{itemize}
\item{The spin-independent elastic scattering cross sections of the lightest neutralino with nucleons}
\item{The spin-dependent elastic scattering cross sections of the lightest neutralino with nucleons}
\item{The relative fraction of neutralino annihilations which result in $\gamma \gamma$ or $\gamma Z$ final states (at low velocities)} 
\item{The magnitude of the lightest neutralino's annihilation cross section (at low velocities)} 
\end{itemize}
Using this set of astrophysical observables, it may be possible to infer some of the properties of the underlying supersymmetric model. In this section, we will consider how these phenomenological quantities may be measured in near future astrophysical observations, and what these measurements may be able to tell us about the nature of supersymmetry.

\subsection{The Neutralino's Elastic Scattering Cross Section}

There are several experiments currently in operation that hope to detect neutralino dark matter particles by observing them elastically scattering off of a detector. This class of techniques, known as direct detection, has become very exciting in recent years, reaching the level of sensitivity needed to explore regions of the supersymmetric parameter space. Furthermore, a large fraction of the remaining supersymmetric parameter space is within the projected reach of direct detection experiments planned for the near future.

Currently, the strongest constraint on the neutralino's spin-independent elastic scattering cross section come from the CDMS experiment \cite{cdms}. For a 50--100 GeV neutralino, CDMS has excluded $\sigma_{\chi N} \gsim 2 \times 10^{-7}$ pb, assuming a standard local dark matter density ($\rho_{\chi^0} \approx 0.3$ GeV/cm$^3$) \cite{local}. For a 1 TeV neutralino, this constraint is about an order of magnitude weaker. Constraints from the Edelweiss \cite{edelweiss}, Zeplin \cite{zeplin} and WARP \cite{liquid} experiments are not very much weaker than those of CDMS. 

\begin{figure}[tbp]
\includegraphics[width=3.2in,angle=-90]{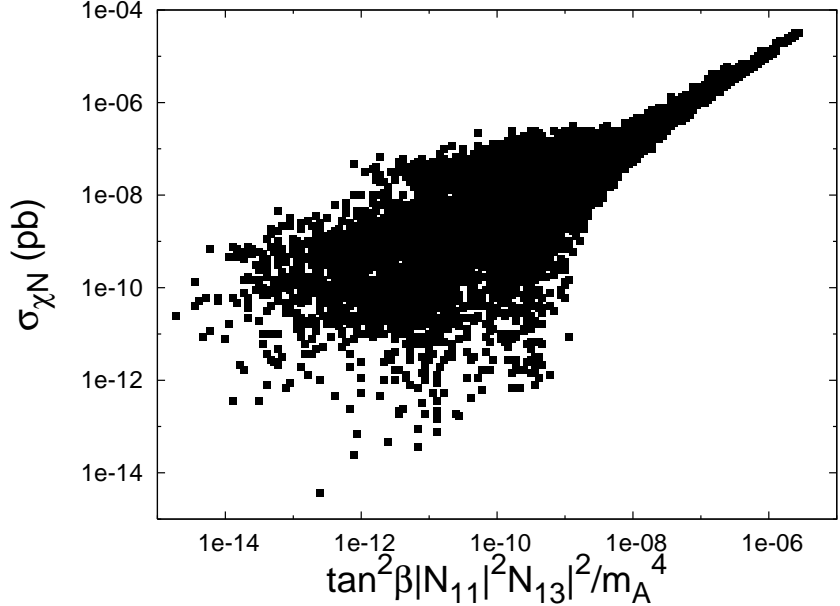}
\caption{The relationship between the quantity  $|N_{11}|^2 |N_{13}|^2 \tan^2 \beta/m^4_A$ and the spin-independent neutralino-nucleon elastic scattering cross section. For models with a large cross section ($ \gsim 10^{-7}$ pb), the correlation is quite strong. Each point shown represents a set of parameters within the MSSM that is not in violation of direct collider constraints and that generates a thermal relic density within the $2\sigma$ range measured by WMAP ($0.119 > \Omega_{\chi^0} > 0.0787$)\cite{wmap}. The GUT relationship between the gaugino masses has been adopted.}
\label{dir}
\end{figure}

The spin-independent neutralino-nucleon elastic scattering cross section is a function of many supersymmetric parameters, including the composition of the lightest neutralino, the masses and mixings of the exchanged Higgs bosons and squarks, and $\tan \beta$. This can make it difficult to extract much information about the supersymmetric model from a measurement of this quantity. If the cross section is measured to be quite large, however, it would indicate that this process is likely dominated by the exchange of the heavy CP-even Higgs boson, $H_1$, with a large value of $\tan \beta$ and a non-negligible higgsino fraction. In this limit, the cross section is roughly proportional to: $\sigma_{\chi N} \propto |N_{11}|^2 |N_{13}|^2 \tan^2 \beta/m^4_A$ (see section~\ref{si}). In Fig.~\ref{dir}, we plot the relationship between these quantities, and demonstrate that in models with a large cross section ($\sim 10^{-7}$ pb or larger) that the correlation is very strong.

If a positive detection is made by CDMS or another direct detection experiment, such a measurement could be used to constrain both the neutralino's elastic scattering cross section and its mass. With an accumulation of $\sim100$ events, the neutralino's mass could be measured to $\sim 25\%$ accuracy, and the cross section even more accurately (up to uncertainties in the local halo profile) \cite{baltz}. With fewer events, these parameters will be considerably less constrained.

\subsection{The Flux of High-Energy Neutrinos From The Sun}
\label{neutrino}
Neutralinos traveling through the solar system can elastically scatter off of massive bodies such as the Sun, potentially becoming gravitationally bound. If this occurs sufficiently frequently, the neutralino annihilation rate in the Sun's core may be large enough to generate an observable flux of high-energy neutrinos \cite{neutrino}.

Unlike rates from other methods of indirect detection, the neutrino flux from the Sun depends primarily on the neutralino's elastic scattering cross section with nuclei in the Sun rather than on its annihilation cross section. The annihilation cross section drops out of the calculation in the case that equilibrium is reached between the solar capture and annihilation rates. A measurement of the neutrino flux from the Sun, therefore, can be used to infer the elastic scattering cross section of the lightest neutralino, much as direct detection experiments do. Unlike the event rates produced in CDMS, Zeplin or Edelweiss, however, the neutralino capture rate in the Sun is the result of both axial-vector (spin-dependent) and scalar (spin-independent) interactions. Although direct detection experiments constrain the spin-independent elastic scattering of a neutralino, spin-dependent scattering is not yet constrained at a practical level. Furthermore, since it will be a number of years before next generation neutrino telescopes IceCube \cite{icecube} and Antares \cite{antares} are fully operational, the limits on spin-independent scattering are likely to becomes stronger by at least an order of magnitude and probably more by the time the neutrino flux from the Sun will be measured. In models in which direct detection experiments will not discover neutralino dark matter in the near future, neutralinos will be potentially observable at next generation neutrino telescopes only if they are captured in the Sun primarily though spin-dependent elastic scattering.  By combining information provided by direct detection experiments with the rates from the Sun observed by neutrino telescopes, it may be possible to measure the neutralino-proton spin-dependent elastic scattering cross section.\footnote{It is also possible that future direct detection experiments will be sensitive to spin-dependent neutralino elastic scattering. The COUPP collaboration, for example, is developing bubble chamber detector technology that could potentially reach the level of sensitivity needed to probe parameter space of the MSSM through spin-dependent neutralino scattering~\cite{coupp}.}

\begin{figure}[tbp]
\includegraphics[width=3.2in,angle=-90]{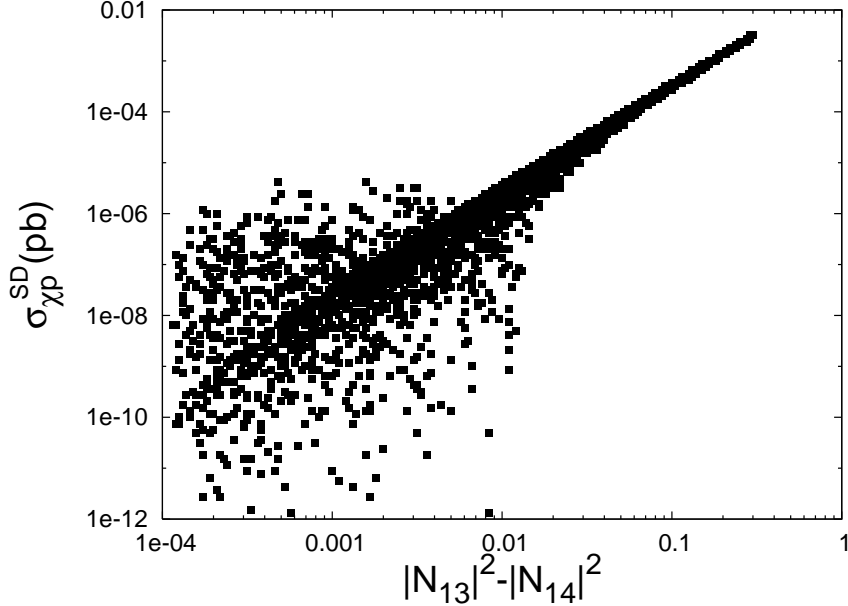}
\caption{The relationship between the neutralino-proton spin-dependent (axial-vector) elastic scattering cross section and the quantity $|N_{13}|^2-|N_{14}|^2$.   For models along the diagonal line extending from the lower left to upper right portions of the figure, the process is dominated by $Z$ exchange. Those points scattered from this line obtain their spin-dependent elastic scattering cross section significantly from squark exchange. Each point shown represents a set of parameters within the MSSM that is not in violation of direct collider constraints and that generates a thermal relic density within the $2\sigma$ range measured by WMAP ($0.119 > \Omega_{\chi^0} > 0.0787$)\cite{wmap}. The GUT relationship between the gaugino masses has been adopted.}
\label{sd}
\end{figure}

The spin dependent neutralino-proton elastic scattering cross section is generated through two classes of diagrams: t-channel $Z$ exchange and s-channel squark exchange. The contribution from squark exchange, however, is invariably small (typically $10^{-6}$ pb or less) compared to the value needed to generate an observable flux of neutrinos from the Sun. For any model that is potentially observable in existing or planned neutrino telescopes, neutralino spin-dependent elastic scattering will, therefore, be dominated by $Z$ exchange. This cross section, following from the $\chi^0_1-\chi^0_1-Z$ coupling, is proportional to $|N_{13}|^2-|N_{14}|^2$. In Fig.~\ref{sd}, we plot the relationship between the neutralino-proton spin-dependent elastic scattering cross section and the quantity $|N_{13}|^2-|N_{14}|^2$. 

\begin{figure}[tbp]
\includegraphics[width=2.2in,angle=-90]{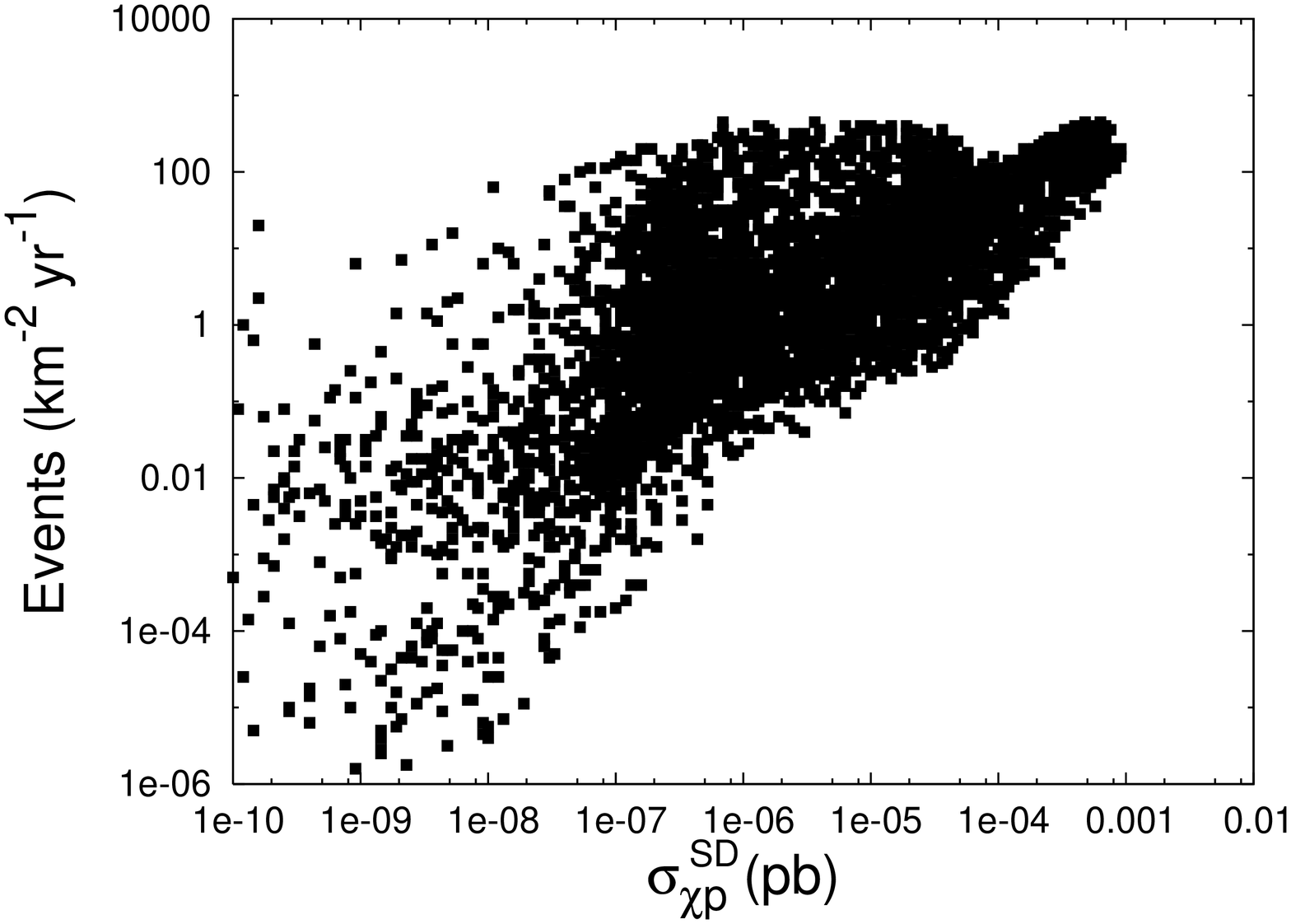}
\includegraphics[width=2.2in,angle=-90]{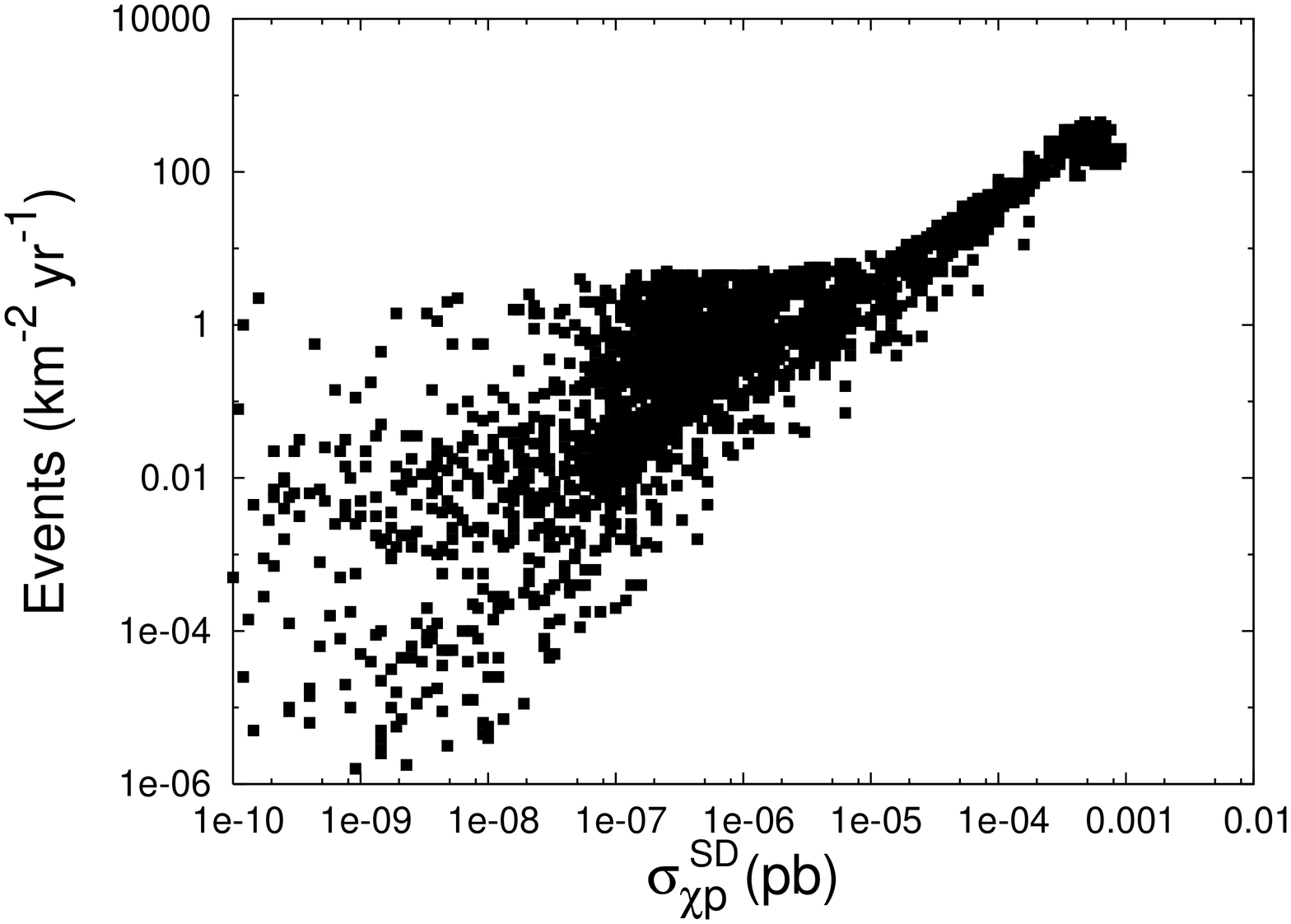}\\
\includegraphics[width=2.2in,angle=-90]{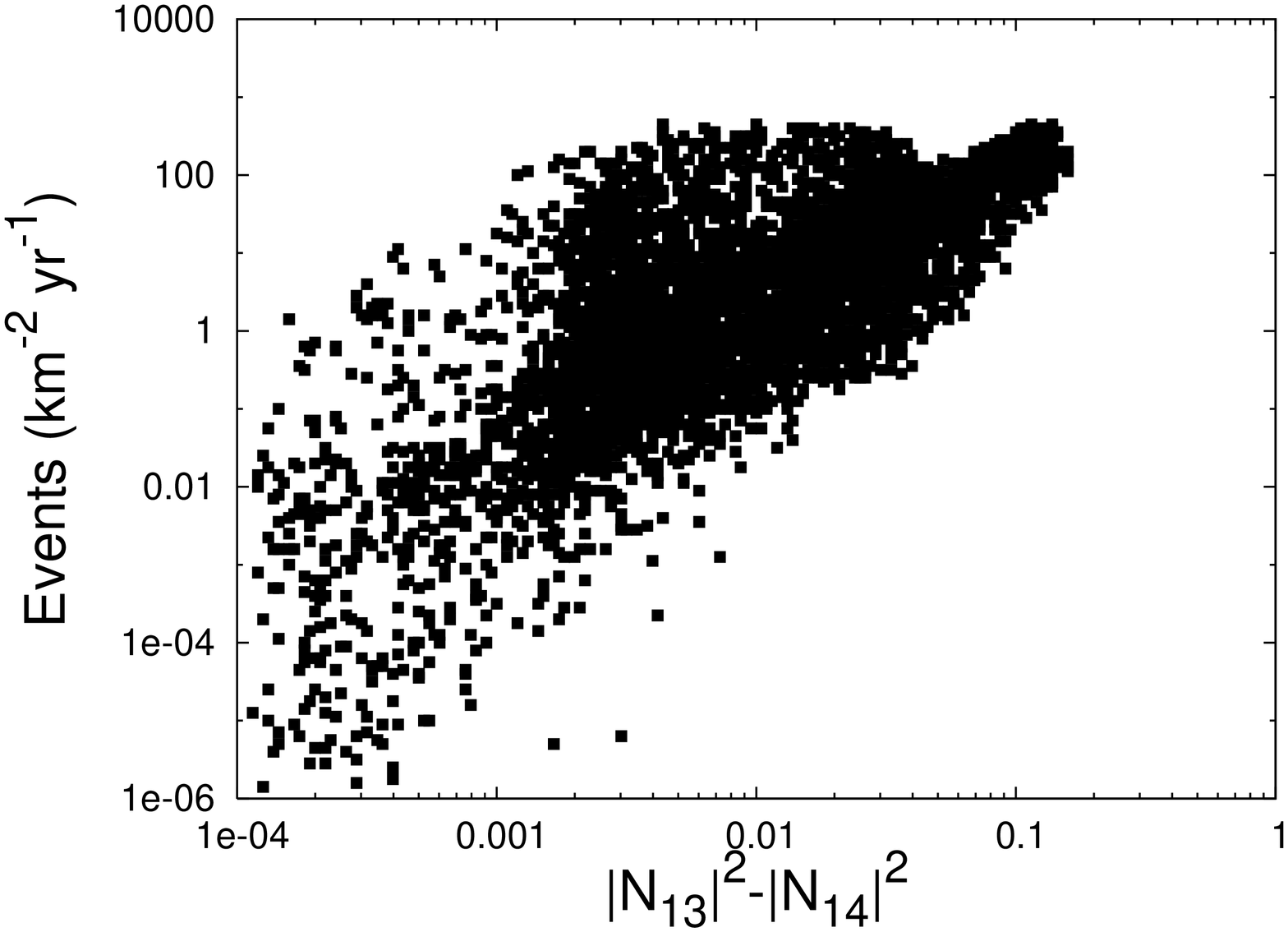}
\includegraphics[width=2.2in,angle=-90]{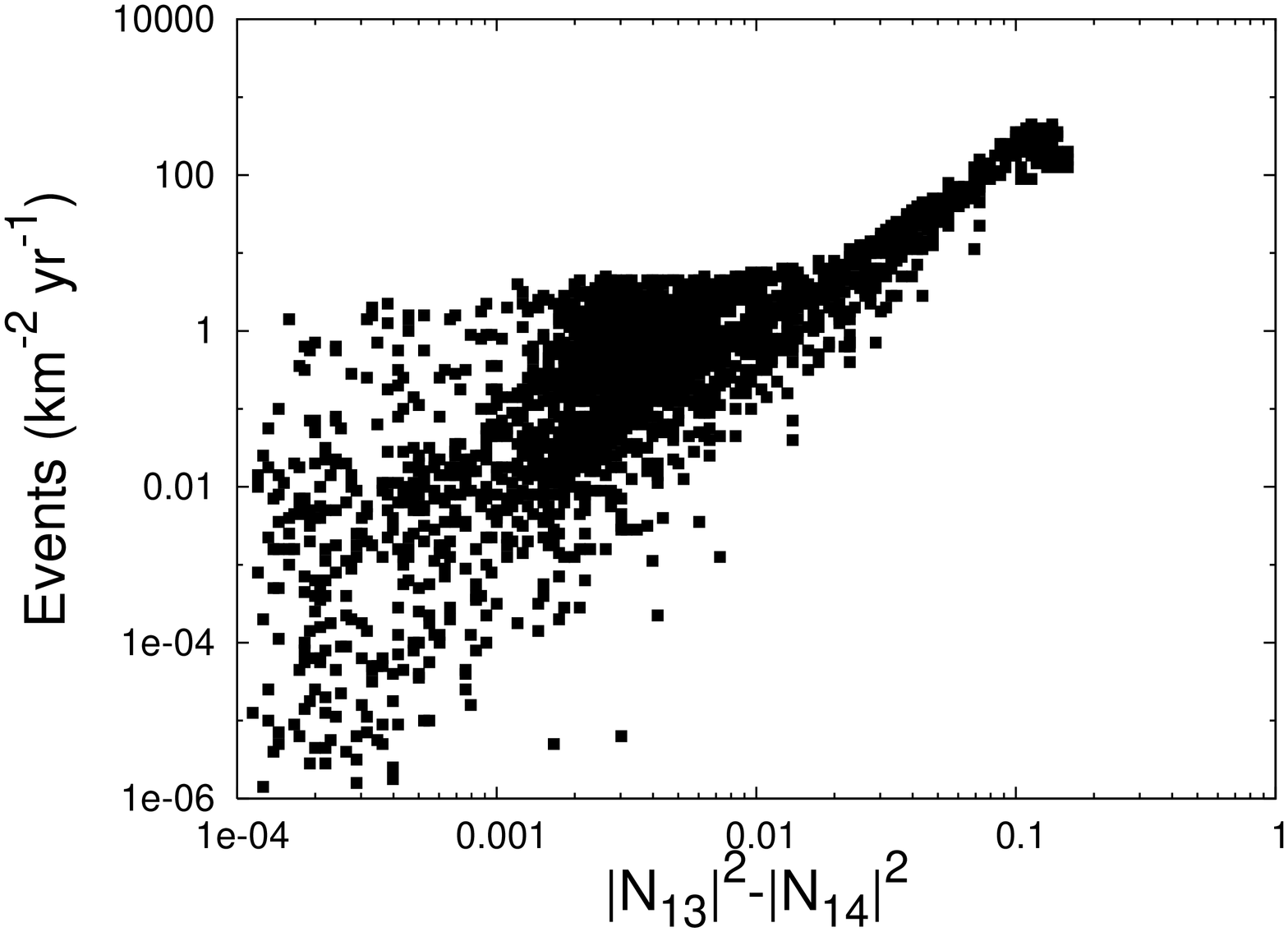}
\caption{The rate of neutrinos detected in a kilometer scale neutrino telescope, such as IceCube, from neutralino annihilations in the Sun as a function of the neutralino's spin-dependent elastic scattering cross section (upper), and as a function of the quantity $|N_{13}|^2-|N_{14}|^2$ (lower).  While the current CDMS constraint on the spin-independent elastic scattering cross section \cite{cdms} has been applied in the left frames, a constraint 100 times more stringer has been applied in the right frames in anticipation of increased sensitivity from direct detection experiments in the coming few years. Each point shown represents a set of parameters within the MSSM that is not in violation of direct collider constraints and that generates a thermal relic density within the $2\sigma$ range measured by WMAP ($0.119 > \Omega_{\chi^0} > 0.0787$)\cite{wmap}. The GUT relationship between the gaugino masses has been adopted. Each model shown contains a LSP heavier than 100 GeV, in order to avoid strong suppression from the muon energy threshold of 50 GeV that has been used.}
\label{neutrinorate}
\end{figure}

\begin{figure}[tbp]
\includegraphics[width=3.2in,angle=0]{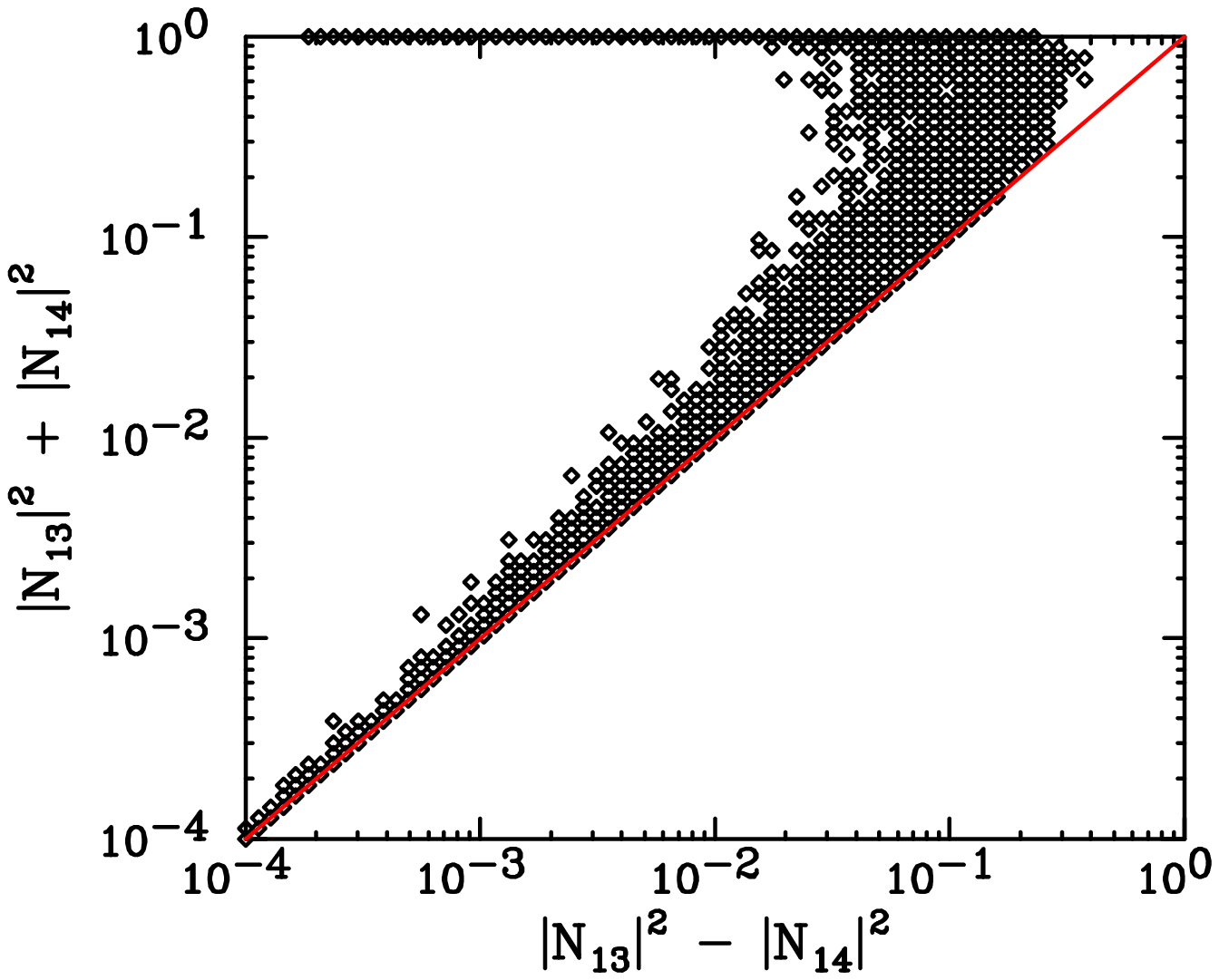}
\caption{The higgsino fraction of the lightest neutralino ($|N_{13}|^2+|N_{14}|^2$) verses the quantity $|N_{13}|^2-|N_{14}|^2$. For neutralinos with less than roughly 10\% higgsino fraction, a measurement of $|N_{13}|^2-|N_{14}|^2$ can act as a measurement of the neutralino's higgsino fraction. Each point shown represents a set of parameters within the MSSM that is not in violation of direct collider constraints and that generates a thermal relic density within the $2\sigma$ range measured by WMAP ($0.119 > \Omega_{\chi^0} > 0.0787$)\cite{wmap}. The GUT relationship between the gaugino masses has been adopted.}
\label{higgsino}
\end{figure}

In figure~\ref{neutrinorate} we plot the event rate in a kilometer-scale neutrino telescope as a function of the spin-dependent neutralino-proton elastic scattering cross section and as a function of $|N_{13}|^2-|N_{14}|^2$. We can see from this figure that a measurement of the high-energy neutrino flux from the Sun can be used as a rough measurement of the quantity $|N_{13}|^2-|N_{14}|^2$. Furthermore, unless the higgsino fraction of the lightest neutralino is very large, the quantity  $|N_{13}|^2-|N_{14}|^2$ translates fairly accurately to the neutralino's higgsino fraction,  $|N_{13}|^2+|N_{14}|^2$. This is demonstrated in figure~\ref{higgsino}. In this way, neutrino telescopes may provide a rough measurement of the lightest neutralino's higgsino fraction \cite{darkhalzen}. By studying the spectrum of neutrinos, information regarding the lightest neutralino's mass may also be inferred \cite{neutrinomass}.

Measurements of the neutrino flux from the Sun will be limited by the background of atmospheric neutrinos. Atmospheric neutrinos generate roughly $\sim500$ muons per square kilometer per year above 50 GeV in the angular window corresponding to the Sun. Over ten years, a 3$\sigma$ detection would therefore require $10 \times R_{\nu} \gsim 3 \times \sqrt{10 \times 500}$, or an annual rate in IceCube or KM3 of $\sim 20$ events. If a muon energy threshold of 100 GeV were imposed, a rate of 8 events per year would constitute a 3-$\sigma$ detection. To be observable in a kilometer-scale neutrino telescope such as IceCube (or a kilometer-scale version of Antares, such as KM3), an event rate on the order of $\sim$10 events per square kilometer per year will be required. If such a rate is observed, it would indicate a neutralino with a substantial higgsino component ($\sim$1\% or greater). The absence of any signal of this magnitude could be used to constrain $|N_{13}|^2-|N_{14}|^2 \lsim 0.05$.


\subsection{The Shape of the Gamma Ray Spectrum}

Gamma-rays can be produced in dark matter annihilations through a variety of annihilation modes. If annihilations occur sufficiently frequently in nearby regions, these gamma-rays may be observable by operating or upcoming Atmospheric Cerenkov Telescopes such as MAGIC \cite{magic}, HESS \cite{hess} or VERITAS \cite{veritas}, or by satellite-based experiments such as GLAST \cite{glast}. Among the most promising sources of dark matter annihilation radiation are the center of our galaxy \cite{center}, dwarf spheroidal galaxies (such as Draco, Sagitarius, or Canis Major) \cite{dwarfs}, external galaxies \cite{external}, galactic intermediate mass black holes \cite{imbh}, and small scale dark matter substructure \cite{earthmass}.

\begin{figure}[!]
\includegraphics[width=3.2in,angle=0]{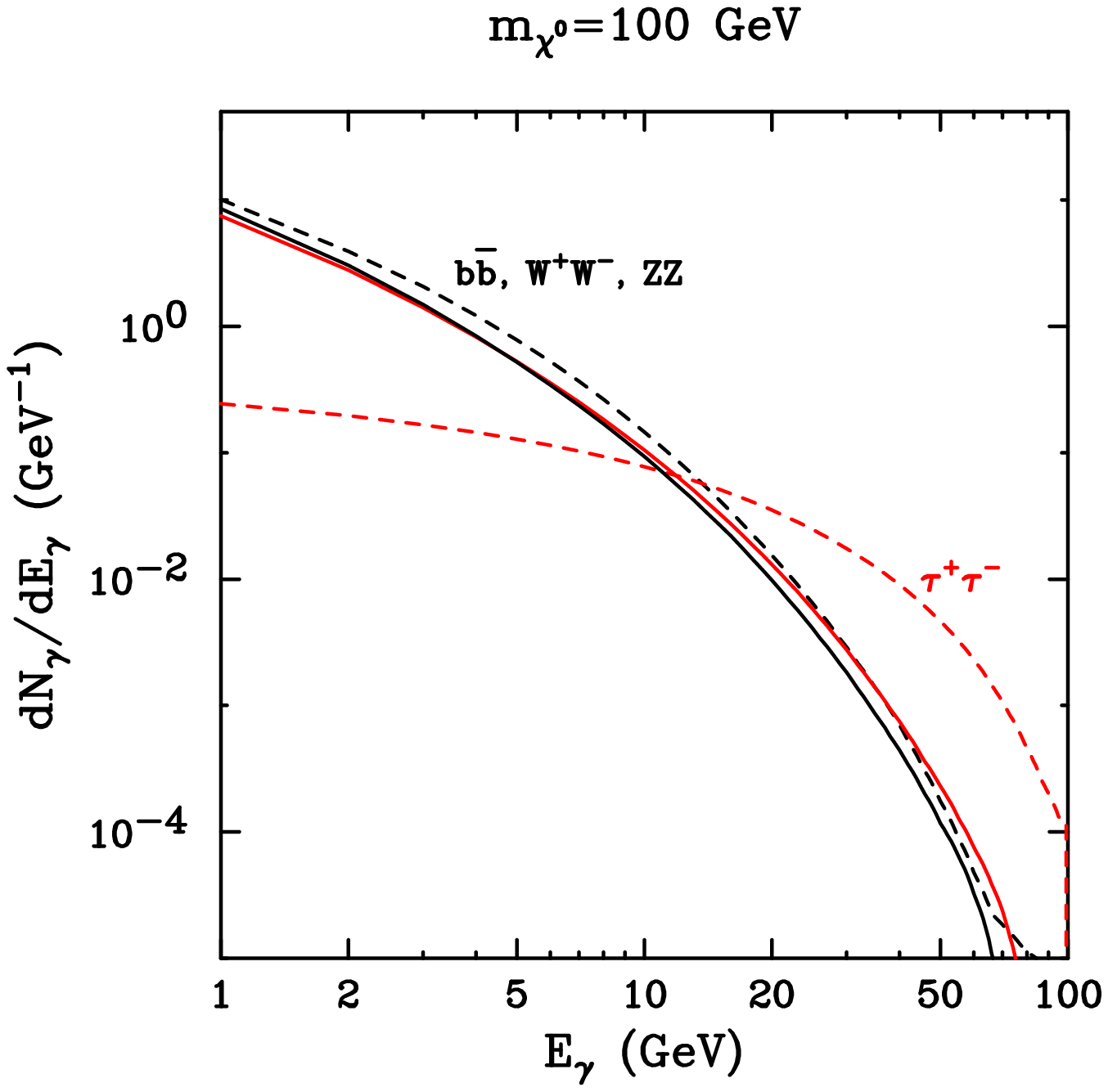}
\includegraphics[width=3.2in,angle=0]{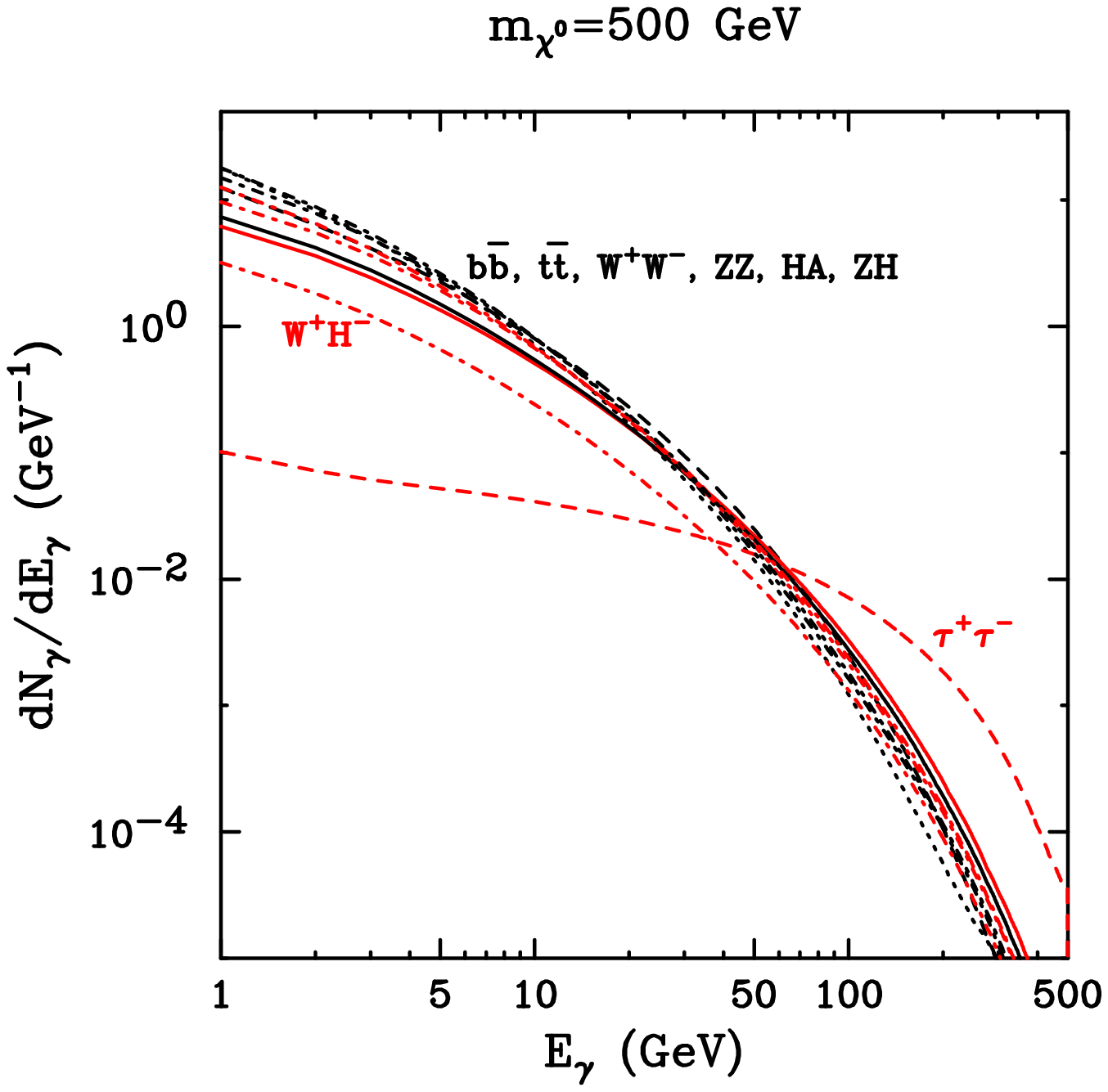}
\caption{The gamma ray spectrum per neutralino annihilation for various modes and two choices of the neutralino mass. With the exception of the $\tau^+\tau^-$ mode (and to a lesser extent, $W^{\pm}H^{\mp}$), the spectrum produced is very similar. The harder spectrum produced through taus is potentially distinguishable from other modes, however. Masses of $m_{A, H_1, H^{\pm}}\simeq$ 300 GeV and $m_{H_2}\simeq 115$ GeV were used.}
\label{gammaspec}
\end{figure}

If one or more of these sources is identified, the spectrum of gamma-rays could (in principle) be used to learn about the modes through which neutralinos annihilate. In figure~\ref{gammaspec}, we plot the gamma-ray spectrum, per neutralino annihilation, for several of the annihilation modes likely to contribute significantly. From this figure, it is clear that most of annihilation modes result in a similar spectrum. The primary exception to this is the spectrum from annihilations to tau lepton pairs, which has a considerably harder slope than the other modes. The spectrum is also slightly harder for annihilations to $W^{\pm} H^{\mp}$.

A gamma-ray telescope, at best, might hope to identify the fraction of neutralino annihilation which go to the distinctive $\tau^+ \tau^-$ mode. The ability to make such a measurement, of course, depends on the brightness of the annihilation source (or sources). The number of gamma-rays from dark matter annihilations to be seen by GLAST is constrained by the lack of such detections by EGRET. Although GLAST will be considerably more sensitive than EGRET, it is unlikely that it will detect more than $\sim$10--20 gamma-rays from a source of neutralino annihilation. If multiple such sources are discovered by GLAST, this number could be larger.

\begin{figure}[tbp]
\includegraphics[width=3.2in,angle=-90]{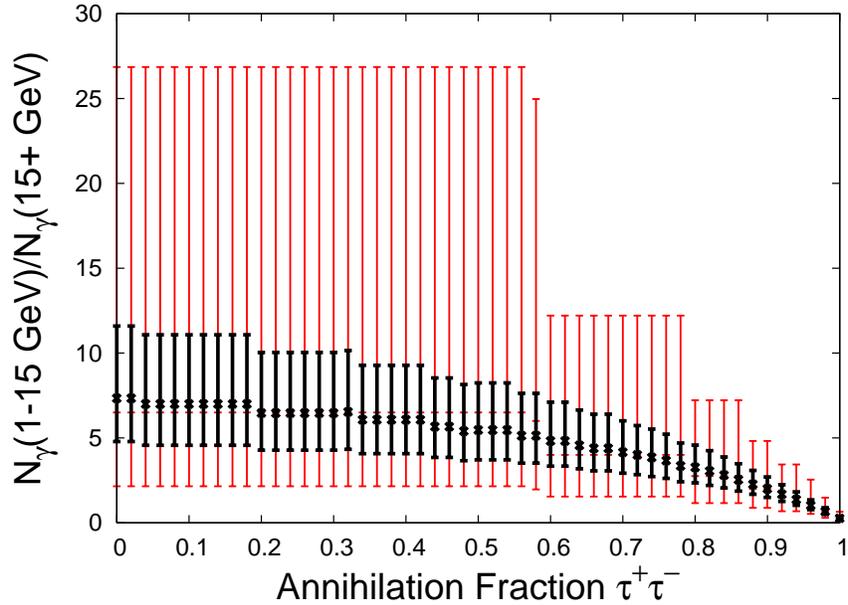}
\caption{The ratio of gamma-rays above and below 15 GeV, as a function of the annihilation fraction to $\tau^+ \tau^-$ (with the remaining annihilations to $b\bar{b}$), for a 500 GeV neutralino. The large and small error bars are for a total number of events of 15 and 100 in GLAST, respectively. A very large number of events would be needed before the fraction of neutralinos to $\tau^+ \tau^-$ could be measured.}
\label{gammaratio}
\end{figure}

In figure~\ref{gammaratio}, we show how difficult it is to identify a $\tau^+ \tau^-$ component in the gamma-ray spectrum from neutralino annihilations. This is done by comparing the number of gamma-rays observed at low energies (1--15 GeV) to those observed at higher energies (above 15 GeV). For as neutralino annihilating purely to b quarks, the ratio of low energy to high energy gamma-rays is $\sim 7.2$, while for annihilations to tau pairs yields a much smaller ratio ($\sim 0.3$). Many events will be required to accurately measure this quantity, unfortunately. The outer and inner error bars shown correspond to a total number of 15 and 100 events, respectively. With 15 events, a neutralino annihilating only to $b\bar{b}$ is indistinguishable from a WIMP annihilating most of the time to $\tau^+ \tau^-$. The situation is not all that much better with 100 events. It appears that gamma-ray measurements will not be able to tell us much about the annihilation modes of a neutralino unless it annihilates almost entirely to taus.

In addition to the continuum emission, monoenergetic gamma-ray lines can be generated through loop diagrams in which neutralinos annihilate directly to final states including photons, $\chi^0_1 \chi^0_1 \rightarrow \gamma \gamma$ and  $\chi^0_1 \chi^0_1 \rightarrow \gamma Z$ \cite{lines}. The cross sections to these final states are only a small fraction of the total neutralino annihilation cross section, however, making lines more difficult to observe. If the dark matter density is large in the galactic center region, ({\it ie.} a cusped \cite{cusp} or adiabatically compressed \cite{ac} profile, or a density spike resulting from the adiabatic accretion of dark matter onto the central supermassive black hole \cite{spike}) it is possible that such lines could be detected. Observations of this region by HESS \cite{hessgc} and MAGIC \cite{magicgc} have revealed a background from this region which will make gamma-ray lines difficult to detect, however \cite{gabi}. In light of this, other regions (such as dwarf spheriodal galaxies \cite{dwarfs}) appear to be more likely sources of observable neutralino annihilation lines.

The observation of gamma-ray lines from neutralino annihilations would provide both a measurement of the neutralino's mass, and information regarding the neutralino's composition. Although the processes which lead to $\gamma \gamma$ and $\gamma Z$ final states include a wide variety of loop diagrams (involving sfermions, charginos, neutralino and Higgs bosons), making it difficult to determine the responsible supersymmetric parameters from so few observables, we have shown in Fig.~\ref{linesplot} that neutralinos with a substantial higgsino (or wino) component generate brighter lines than highly bino-like neutralinos.

\subsection{The Cosmic Positron Flux}

If gamma-rays are detected from neutralino annihilations, they will not tell much about the magnitude of the responsible cross section. This is because little is known about the density of dark matter in the inner region of halo profiles, which is critically important in determining the resulting annihilation rate. The annihilation rate anticipated in the central region of our galaxy, for example, can vary by at least $\sim$8 orders of magnitude depending on whether a profile with a flat core is present, or instead an adiabatically compressed cusp \cite{ac}, or a density spike generated through the accretion of dark matter onto the galaxy's supermassive black hole \cite{spike}.  Given this uncertainty, a measurement of the annihilation rate in the region of the galactic center will tell us little about the neutralino's annihilation cross section. The uncertainties associated with the halo profiles of other objects (dwarf spheroidal galaxies, for example), although less severe, are still sufficiently large to make cross section measurements impractical.

If the annihilation rate could be measured in a region without such enormous uncertainties in the halo profile, however, perhaps a determination of the neutralino's annihilation cross section could be made. The cosmic positron flux is, perhaps, the most useful tool for making such a determination.

Unlike with gamma-ray measurements, cosmic positron (as well as anti-proton) measurements potentially observe the annihilation products of dark matter produced over large volumes of space (several kpc$^3$). Such a measurement, therefore, can be used to determine the product of the neutralino's annihilation cross section and the neutralino density squared, averaged over the sampled volume. The volume sampled roughly corresponds to the distance a typical positron travels from its point of origin before losing the majority of its energy; a few kiloparsecs for positrons in the relevant energy range. As a result of this limited range, only the dark matter distribution in the local halo is relevant to the observed cosmic positron flux. Assuming there are no very large and unknown clumps of dark matter in the surrounding kiloparsecs (which, although not impossible, is very unlikely~\cite{hoopertaylorsilk}), a measurement of the cosmic positron spectrum could be used to infer the neutralino's annihilation cross section (in the low velocity limit) with only a relatively modest degree of astrophysical uncertainty. 

\begin{figure}[t]
\includegraphics[width=2.3in,angle=-90]{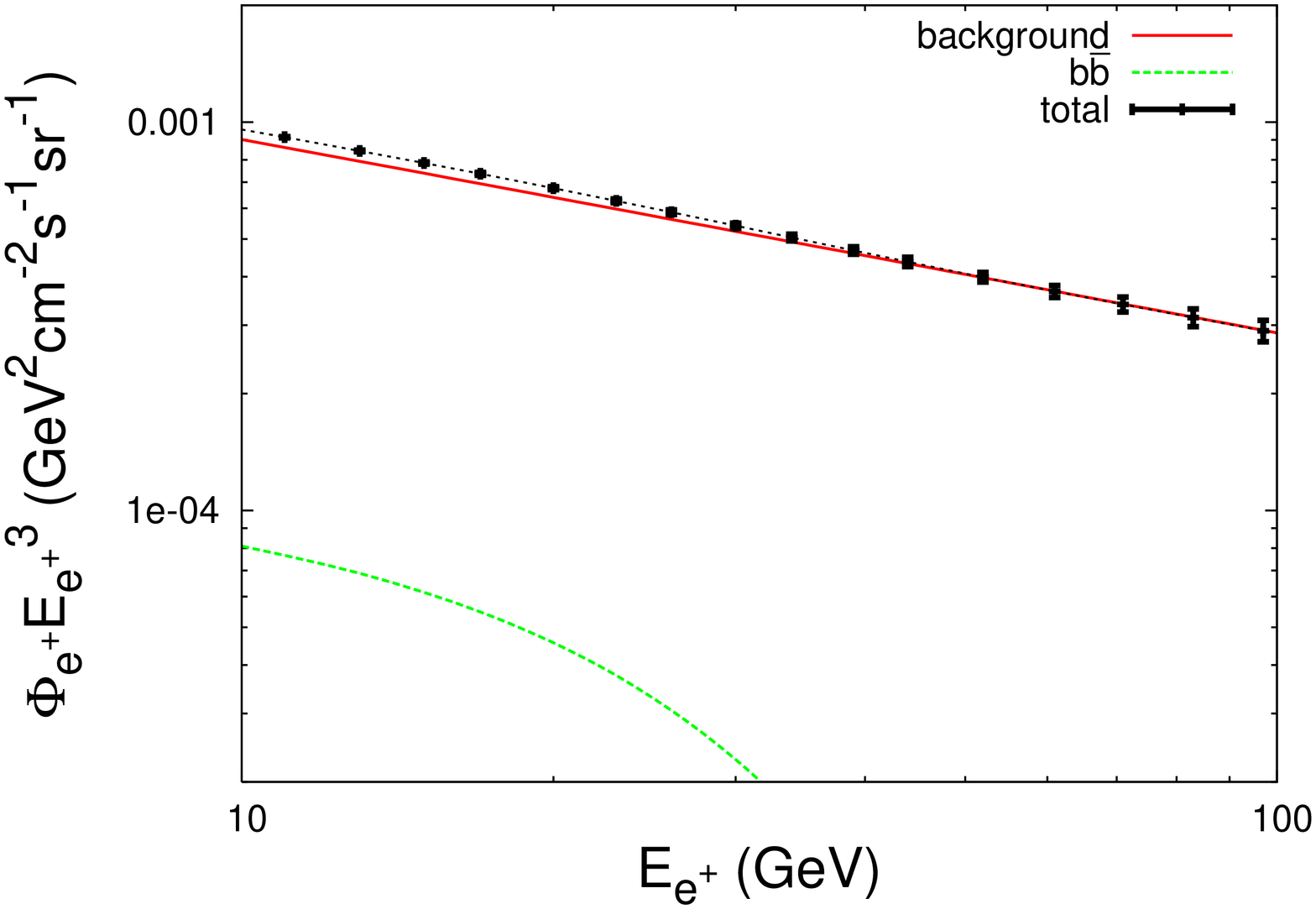}
\includegraphics[width=2.3in,angle=-90]{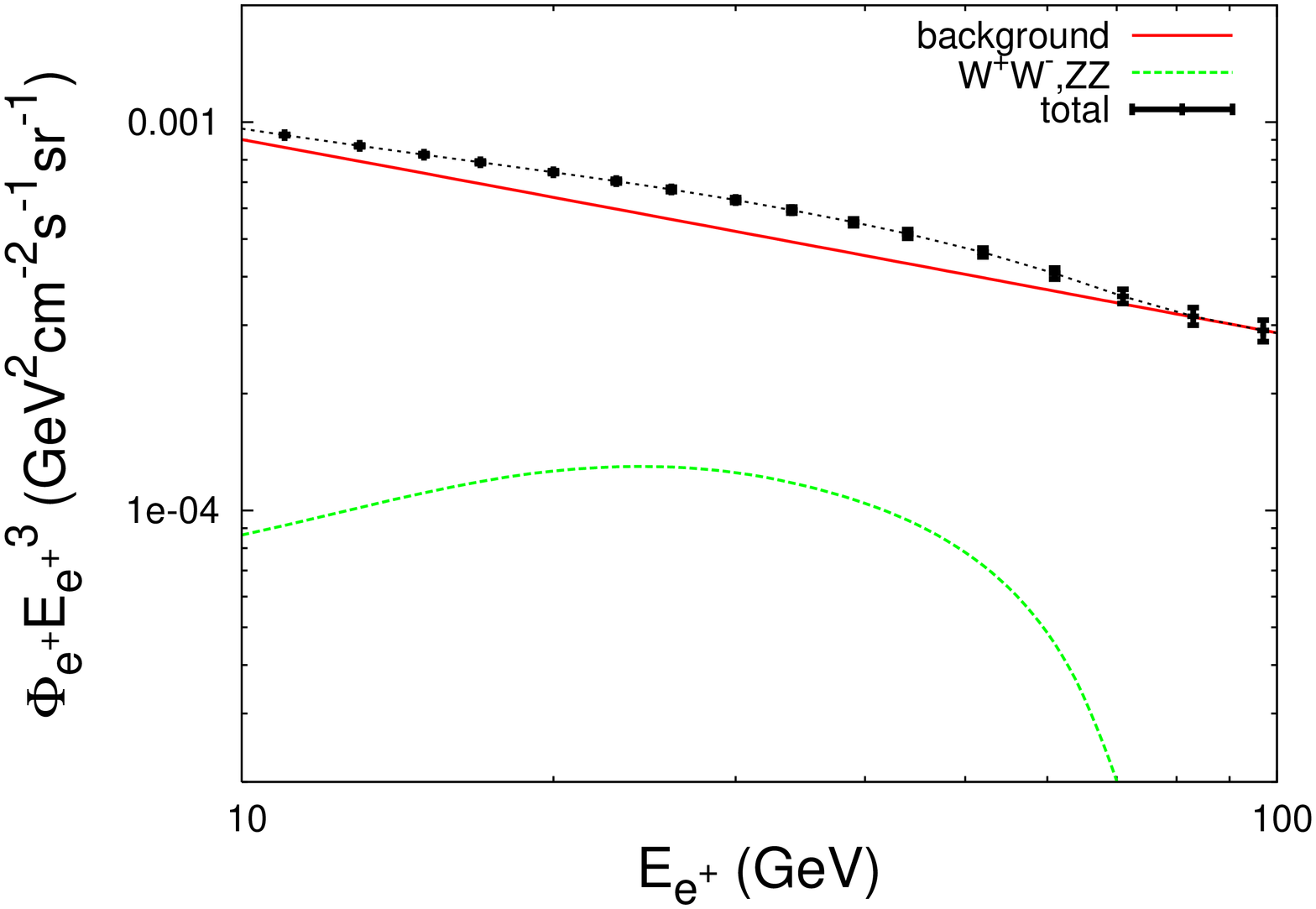}
\caption{The positron spectrum for a 100 GeV neutralino annihilating to $b \bar{b}$ (left) and to $W^+ W^-$, $ZZ$ (right). Also shown as error bars is the projected ability of AMS-02 to measure this spectrum. An annihilation cross section of $3\times 10^{-26}$ cm$^3$/s, and a boost factor of 5 have been used.}
\label{posdetection}
\end{figure}

\begin{figure}[t]
\includegraphics[width=3.2in,angle=-90]{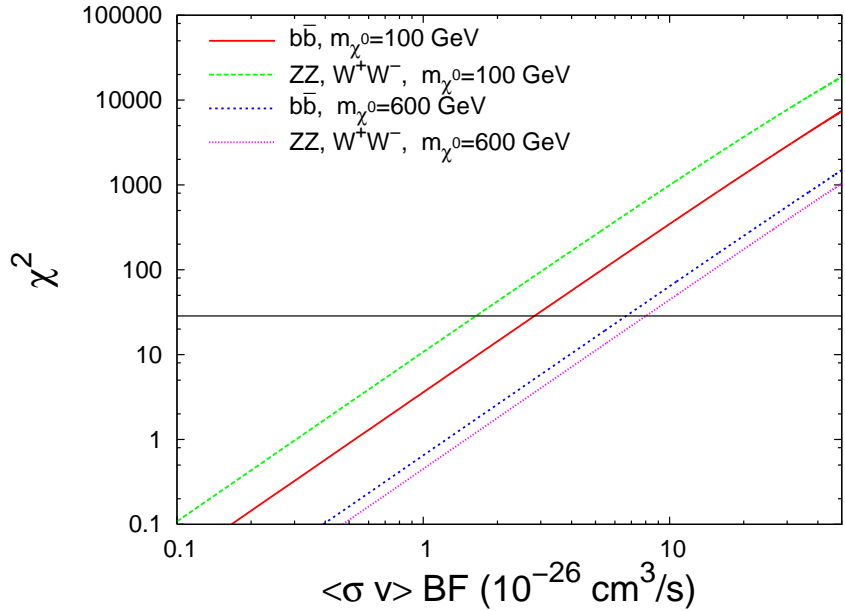}
\caption{The $\chi^2$ over the expected background of a detection of cosmic positrons generated in neutralino annihilations, as projected for AMS-02. Along the x-axis is the neutralino annihilation cross section multiplied by the boost factor resulting from local inhomogeneities in the dark matter distribution. For somewhat light neutralinos, $<\sigma v> \times\,\, \rm{BF} \gsim 2 \times 10^{-26}$ cm$^3$/s could be detected by AMS-02~\cite{silkpos}.}
\label{posdetection2}
\end{figure}

Future measurements of the cosmic positron spectrum will potentially reach the level of sensitivity needed to detect a contribution from neutralino annihilations \cite{silkpos}. In Figs.~\ref{posdetection} and \ref{posdetection2}, we show the ability of the AMS-02 experiment to detect the presence of positrons generated in neutralino annihilations.

\section{Input From Colliders}

If TeV-scale supersymmetry exists in nature, it will very likely be discovered at the Large Hadron Collider (LHC), or perhaps even earlier at the Tevatron. Although the work presented here is not a collider study, we will briefly discuss here some of the measurements that could be made at the LHC or Tevatron that would be complementary to the information ascertained through astrophysical experiments. For further discussions regarding neutralino dark matter at collider experiments see, for example, Refs.~\cite{baltz,drees,allan,baer,kane,nojiri,andreas}.

Of the searches for new physics to be conducted at the LHC, those for squark and gluinos will be among the most easy to perform. Squarks and gluinos are each produced with very large cross sections \cite{sqgl}, and their subsequent decays result in distinctive jets plus missing energy signatures (assuming R-parity is conserved). After about one month of running (at the first year design luminosity), the LHC should be able to discover squarks or gluinos if either are lighter than about 1.5 TeV. Of course, this is dependent upon understanding the behavior of the experiment, which could very well take a year or so to achieve. Ultimately, the LHC will be sensitive to squarks and gluinos with masses up to $\sim3$ TeV \cite{lhc}.

By studying the decays of squarks and/or gluinos, it will also be possible to discover others superpartners at the LHC. For example, in many models, decays of the variety, $\tilde{q} \rightarrow \chi^0_2 q \rightarrow \tilde{l}^{\pm} l^{\mp} q \rightarrow  \chi^0_1 l^+ l^- q$, provide a clean signal of supersymmetry in the form of $l^+l^- +\, \rm{jets}\, + \,\,\rm{missing}\,\, E_T$. By studying the kinematics of these decays, the quantities $m_{\tilde{q}}$, $m_{\chi^0_2}$, $m_{\tilde{l}}$ and $m_{\chi^0_1}$ can each be reconstructed~\cite{recon,drees,fitter}, that is if the sleptons are sufficiently light for this decay chain to take place. More generally speaking, the LHC is in most models likely to measure the mass of the lightest neutralino to roughly 10\% accuracy, and may also be able to determine the masses of one or more of the other neutralinos, and any light sleptons \cite{lhc}. Charginos are much more difficult to study at the LHC. 

Production of heavy, neutral Higgs bosons ($A$, $H_1$), can also be studied at the LHC. In particular, in models with large $\tan \beta$, heavy Higgs bosons have enhanced couplings to down-type fermions, thus leading to potentially observable di-tau final states. If enough of these events are observed, the masses of the heavy Higgs bosons could be potentially reconstructed, and $\tan \beta$ measured \cite{ditau,higgsmeasure}.

Prospects for the discovery of supersymmetry at the Tevatron, although not nearly as strong as at the LHC, are also very exciting. The most likely discovery channel at the Tevatron is probably through clean tri-lepton plus missing energy events originating from the production of a chargino and a heavy neutralino, followed by a decay of the form, $\chi^{\pm} \chi^0_2 \rightarrow \tilde{\nu} l^{\pm} l^{\pm} \tilde{l^{\mp}} \rightarrow  \nu \chi^0_1 l^{\pm} l^+ l^- \chi^0_1$ \cite{fermilab}. Only models with rather light gaugino masses (neutralinos and charginos) can be discovered in this way, however. For some of the recent results from supersymmetry searches at the Tevatron, see Ref.~\cite{recenttevatron}.

Determining the structure of an underlying supersymmetric theory from collider observables is by no means a simple task. Considerable effort has been recently been put into the ``inverse problem'' \cite{inverse,sfitter,fittino,lhc,recon} of extracting supersymmetric parameters from LHC data. Computational tools such as Sfitter \cite{sfitter} and Fittino \cite{fittino} have been developed for this purpose.

We are not going to attempt a detailed collider analysis in this paper. Instead, we will simply adopt the position that, at a future time in which astrophysical dark matter experiments may be measuring the neutralino's phenomenological characteristics relevant to direct and indirect detection, collider experiments such as the LHC will be likely to have determined some of the basic features of the supersymmetric spectrum, such as the squark and gluino masses, the lightest neutralino's mass (and possibly one or more other neutralino masses) and possibly $\tan\beta$ and $m_A$ (in the case of large $\tan \beta$ and light $m_A$). In the more distant future, an International Linear Collider (ILC) could measure many more of these quantities, and at far greater precision \cite{ilc}, revealing a great deal about supersymmetry and neutralino dark matter~\cite{baltz}.

\begin{figure}[tbp]
\includegraphics[width=6.5in,angle=0]{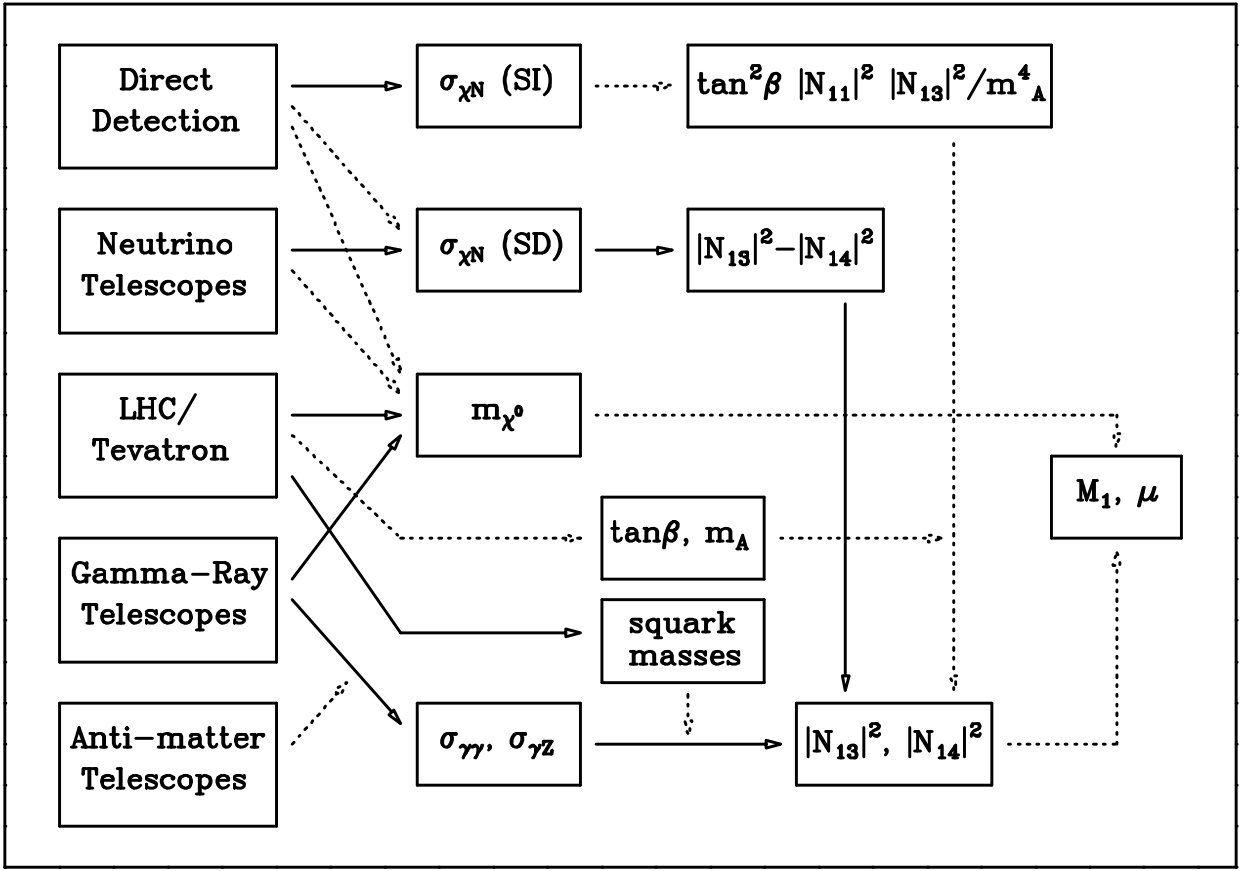}
\caption{An illustration of how supersymmetric parameters can be determined though astrophysical observations, aided by collider measurements. Solid lines represent methods that are more reliable or likely, whereas dotted lines are (qualitatively) less so. See text for further discussion.}
\label{flowchart}
\end{figure}
\section{Putting It All Together: Constraining The SUSY Model}

Thus far, we have discussed several ways of constraining supersymmetric parameters with astrophysical and collider measurements. We turn our attention now to combining these observables in an effort to learn as much as possible about the characteristics of supersymmetry. 

In Fig.~\ref{flowchart}, we illustrate how supersymmetric parameters, such as $M_1$ and $\mu$ (or alternatively the mass and composition of the lightest neutralino), could be ascertained from astrophysical observations, aided by measurements at the Tevatron and/or LHC. In particular, we emphasize that information regarding the neutralino's composition is included in the elastic scattering rate, the neutrino flux from the Sun and in the brightness of gamma-ray lines. If we assist this process with measurements of the lightest neutralino's mass and squark masses at the LHC (as well as $\tan \beta$ and $m_A$ in some cases), we can combine these observables to hopefully arrive at a determination of supersymmetric parameters which are not well constrained by collider measurements alone.

The complex interplay between these different observables is very difficult to discuss in a model independent fashion. To further study the collective reach of these experiments, we will consider a set of representative supersymmetric benchmark models.

\section{Benchmark Models}

In an effort to evaluate how astrophysical measurements will be able to help constrain the nature of supersymmetry, in this section we will consider a series of benchmark models. We intend for these representative models to illustrate how various astrophysical observations could be used collectively to infer information about the composition of the lightest neutralino and other characteristics of supersymmetry.  Although many sets of supersymmetric benchmark scenarios have been proposed elsewhere (see for example, Refs.~\cite{snowmass,michigan}), we have chosen to introduce our own here. The input parameters of these benchmark models are given in table~\ref{t1}.

\vspace{0.5cm}
 \begin{table}[!ht]
 \hspace{0.0cm}
 \begin{tabular} {c c c c c c c c c c c c c c c c c c c c c c} 
\hline \hline
 Model &\vline& $M_2$ & \,\, &  $\mu$ &\,\,&  $\tan \beta$ &\,\,&  $m_{\tilde{f}}$ &  \,\,&  $m_A$  &  \,\,&  $A_t$ &  \,\,&  $A_b$ &\vline& $\Omega_{\chi^0} h^2$ && $\delta a_{\mu}$  &\vline& Notes  \\
 \hline \hline
LT1 &\vline& 120.2 && 302.0 && 55.7 && 1715. && 352.2 && 2538. && 5145. &\vline& 0.097 && $8 \cdot 10^{-10}$ &\vline& \\
 \hline
LT2  &\vline& 167.9 &&  350.6 &&  56.1 && 430.4 &&  326.2 && 761.5 && 1129. &\vline& 0.111 && $7 \cdot 10^{-9}$&\vline& \\
\hline
LT3  &\vline& 214.5 &&  232.3 &&  14.9 && 350.5 &&  296.0 && 34.5 && -76.8 &\vline& 0.090 && $3 \cdot 10^{-9}$&\vline& bino-higgsino \\
\hline
LT4  &\vline& 123.4 &&  574.2 &&  58.2 && 810.1 &&  236.6 && 1784. && -890.0 &\vline& 0.109 && $2 \cdot 10^{-9}$&\vline& \\
\hline
\hline
IM1  &\vline& 550.8 &&  1318. &&  6.8 && 2239. &&  580.2 && 1308. && 2902. &\vline& 0.079 && $4 \cdot 10^{-11}$&\vline& $A$-funnel \\
\hline
IM2  &\vline& 468.1 &&  296.1 &&  7.1 && 2820. &&  745.1 && 1452. && 329.3 &\vline& 0.108 && $3 \cdot 10^{-11}$&\vline& bino-higgsino \\
\hline
IM3  &\vline& 472.9 &&  619.2 &&  50.6 && 2130. &&  396.5 && 4343. && -3717. &\vline& 0.098 && $4 \cdot 10^{-10}$&\vline& $A$-funnel \\
\hline
IM4  &\vline& 627.6 &&  380.4 &&  12.2 && 505.8. &&  317.8 && 788.1 && 1297. &\vline& 0.097 && $6 \cdot 10^{-10}$&\vline& bino-higgsino \\
\hline
IM5  &\vline& 463.0 &&  862.1 &&  10.0 && 323.3. &&  908.1 && 123.1 && 331.5 &\vline& 0.103 && $8 \cdot 10^{-10}$&\vline& $\tilde{q}$, $\tilde{l}$ coannihilation \\
\hline
\hline
HV1  &\vline& 825.9 &&  1073. &&  4.2 && 2100. && 873.7 &&  3471. && -2329. &\vline& 0.103 && $2 \cdot 10^{-11}$&\vline& $A$-funnel \\
\hline
HV2  &\vline& 1296. &&  671. &&  35.0 && 3046. && 817.9 &&  2352. && 6506. &\vline& 0.090 && $1 \cdot 10^{-10}$&\vline& bino-higgsino \\
\hline
HV3  &\vline& 1123. &&  3068. &&  25.8 && 863.3 && 816.7 &&  133.8 && 8567. &\vline& 0.088 && $3 \cdot 10^{-10}$&\vline& $\tilde{t}$ coannihilation \\
 \hline \hline
 \end{tabular}
 \caption{The set of benchmark models we have adopted in our study. In addition to the input parameters of the model, the relic density and the amplitude of the contribution to the muon's magnetic moment are given. Each point shown is in agreement with the constraint on the branching fraction $B \rightarrow X_s \gamma$ \cite{bsg} as well as with all direct collider constraints. The labels LT, IM and HV denote models with light, intermediate mass and heavy neutralinos.}
\label{t1}
 \end{table}

\vspace{0.5cm}
 \begin{table}[!ht]
 \hspace{-2.0cm}
 \begin{tabular} {c c c c c c c c c c c c c c c c c c c c c c c c} 
\hline \hline
Model &\vline& $m_{\chi^0_1}$ && $|N_{11}|^2$  && $|N_{12}|^2$ && $|N_{13}|^2$ && $|N_{14}|^2$ &\vline& $\sigma_{\chi N}$ (SI) && $\sigma_{\chi p}$ (SD) && $\sigma_{\rm{tot}}v$ && $\sigma_{\gamma \gamma}v$ && $\sigma_{\gamma Z}v$ && $R_{\nu}$ \\
 \hline \hline
 LT1 &\vline& 58.8 && 0.97 && 0.002 && 0.03 && 0.001 &\vline& $6.7 \cdot 10^{-8}$  && $2.2 \cdot 10^{-5}$ && $4\cdot 10^{-27}$ &&  $2\cdot 10^{-33}$ &&  $8\cdot 10^{-34}$ && $< 10^{-3}$  \\
 \hline
 LT2 &\vline&  82.8 &&  0.98 &&  0.001 &&  0.02 &&  0.001 &\vline& $1.2 \cdot 10^{-7}$ && $7.0 \cdot 10^{-6}$&& $2\cdot 10^{-26}$ && $2 \cdot 10^{-31}$&& $3\cdot 10^{-32}$&& $0.06$ \\
\hline
 LT3 &\vline&  100.9 &&  0.90 &&  0.01 &&  0.08 &&  0.02 &\vline& $9.7 \cdot 10^{-8}$ && $9.2 \cdot 10^{-5}$&& $2\cdot 10^{-26}$ && $8 \cdot 10^{-32}$&& $2\cdot 10^{-30}$&& $1.8$ \\
\hline
 LT4 &\vline&  61.6 &&  0.993 &&  0.0002 &&  0.006 &&  $1 \times 10^{-5}$ &\vline& $7.8 \cdot 10^{-8}$ && $8.4 \cdot 10^{-7}$&& $8\cdot 10^{-27}$ && $2 \cdot 10^{-32}$&& $2\cdot 10^{-34}$&& $0.0001$ \\
\hline
\hline
 IM1 &\vline&  276.4 &&  0.999 &&  $3\cdot 10^{-5}$ &&  0.001 &&  0.0002 &\vline& $7.2 \cdot 10^{-11}$ && $3.3 \cdot 10^{-8}$&& $1\cdot 10^{-27}$ && $7 \cdot 10^{-33}$&& $4\cdot 10^{-33}$&& $0.002$ \\
\hline
 IM2 &\vline&  217.9 &&  0.75 && 0.012 &&  0.14 && 0.090 &\vline& $2.2 \cdot 10^{-8}$ && $9.9 \cdot 10^{-5}$&& $2\cdot 10^{-26}$ && $9 \cdot 10^{-31}$&& $1\cdot 10^{-29}$&& $68$ \\
\hline
 IM3 &\vline&  236.4 &&  0.991 && 0.0001 &&  0.007 && 0.001 &\vline& $9.6 \cdot 10^{-9}$ && $1.2 \cdot 10^{-6}$&& $3\cdot 10^{-26}$ && $7 \cdot 10^{-33}$&& $4\cdot 10^{-34}$&& $0.7$ \\
\hline
 IM4 &\vline&  300.6 &&  0.77 && 0.006 &&  0.12 && 0.08 &\vline& $7.4 \cdot 10^{-8}$ && $4.8 \cdot 10^{-5}$&& $2\cdot 10^{-26}$ && $2 \cdot 10^{-30}$&& $6\cdot 10^{-30}$&& $44$ \\
\hline
 IM5 &\vline&  231.8 &&  0.996 && $8 \times 10^{-5}$ &&  0.003 && 0.0004 &\vline& $1.4 \cdot 10^{-9}$ && $1.1 \cdot 10^{-6}$&& $4\cdot 10^{-27}$ && $4 \cdot 10^{-30}$&& $8\cdot 10^{-30}$&& $0.7$ \\
\hline
\hline
 HV1 &\vline&  413.8 &&  0.996 && $6 \times 10^{-5}$ &&  0.003 && 0.0009 &\vline& $1.8 \cdot 10^{-10}$ && $9.2 \cdot 10^{-8}$&& $2\cdot 10^{-27}$ && $2 \cdot 10^{-32}$&& $7\cdot 10^{-33}$&& $0.02$ \\
\hline
 HV2 &\vline&  627.0 &&  0.60 && 0.003 &&  0.21 && 0.19 &\vline& $3.6 \cdot 10^{-8}$ && $2.3 \cdot 10^{-5}$&& $3\cdot 10^{-26}$ && $2 \cdot 10^{-30}$&& $1\cdot 10^{-29}$&& 17 \\
\hline
 HV3 &\vline&  565.0 &&  0.9998 && $3\times 10^{-7}$ &&  0.0002 && $1\times 10^{-5}$ &\vline& $4.7 \cdot 10^{-11}$ && $3.3 \cdot 10^{-8}$&& $4\cdot 10^{-28}$ && $5 \cdot 10^{-31}$&& $8\cdot 10^{-32}$&& 0.0004 \\
 \hline \hline
 \end{tabular}
 \caption{Phenomenological quantities for our benchmark models. From left to right are: the lightest neutralino's mass (in GeV), the bino, wino and higgsino fractions of the lightest neutralino, the spin-independent elastic scattering cross section of the neutralino with nucleons (in pb), the spin-dependent elastic scattering cross section of the neutralino with protons (in pb), the annihilation cross section of the lightest neutralino (in the low velocity limit, in cm$^3$/s), the annihilation cross section to $\gamma \gamma$ and $\gamma Z$ final states (in cm$^3$/s), and the rate per square kilometer, per year in a neutrino telescope (with a 50 GeV muon energy threshold).}
\label{t2}
 \end{table}

In addition to the input parameters, in table~\ref{t1} we also show the relic density and the amplitude of the contribution to the muon's magnetic moment for each benchmark model, both calculated using DarkSUSY \cite{darksusy}. For each model, the relic density falls within the 2$\sigma$ range measured by WMAP \cite{wmap}. The value of the muon's magnetic moment, as measured using $e^+ e^-$ data, is larger than the Standard Model prediction by $\delta a_{\mu}=(23.9\pm7.2_{\rm{had-lo}}\pm3.5_{\rm{lbl}}\pm6_{\rm{exp}}) \times 10^{-10}$, where the error bars denote theoretical uncertainties in the leading order hadronic and hadronic light-by-light contributions, as well as the experimental contribution \cite{gminus2}. Considering this data alone, our benchmark models LT2, IM1, IM2, HV1, HV2 and HV3 are each outside of the 2$\sigma$ range of this measurement. Given the unresolved issues involved in this measurement (such as the lower value found using $\tau^+ \tau^-$ data) and the relativity low statistical significance of the measurement, we do not exclude any models on this criterion and leave it to the reader to account for this measurement as they choose. We have checked that each benchmark model is consistent with the measurement of the branching fraction $B \rightarrow X_s \gamma$ \cite{bsg}.

\begin{figure}[tbp]
\includegraphics[width=3.3in,angle=0]{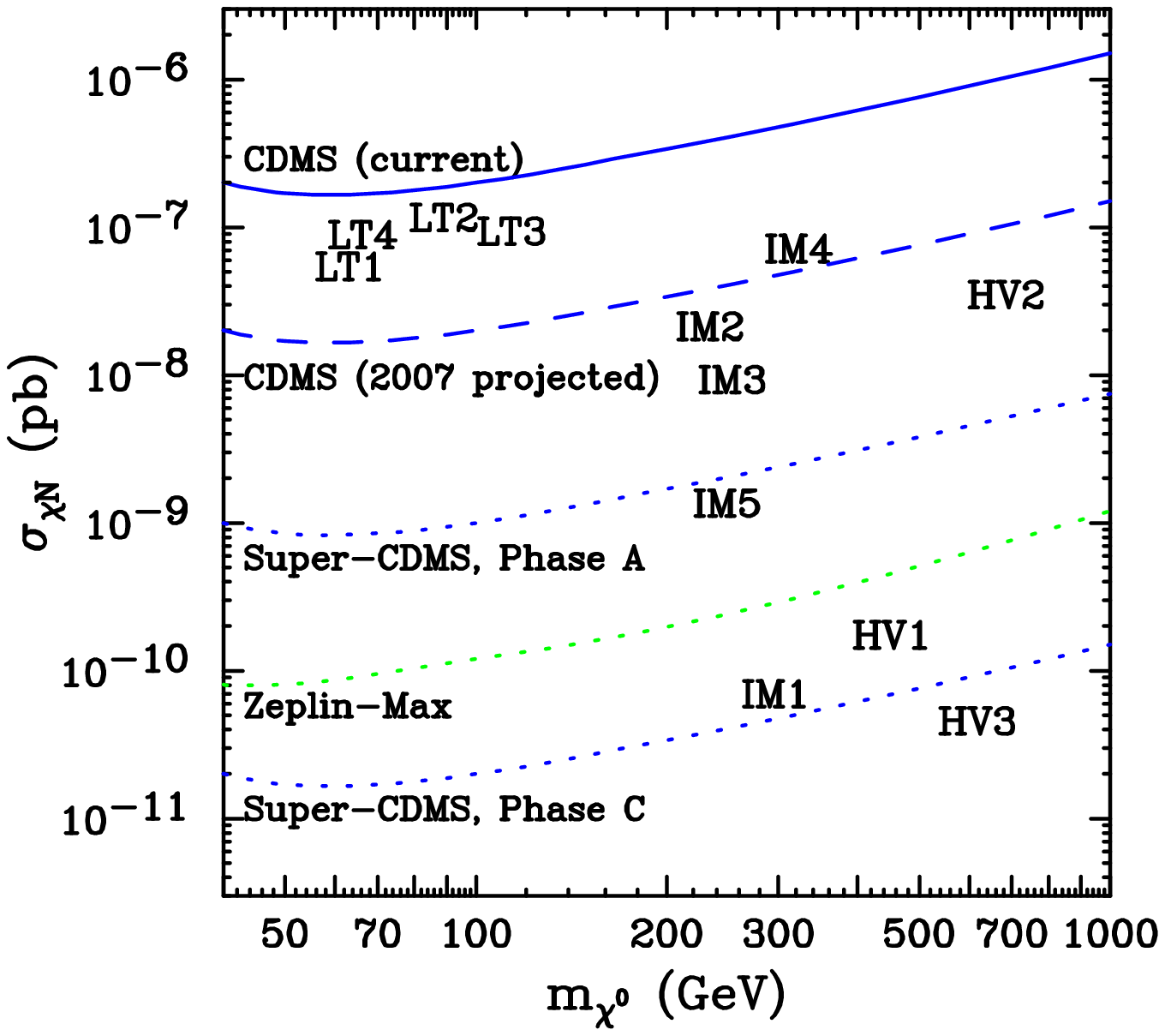}
\includegraphics[width=3.3in,angle=0]{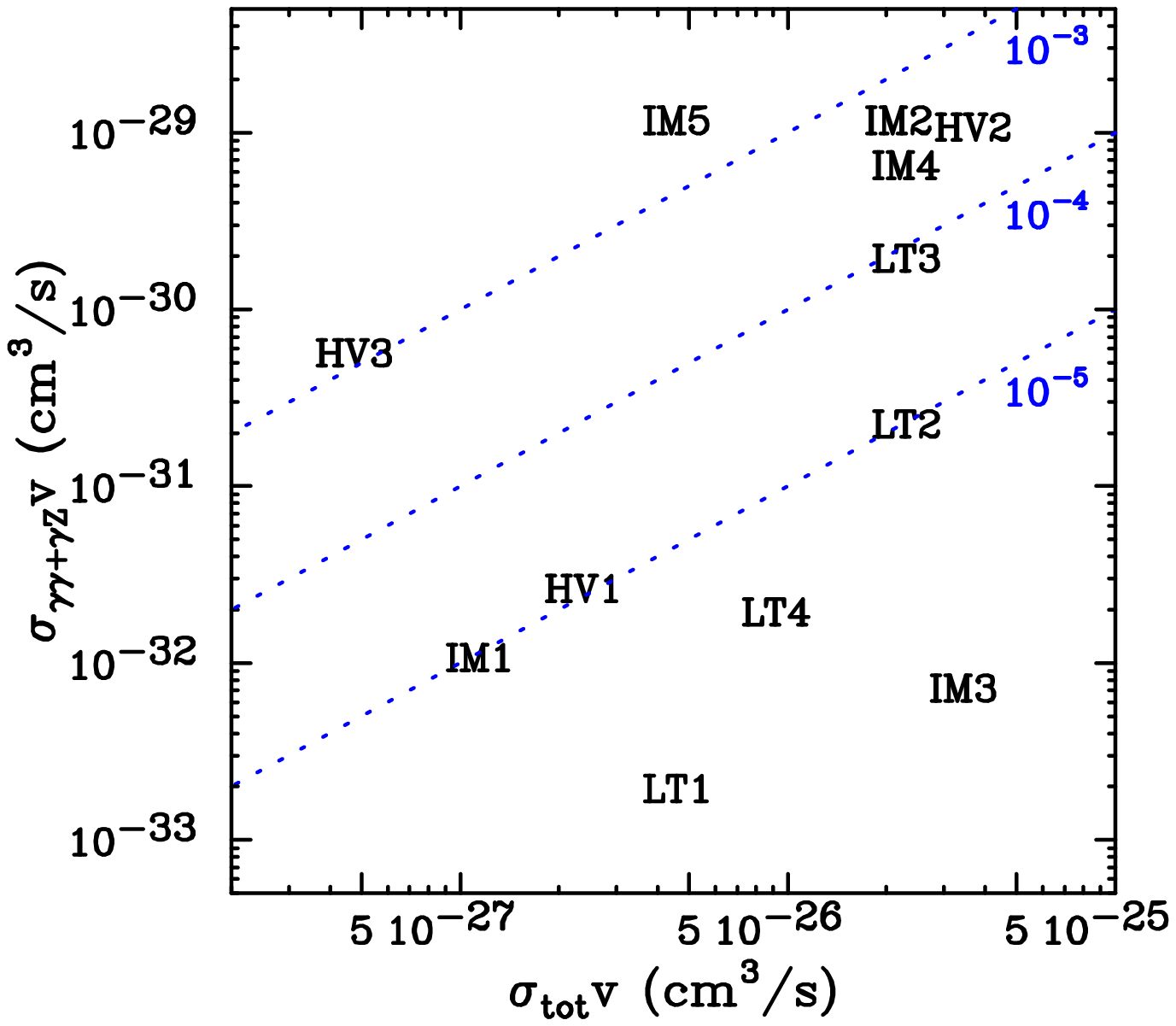}
\caption{In the left frame, we plot the spin-independent elastic scattering cross section for neutralinos in our benchmark models, compared to current and projected constraints. The solid line is the current constraint from the CDMS experiment \cite{cdms}. The dashed line is the projected reach of CDMS (2007), whereas the dotted lines are (top to bottom) the projected reach of Super-CDMS Phase A, Zeplin-Max and Super-CDMS Phase C (see Ref.~\cite{plotter}). In the right frame, we compare the benchmark neutralinos' annihilation cross section to gamma-ray lines ($\gamma \gamma$ plus $\gamma Z$) to the total annihilation cross section. As dotted lines, we plot contours of constant $\sigma_{\gamma \gamma +\gamma Z}/\sigma_{\rm{tot}}$.}
\label{dirben}
\end{figure}

In table~\ref{t2}, we show a number of phenomenological characteristics of the benchmark models, including the mass and composition of the lightest neutralino, as well as the most relevant elastic scattering and annihilation cross sections, and the rate predicted in neutrino telescopes. In the left frame of Fig.~\ref{dirben}, we plot the spin-independent elastic scattering cross sections of the lightest neutralino in these models compared to current and projected direct detection constraints. Although none of the benchmark models are excluded by current constraints, several of the models are within the projected reach of direct detection experiments in the near future. Experiments such as Super-CDMS (phase C), Zeplin-Max, EURECA, or XENON will be capable of testing all but the most heavy supersymmetric models (assuming that a neutralino makes up the dark matter of our universe). In the right frame of the figure, we compare the benchmark neutralinos' annihilation cross section to gamma-ray lines to the total annihilation cross section (in the low velocity limit).

\vspace{0.5cm}
 \begin{table}[tbp]
 \hspace{-1.0cm}
 \begin{tabular} {c c c c c c c c c c c c c c c c c c c c} 
\hline \hline
  &\vline&  &&   LHC&&  && &\vline& CDMS/ZP/ED  &\vline& IceCube/KM3  &\vline& AMS-02 &\vline& GLAST/ACT && \\
\hline
 Model &\vline& $m_{\chi^0_1}$ && $m_A$  && $\tan\beta$ && $m_{\tilde{q}}$ &\vline&  $\sigma_{\chi N}$ (SI) &\vline& $R_{\nu}$ &\vline& $\sigma_{\rm{tot}}v$ &\vline&  $\sigma_{\gamma \gamma+\gamma Z}$/$\sigma_{\rm{tot}}$ && \\
 \hline \hline
 LT1 &\vline& $\pm 10\%$ && YES && $\pm 15\%$ && $\pm 20\%$ &\vline& $\times 2$, $\div 2$ &\vline& $\lsim 10$  &\vline& $\lsim 2\times 10^{-26} $ &\vline& $\lsim 10^{-5}$ &&  \\
 \hline
 LT2 &\vline& $\pm 10\%$ && YES && $\pm 15\%$ && $\pm 10\%$ &\vline& $\times 2$, $\div 2$ &\vline& $\lsim 10$  &\vline& $\times 5$, $\div 5$ &\vline& $\times 10$, $\div 10$ &&  \\
\hline
 LT3 &\vline& $\pm 10\%$ &&  $\pm 20\%$ && $\pm 30\%$ && $\pm 10\%$ &\vline& $\times 2$, $\div 2$ &\vline& $\lsim 10$  &\vline& $\times 5$, $\div 5$ &\vline& $\times 5$, $\div 5$ &&  \\
\hline
 LT4 &\vline& $\pm 10\%$ &&  YES && $\pm 15\%$ && $\pm 10\%$ &\vline& $\times 2$, $\div 2$ &\vline& $\lsim 10$  &\vline& $\times 5$, $\div 5$ &\vline& $\lsim 10^{-5}$ &&  \\
\hline
\hline
 IM1 &\vline& $\pm 10\%$ && NO && NO && $\pm 30\%$ &\vline& $\lsim 10^{-10}$ &\vline& $\lsim 10$  &\vline& $\lsim 2 \times 10^{-26}$ &\vline& $\times 10$, $\div 10$ &&  \\
\hline
 IM2 &\vline& $\pm 10\%$ && NO && NO && $\pm 30\%$ &\vline& $\times 2$, $\div 2$ &\vline& $\times 2$, $\div 2$  &\vline&  $\times 5$, $\div 5$ &\vline& $\times 3$, $\div 3$ &&  \\
\hline
 IM3 &\vline& $\pm 10\%$ && YES && $\pm 15\%$ && $\pm 30\%$ &\vline& $\times 2$, $\div 2$ &\vline& $\lsim 10$  &\vline&  $\times 5$, $\div 5$ &\vline& $\lsim 10^{-5}$ &&  \\
\hline
 IM4 &\vline& $\pm 10\%$ &&  $\pm 20\%$ && $\pm 30\%$ && $\pm 10\%$ &\vline& $\times 2$, $\div 2$ &\vline& $\times2$, $\div 2$  &\vline&  $\times 5$, $\div 5$ &\vline& $\times 3$, $\div 3$ &&  \\
\hline
 IM5 &\vline& $\pm 10\%$ && NO && NO && $\pm 10\%$ &\vline& $\times 2$, $\div 2$ &\vline& $\lsim 10$  &\vline& $\lsim 2 \times 10^{-26}$ &\vline& $\times 3$, $\div 3$ &&  \\
\hline
\hline
 HV1 &\vline& $\pm 10\%$ && NO && NO && $\pm 30\%$ &\vline& $\times 4$, $\div 4$ &\vline& $\lsim 10$  &\vline& $\lsim 2 \times 10^{-26}$ &\vline& $\times 10$, $\div 10$ &&  \\
\hline
 HV2 &\vline& $\pm 10\%$ && $\pm 20\%$ && $\pm 30\%$ && $\gsim$ 2 TeV &\vline& $\times 2$, $\div 2$ &\vline& $\times 2$, $\div 2$  &\vline& $\times 5$, $\div 5$ &\vline& $\times 3$, $\div 3$ &&  \\
\hline
 HV3 &\vline& $\pm 10\%$ && NO && NO && $\pm 10\%$  &\vline& $\lsim 10^{-10}$ &\vline& $\lsim 10$  &\vline& $\lsim 2\times 10^{-26}$ &\vline& $\times 3$, $\div 3$ &&  \\
 \hline \hline
 \end{tabular}
 \caption{A hypothetical set of measurements made by collider and astrophysical experiments for each of our benchmark models. Experiments considered include the Large Hadron Collider (LHC), direct detection experiments (CDMS, Zeplin, Edelweiss, and their planned extensions), neutrino telescopes (IceCube, KM3), antimatter detectors (AMS-02) and gamma-ray telescopes (GLAST, MAGIC, HESS, VERITAS, etc.). $\pm$ percentages and factors shown reflect the precision with which these measurements may plausibly be made. Entries reading as $\lsim X$ simply constrain the observable quantity. An entry of `YES' indicates a measurement of a few GeV accuracy or better. Even if $\tan \beta$ and $m_A$ cannot be measured at the LHC (as denoted by `NO' in the table), the lack of a detection can be used to constrain these quantities roughly as $(\tan\beta/10)(200\, \rm{GeV}/m_A) \lsim 1$.}
\label{t3}
 \end{table}

In table~\ref{t3}, we put forth a set of measurements which could plausibly be made by future experiments for each of our benchmark models. For the LHC, we focus on a few of the easiest to measure and most important quantities: $m_{\chi^0_1}, m_A, \tan\beta$ and $m_{\tilde{q}}$.\footnote{By $m_{\tilde{q}}$, we mean the approximate masses of the first and second generation squarks. The stops and sbottoms, in particular, are sometimes significantly heavier and lighter than this value.} The mass of the lightest neutralino can typically be measured to $\sim$10\% accuracy, as can squark masses if they are sufficiently light (in the case of very heavy squarks, we have estimated less precise measurements of this quantity). $m_A$ and $\tan\beta$ can potentially be measured if sufficiently light and large, respectively, by studying processes such as $A \rightarrow \tau^+ \tau^-$. If this channel is observable, $\tan \beta$ is expected to be measured with a precision of approximately 15\%~\cite{higgsmeasure}. It may be possible to determine $m_A$ within a few GeV at the LHC.

In some of these benchmark models, other supersymmetric particles may be measured at the LHC, especially in the lighter models. Often the masses of one or more additional neutralino and/or light slepton can be determined, for example. To explore this further would require a detailed collider study, which is beyond the scope of this paper.

In addition to quantities measurable at the LHC, a number of astrophysical observables are considered in table~\ref{t3}. Projecting the precision at which these quantities will be measured in the coming years is more difficult as they depend on a number of unknown astrophysical inputs. Measurements of the neutralino's elastic scattering cross section depend on the local dark matter density and, therefore, may be skewed if we happen to live in a dense clump of dark matter, or in a local void. Since these extreme deviations from the dynamically inferred average local density are quite unlikely~\cite{white}, we will proceed under the assumption that, with sufficient exposure, direct detection experiments will measure $\sigma_{\chi N}$ to roughly within a factor of two.

As we stated in section~\ref{neutrino}, the flux of high-energy neutrinos from the Sun will need to generate at least $\sim$10 events per year if it is to be distinguished from the atmospheric neutrino background by experiments such as IceCube or KM3. This channel is even less subject to astrophysical uncertainties than direct detection because the capture rate of neutralinos in the Sun is averaged over the history of the solar system and, therefore, is robust to local fluctuations in the dark matter distribution.

Measurements of the cosmic anti-matter spectrum are subject to uncertainties in the galactic dark matter distribution and corresponding neutralino annihilation rate. Proceeding under the assumption that the positron boost factor is not surprisingly large (BF $\lsim 10$), AMS-02 will either be capable of making a rough measurement of the neutralino's annihilation cross section (in the low velocity limit), or will place an upper limit on this quantity.

It is impossible to know whether the inner cusps of the Milky Way and nearby dwarf spheriodal galaxies will be sufficiently concentrated to generate observable gamma-ray signals from annihilating neutralinos. If they do generate observable gamma-ray fluxes, then the brightness of $\gamma \gamma$ and $\gamma Z$ lines, relative to the continuum emission, could be determined. If the brightness of lines is low compared to the continuum flux, this will be more difficult.

We now will ask how much information regarding the supersymmetric parameters is gained by considering astrophysical measurements in addition to LHC data. We begin by considering the quantity $\mu$. The parameter $\mu$ is difficult to constrain at the LHC in many supersymmetric scenarios. The spectrum and compositions of neutralinos and charginos could be studied to determine $\mu$ at a linear collider, but at the LHC it will likely go largely unconstrained. In contrast, $\mu$ is very important to the phenomenology of neutralino dark matter, critically effecting the composition of the LSP and therefore its couplings and interactions.

\begin{figure}[!tbp]
\hspace{-0.8cm}
\includegraphics[width=2.5in,angle=-90]{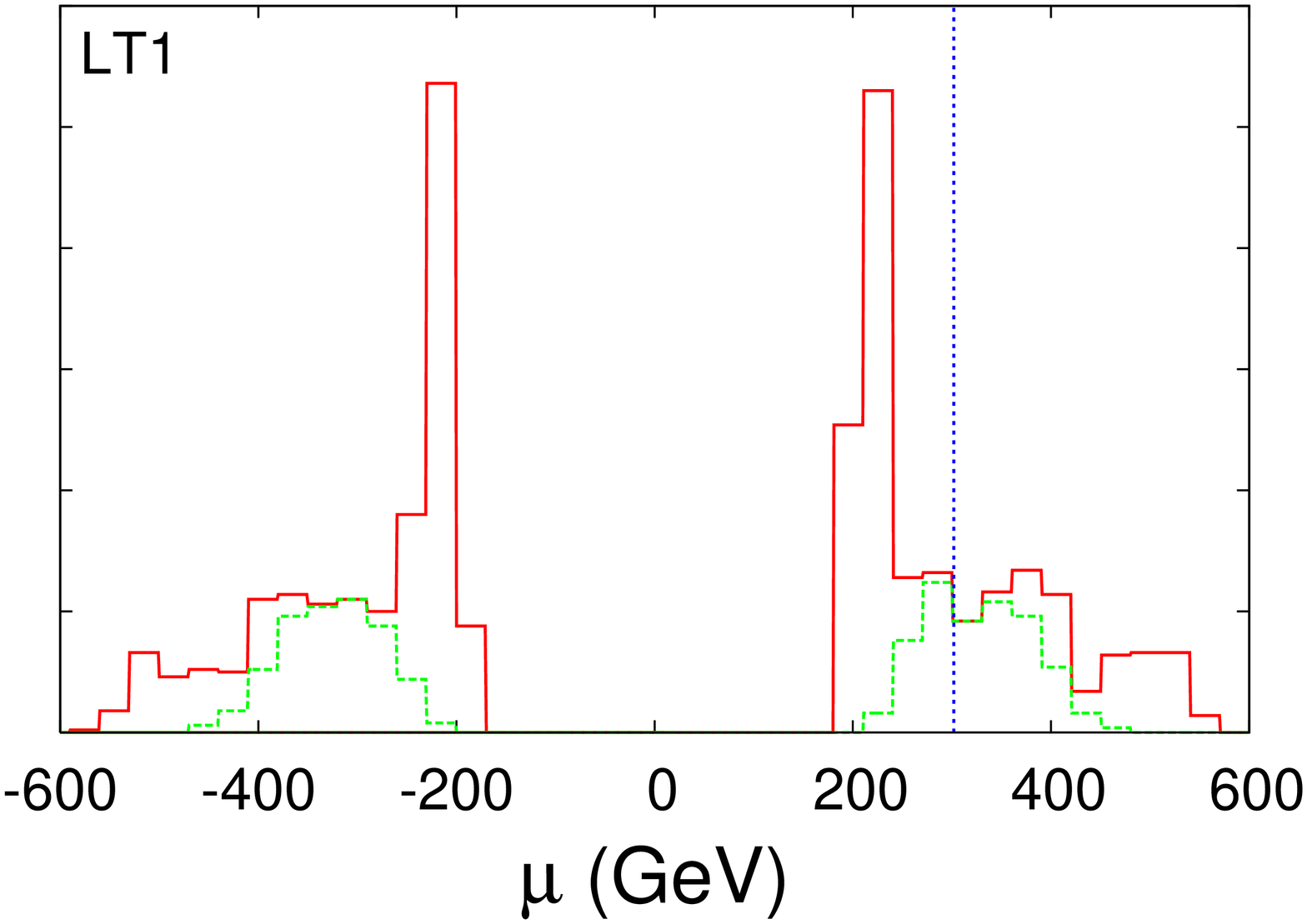}
\hspace{-0.8cm}
\includegraphics[width=2.5in,angle=-90]{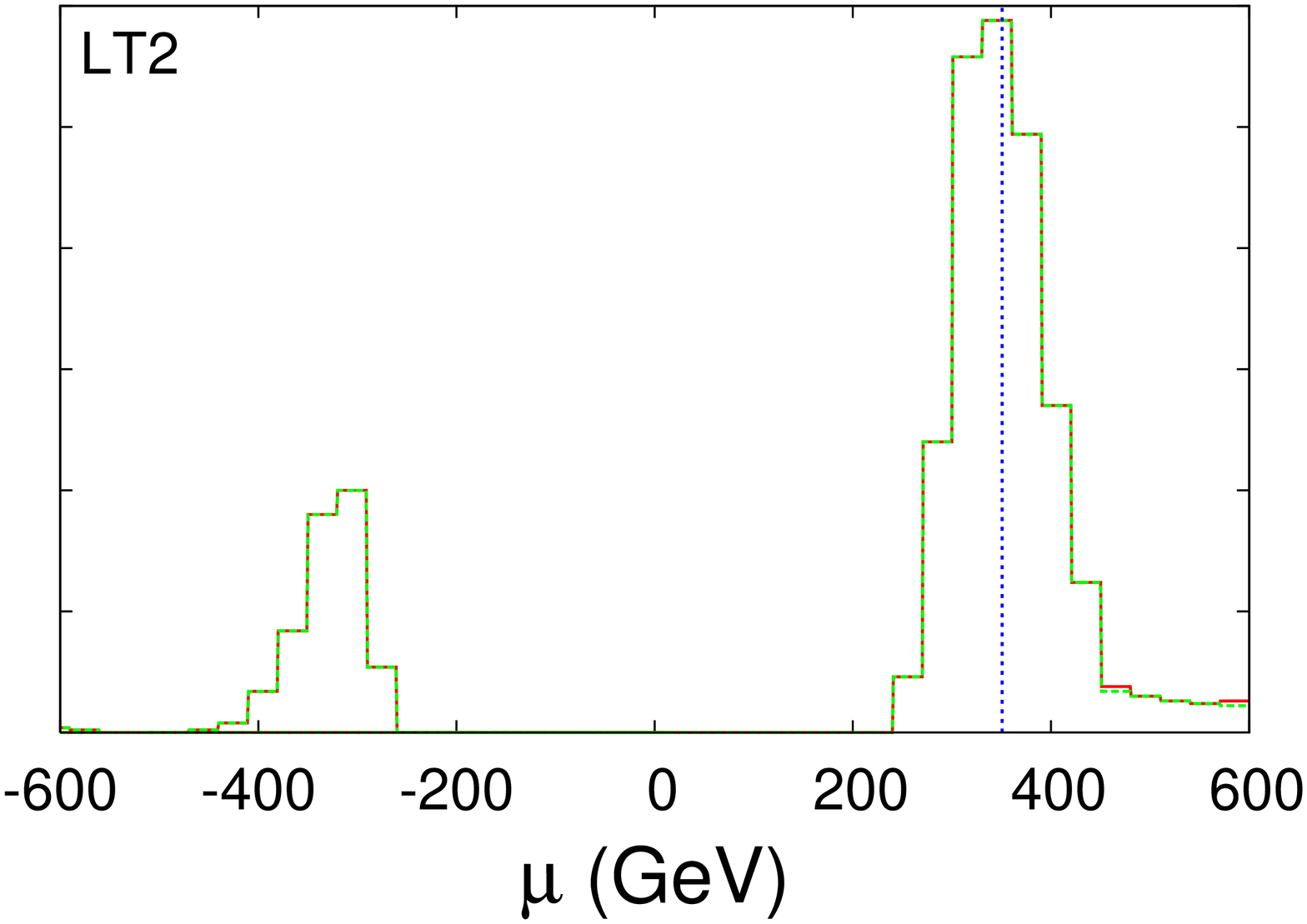}
\\
\hspace{-0.8cm}
\includegraphics[width=2.5in,angle=-90]{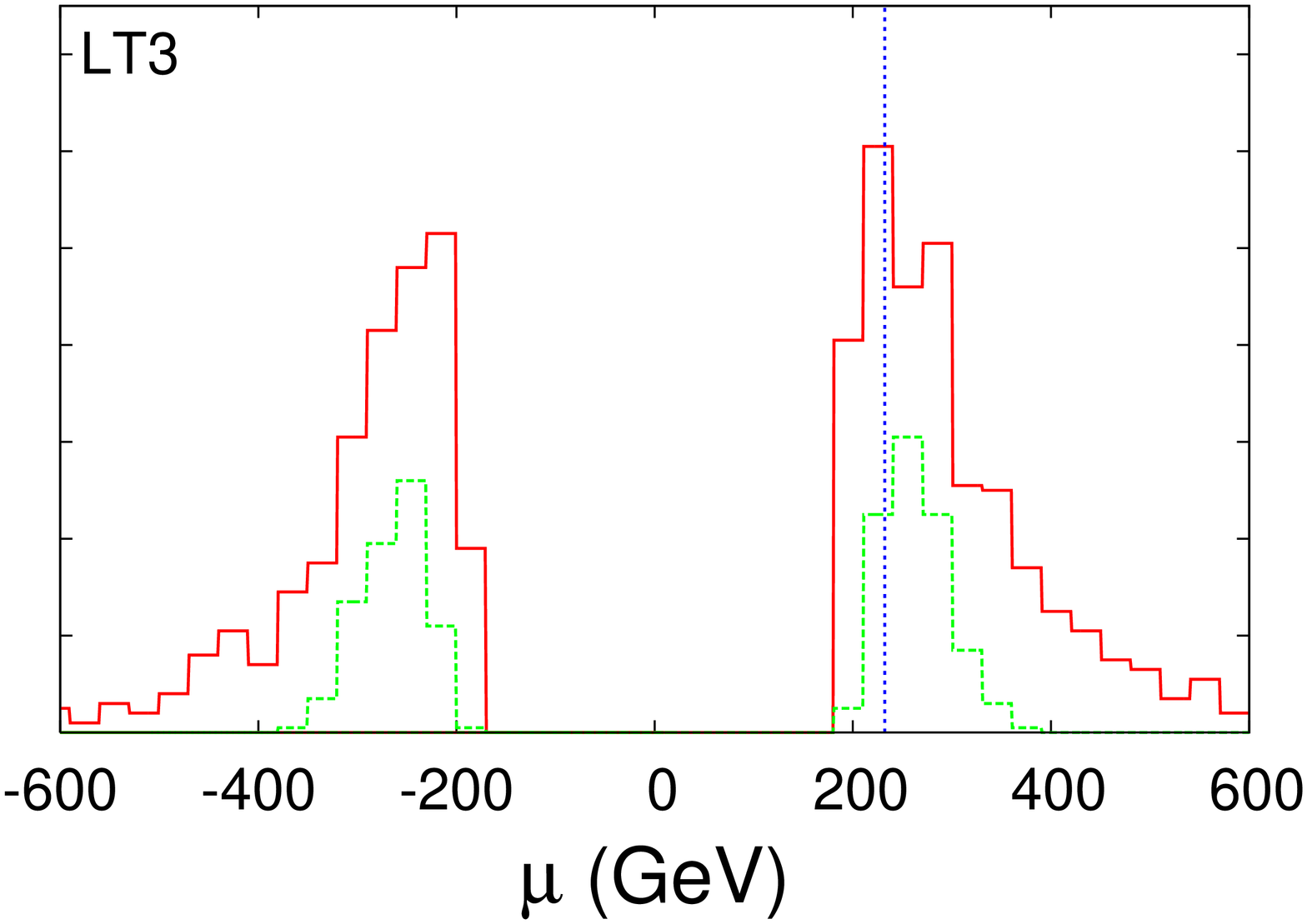}
\hspace{-0.8cm}
\includegraphics[width=2.5in,angle=-90]{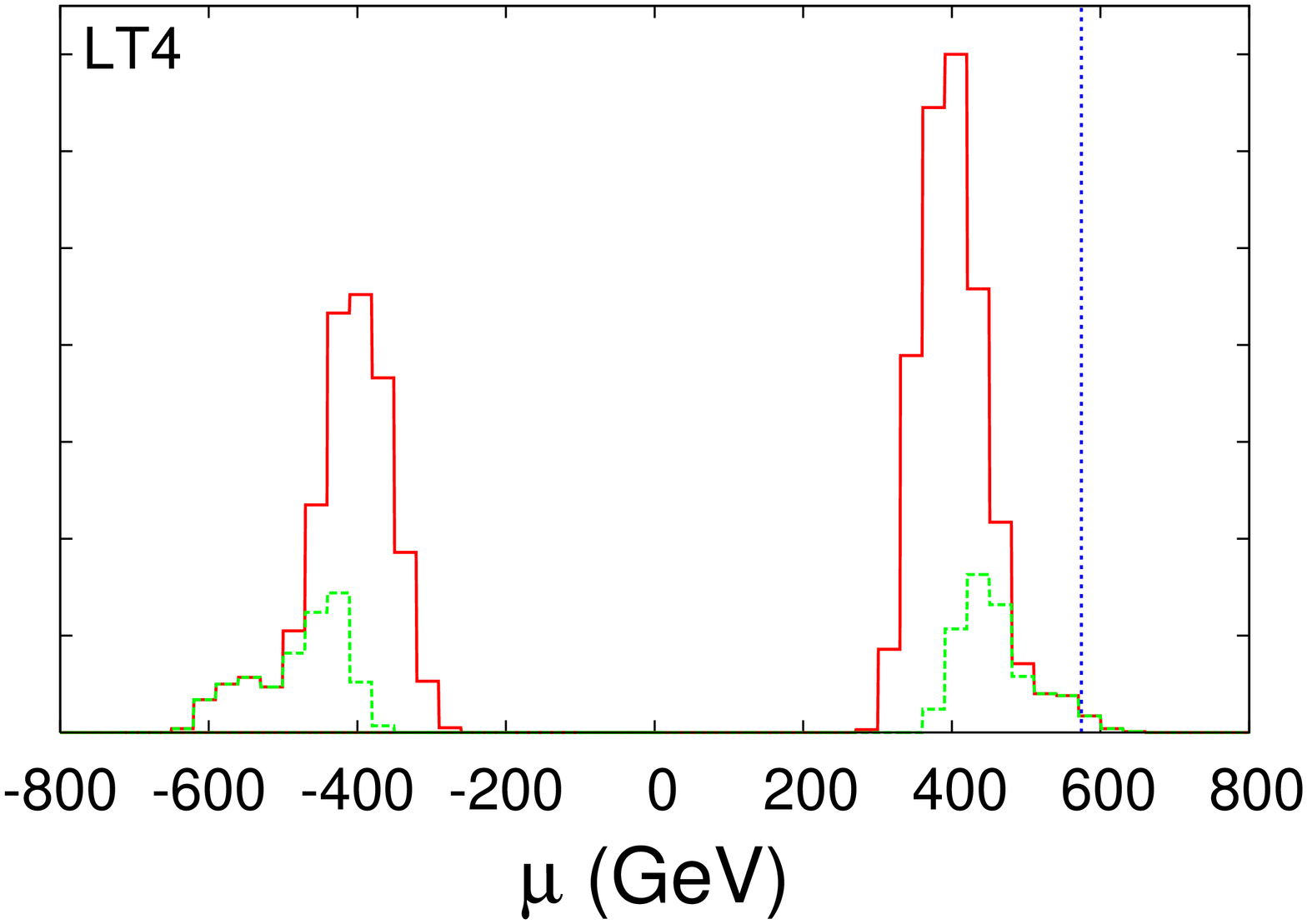}
\\
\hspace{-0.8cm}
\includegraphics[width=2.5in,angle=-90]{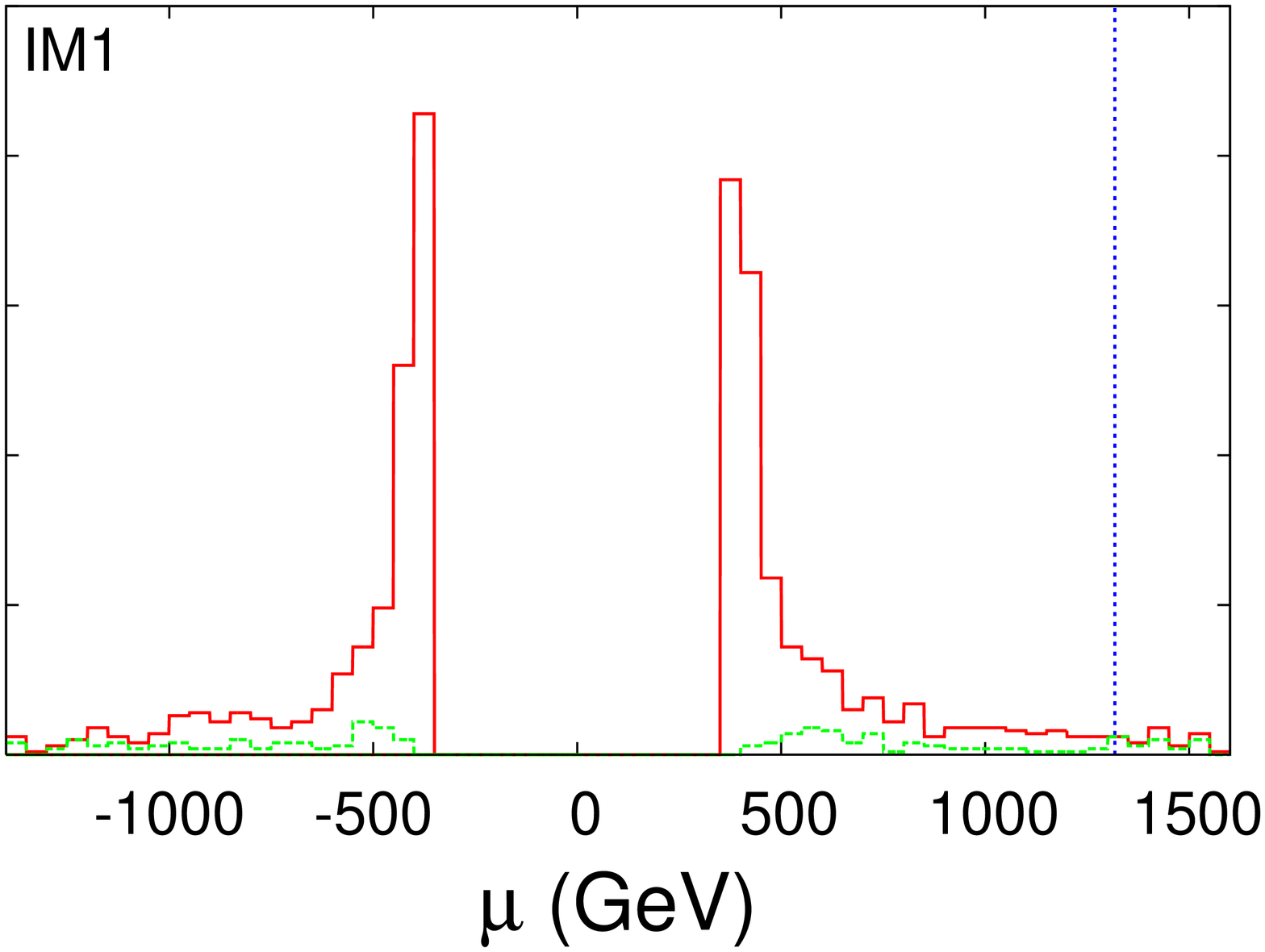}
\hspace{-0.8cm}
\includegraphics[width=2.5in,angle=-90]{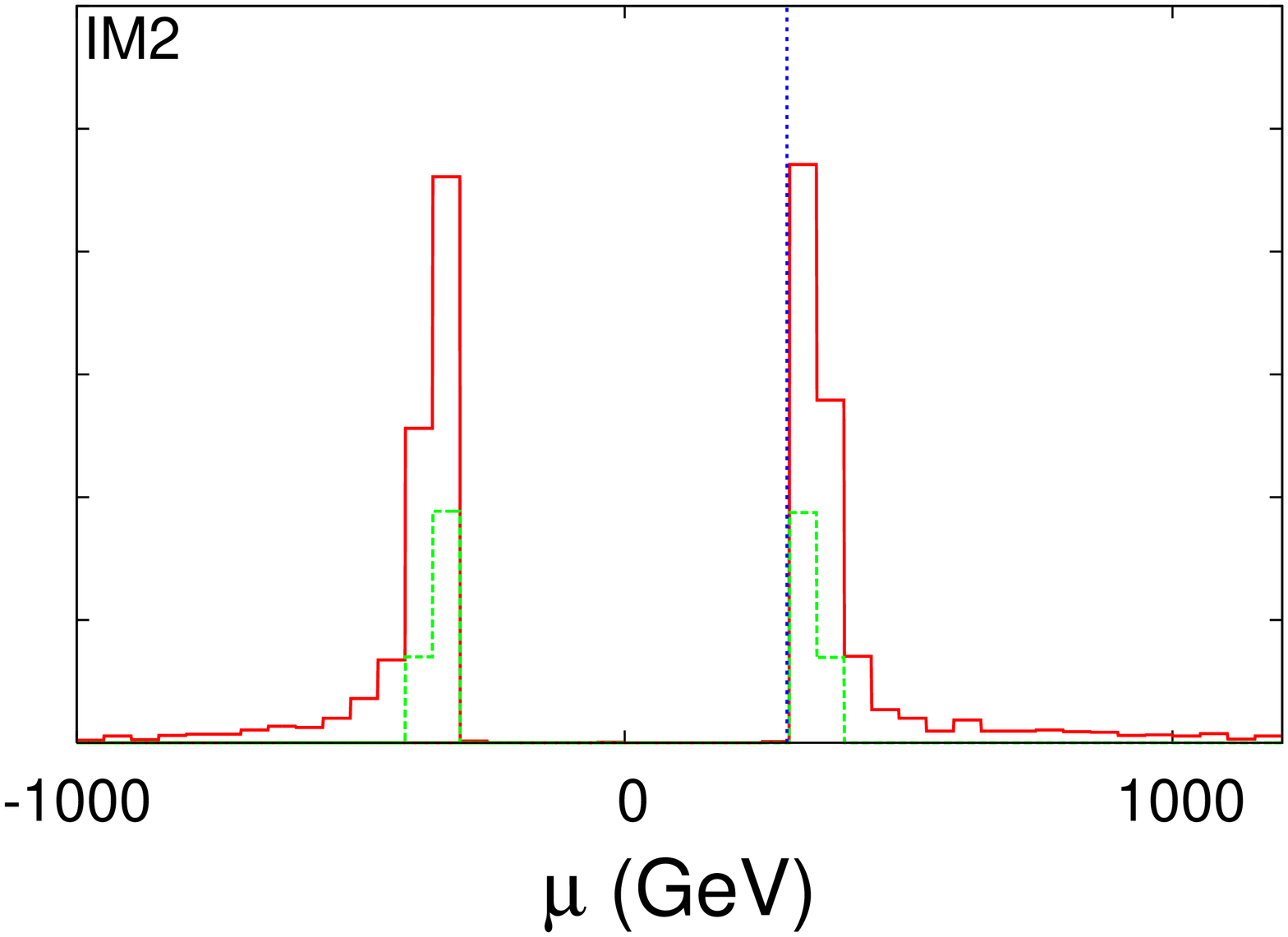}
\\
\caption{The ability of future astrophysical and collider (LHC) data to constrain the parameter $\mu$ in the first six of our benchmark models. In each frame, the upper histogram describes the distribution of models found in our random parameter scan which satisfy the LHC measurements given in table~\ref{t3}, in addition to the relic density constraint. The lower histogram for each benchmark model contains those models which {\it also} satisfy the astrophysical measurements given in table~\ref{t3}. The vertical dotted line in each frame denotes the actual value of $\mu$ for the given benchmark model. For models LT1, LT3, LT4, IM1 and IM2, astrophysical data allows for $\mu$ to be considerably more constrained. The model LT2 benefits far less from astrophysical data. The sign of $\mu$ is not expected to be determined by the techniques discussed here.}
\label{muhist1}
\end{figure}

\begin{figure}[!tbp]
\hspace{-0.8cm}
\includegraphics[width=2.5in,angle=-90]{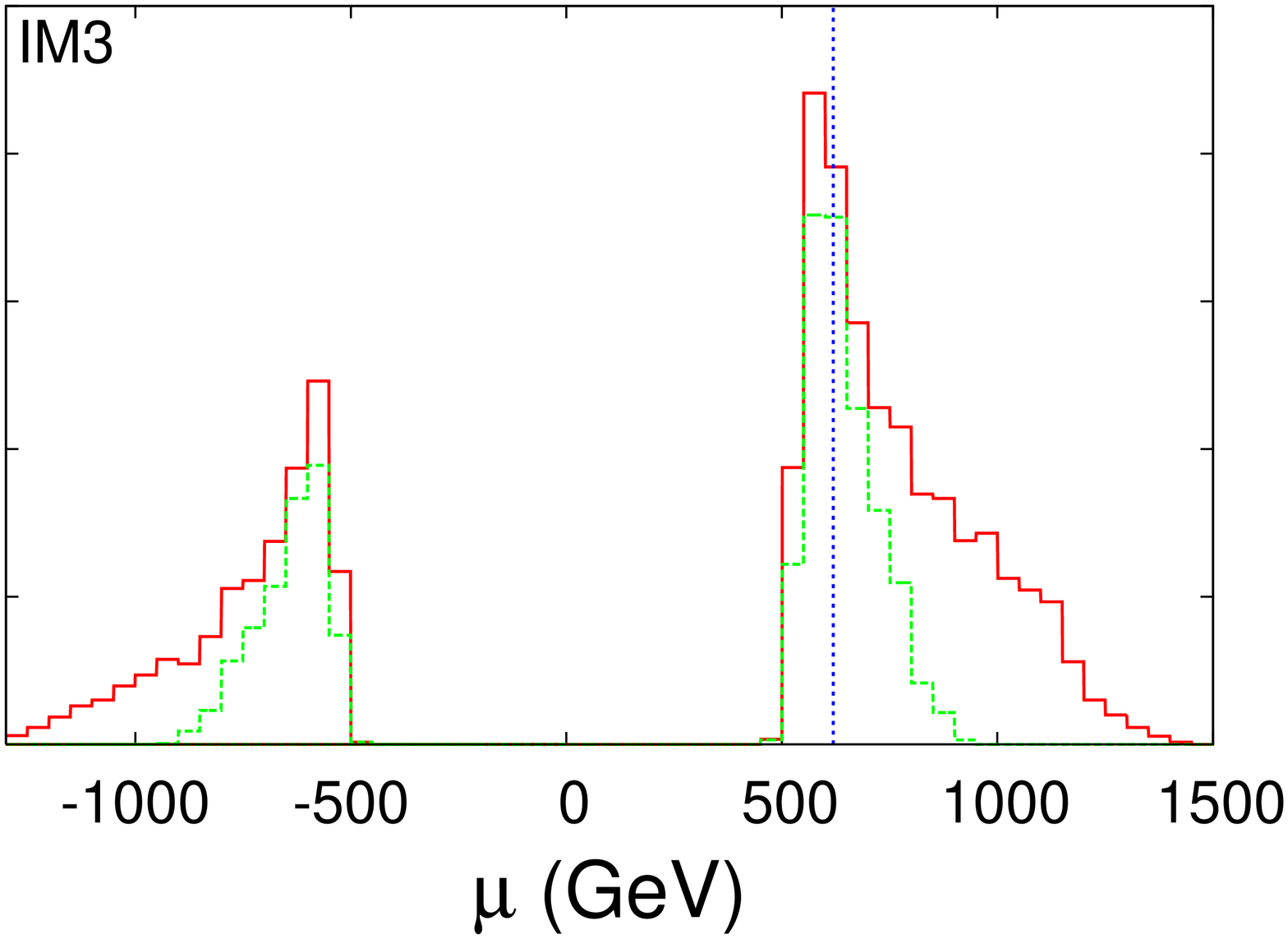}
\hspace{-0.8cm}
\includegraphics[width=2.5in,angle=-90]{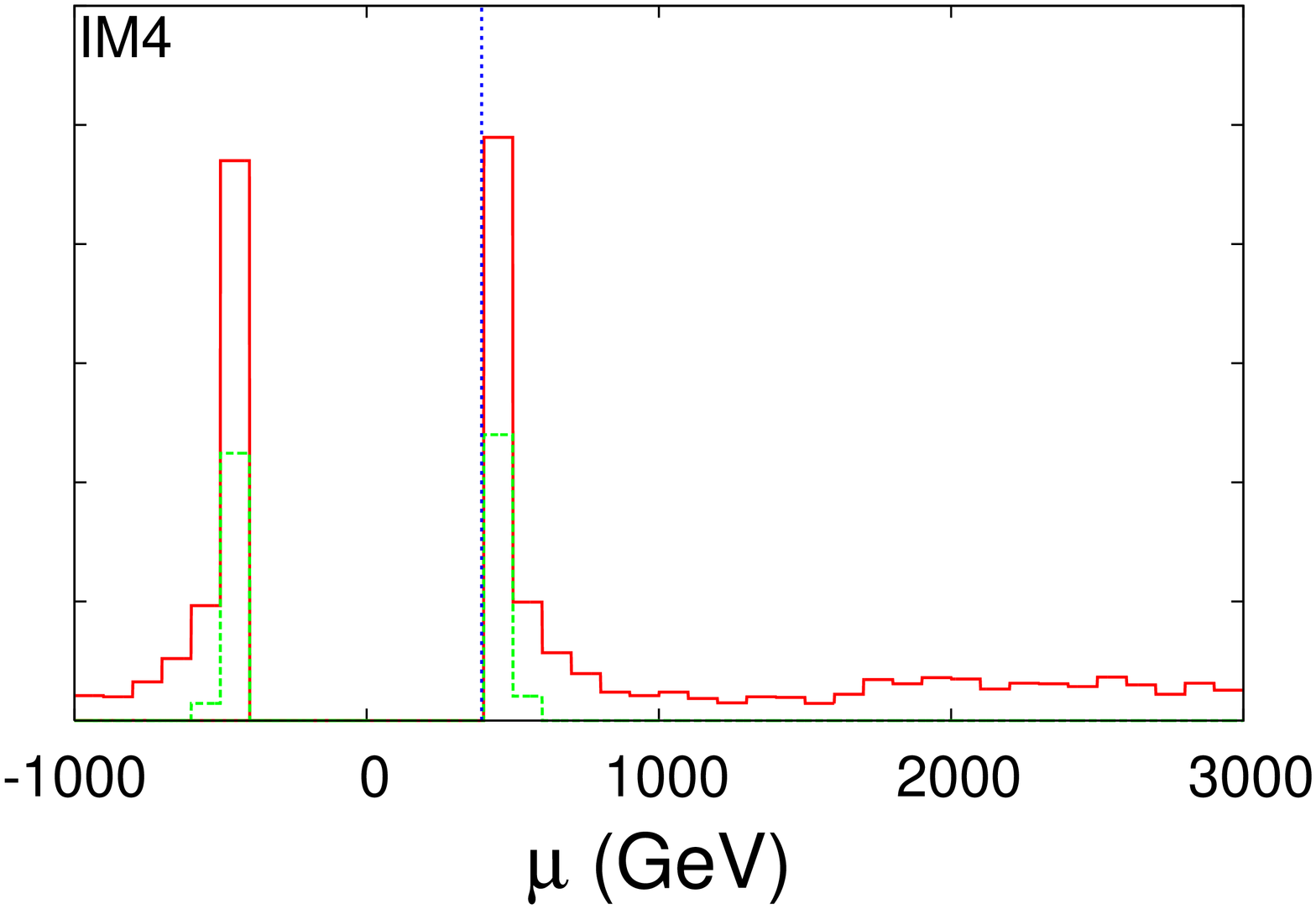}
\\
\hspace{-0.8cm}
\includegraphics[width=2.5in,angle=-90]{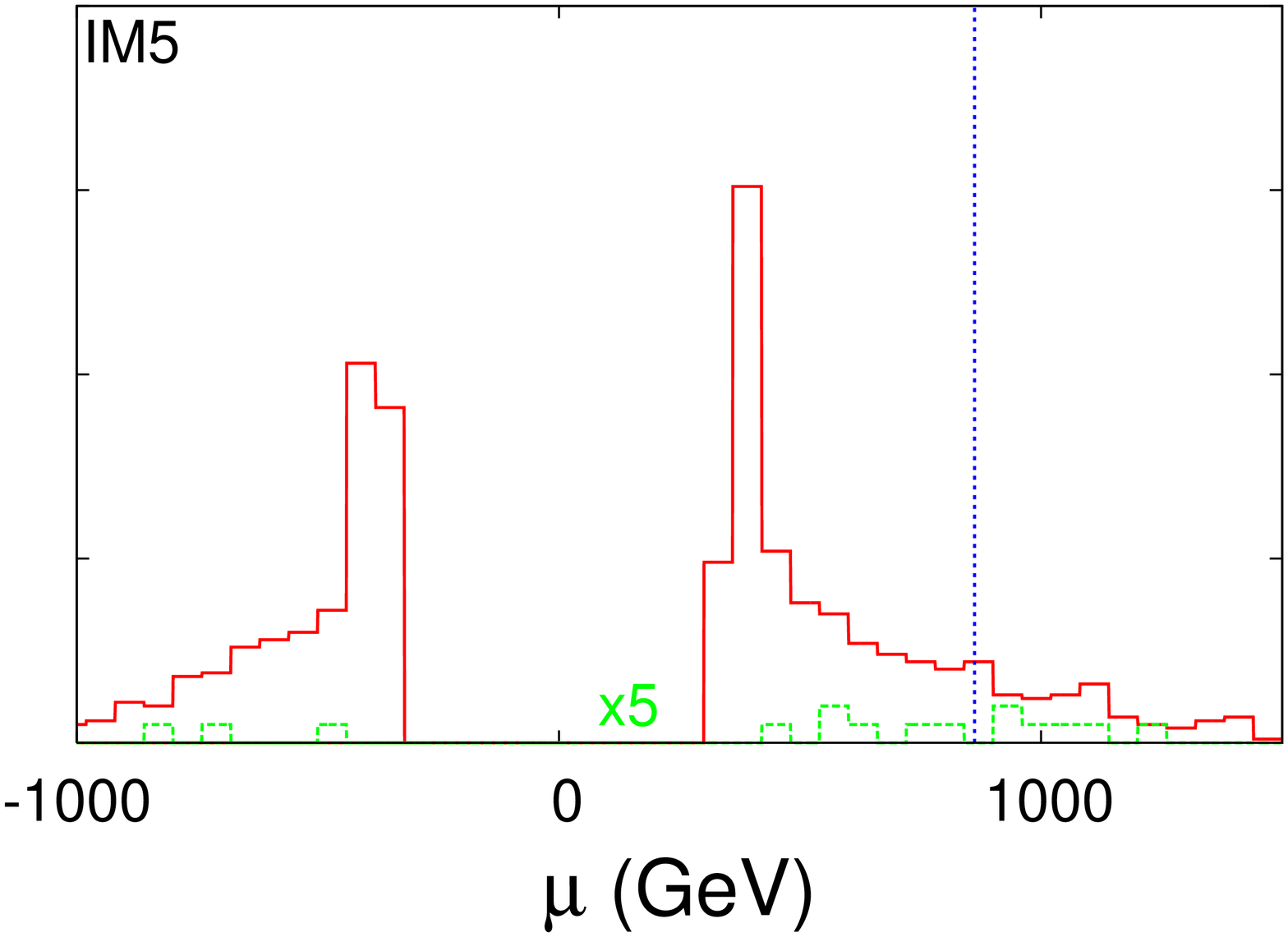}
\hspace{-0.8cm}
\includegraphics[width=2.5in,angle=-90]{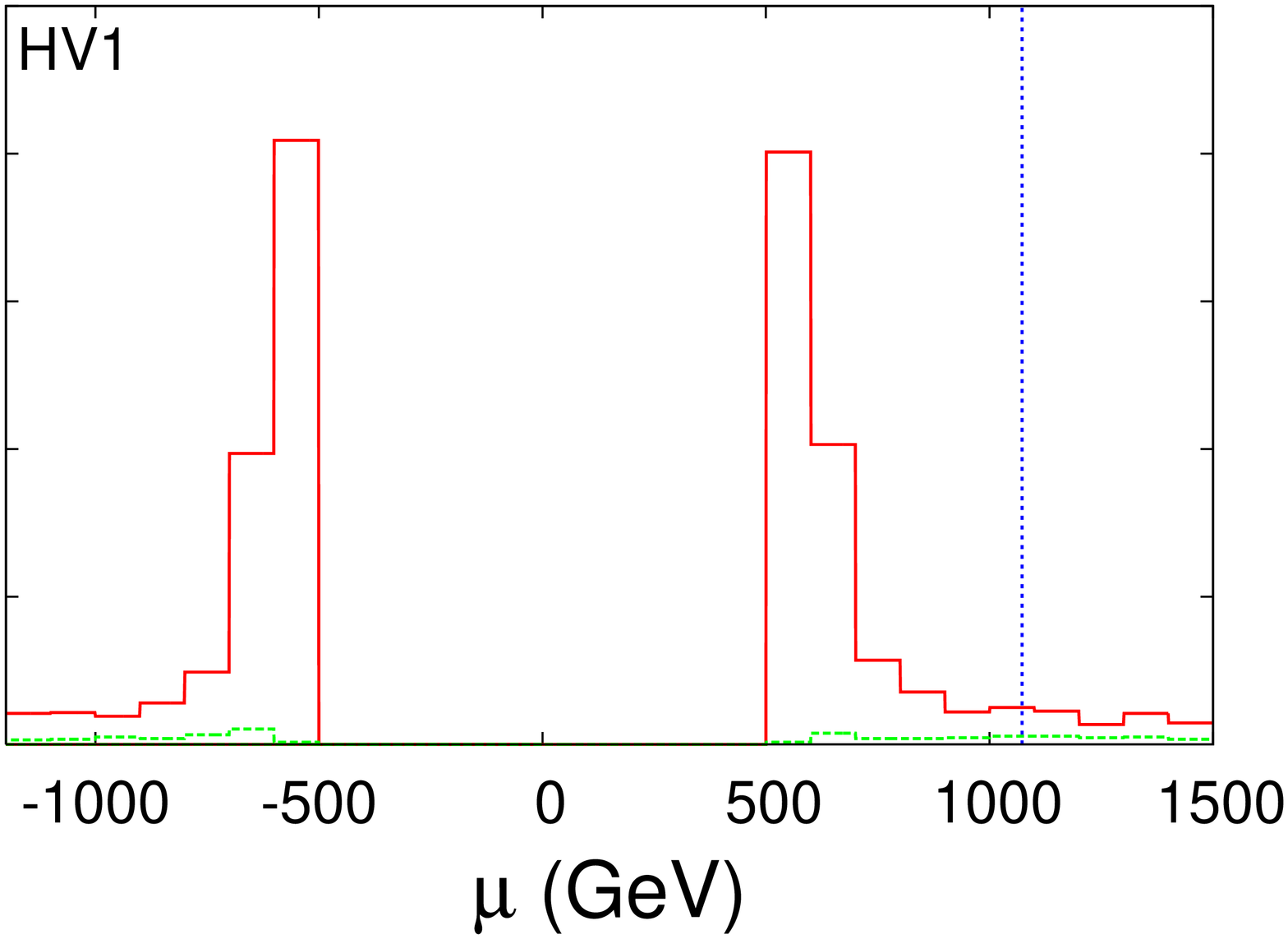}
\\
\hspace{-0.8cm}
\includegraphics[width=2.5in,angle=-90]{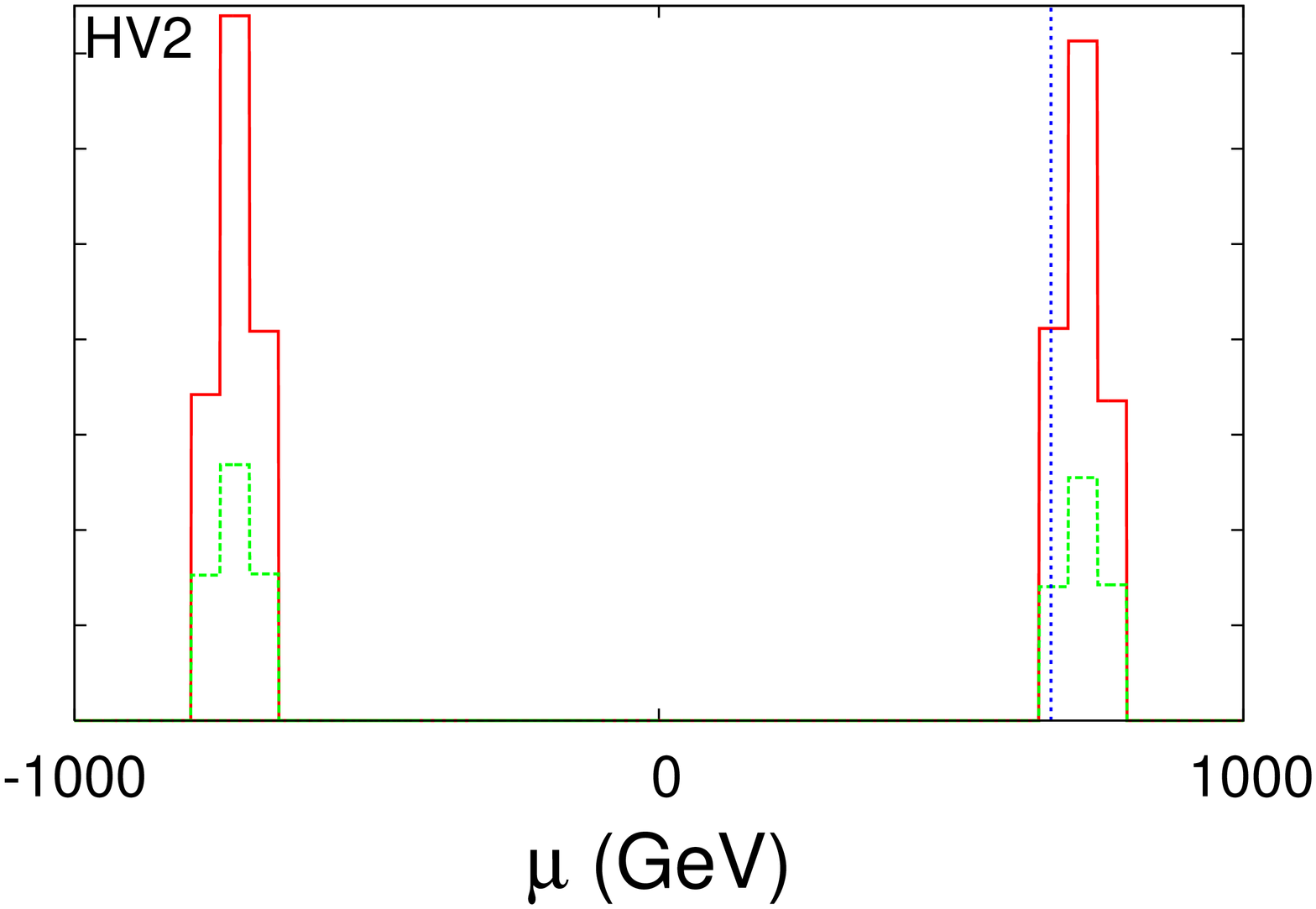}
\hspace{-0.8cm}
\includegraphics[width=2.5in,angle=-90]{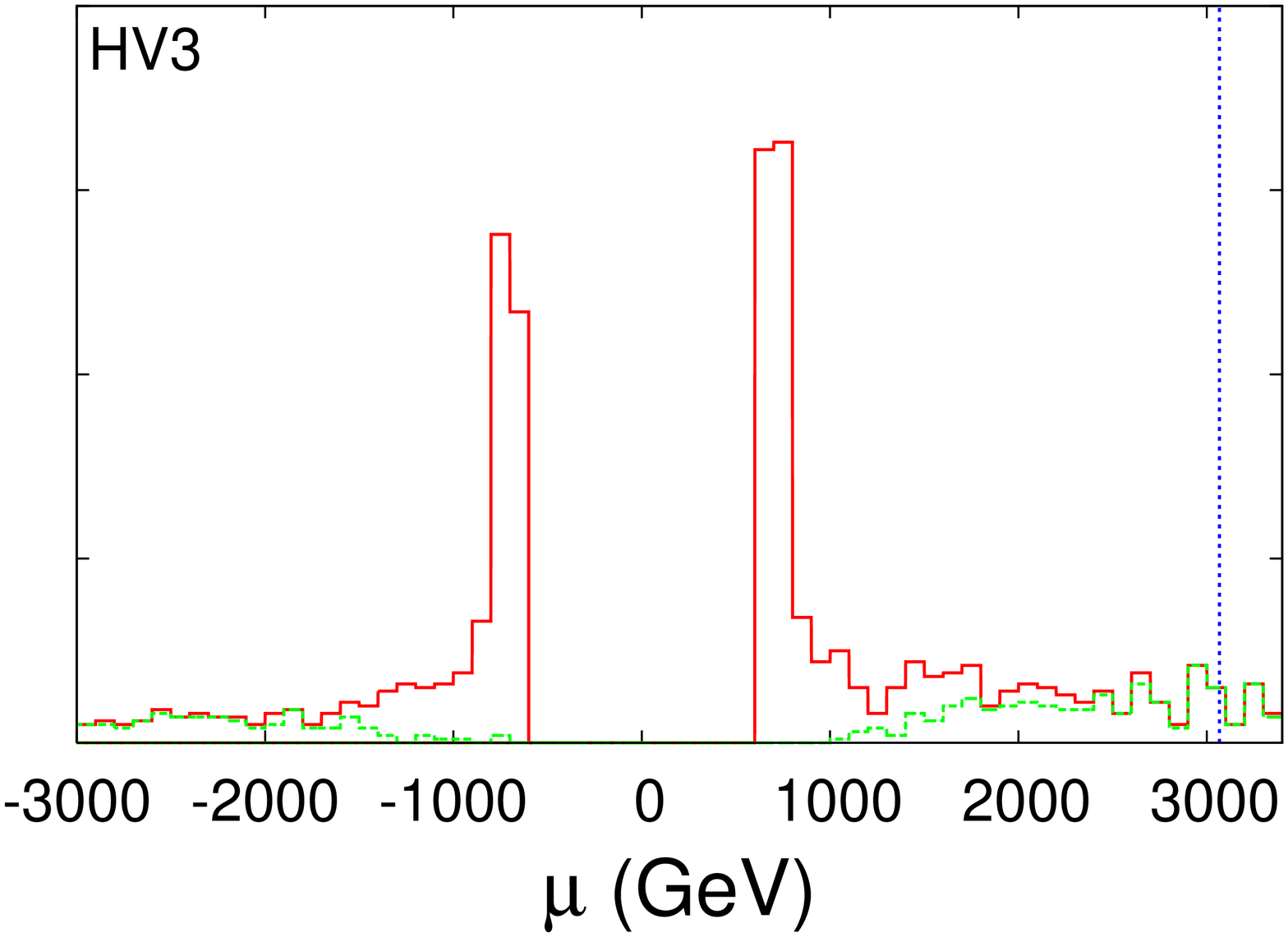}
\\
\caption{The same as Fig.~\ref{muhist1}, but for the last six of our benchmark models. For models IM3, IM4 and HV3, astrophysical data allows for $\mu$ to be considerably more constrained. Models IM5, and HV1 benefit less from astrophysical data. The value of $|\mu|$ is quite well constrained by LHC data alone (plus relic abundance) in model HV2. The sign of $\mu$ is not expected to be determined by the techniques discussed here. For model IM5, the lower curve has been multipled by 5 for illustration.}
\label{muhist2}
\end{figure}

\begin{figure}[!tbp]
\hspace{-0.8cm}
\includegraphics[width=2.5in,angle=-90]{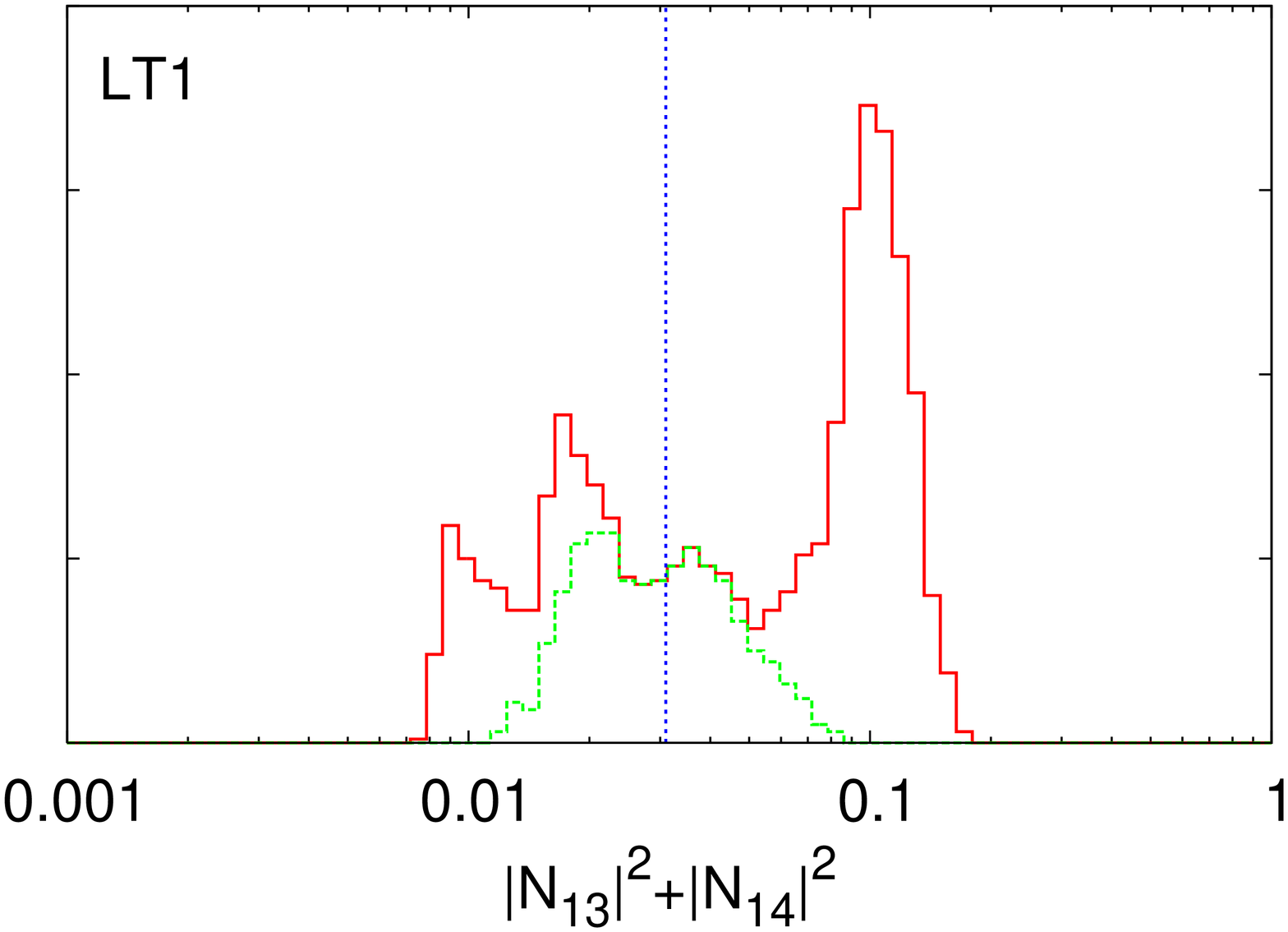}
\hspace{-0.8cm}
\includegraphics[width=2.5in,angle=-90]{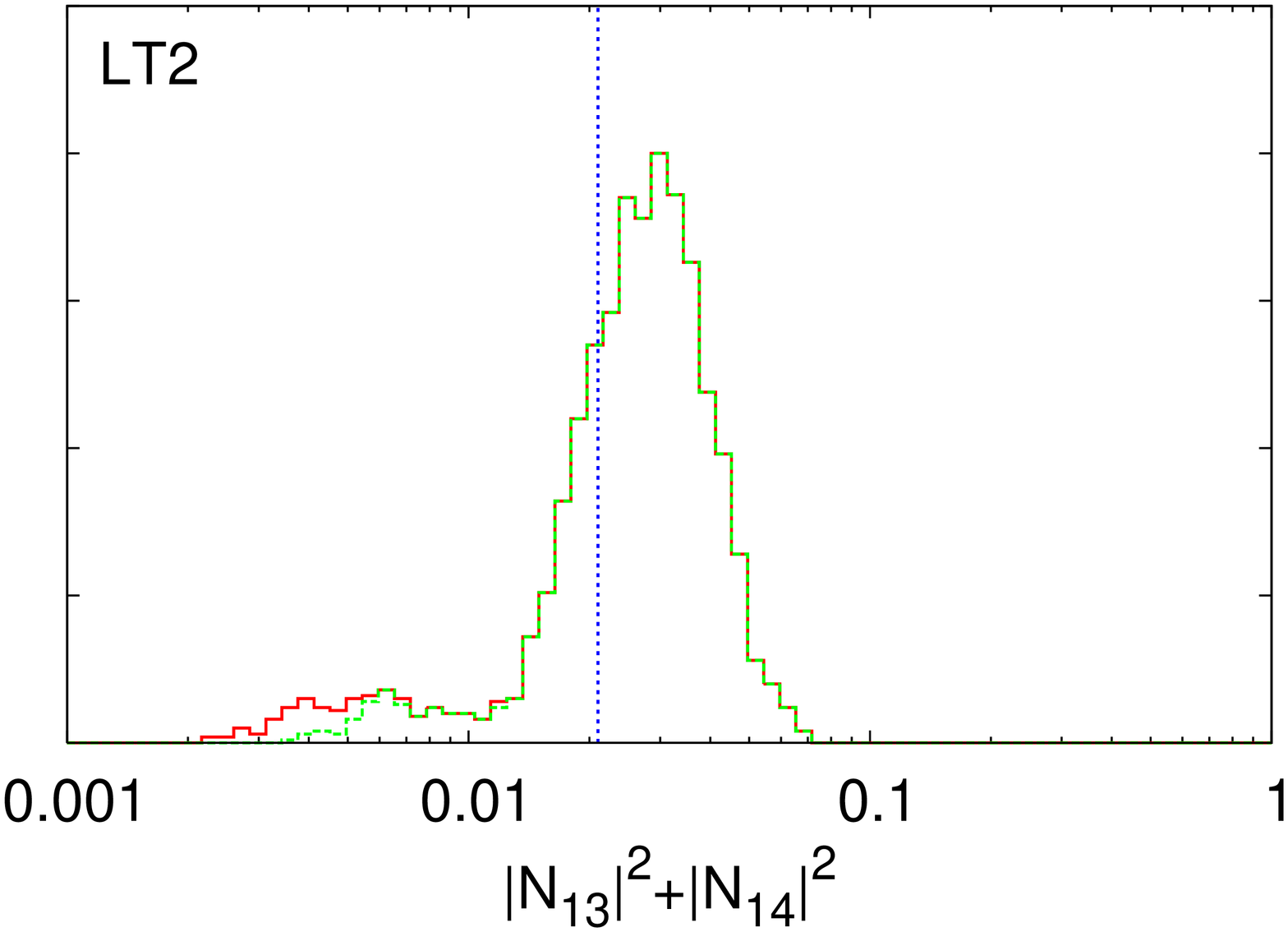}
\\
\hspace{-0.8cm}
\includegraphics[width=2.5in,angle=-90]{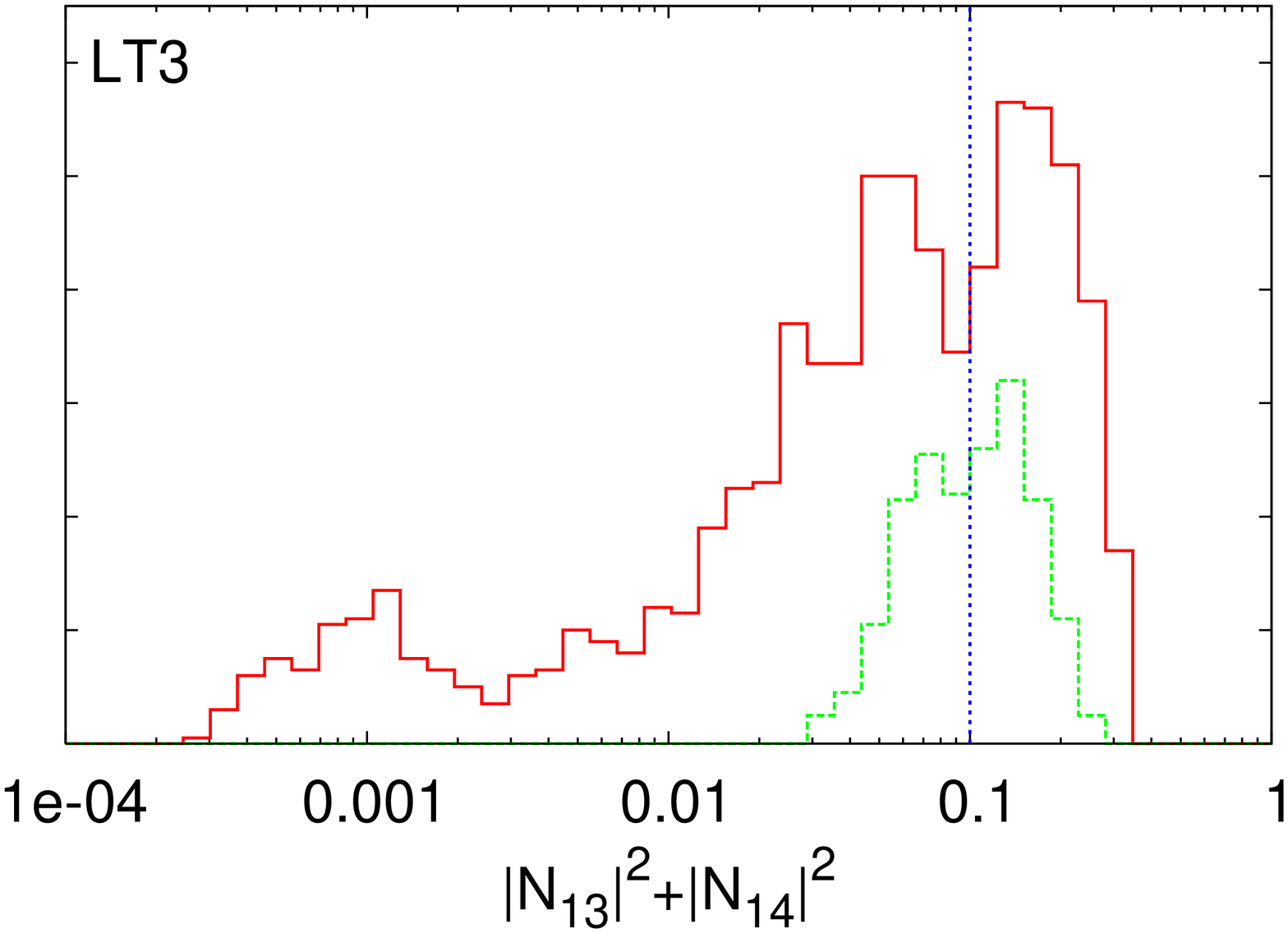}
\hspace{-0.8cm}
\includegraphics[width=2.5in,angle=-90]{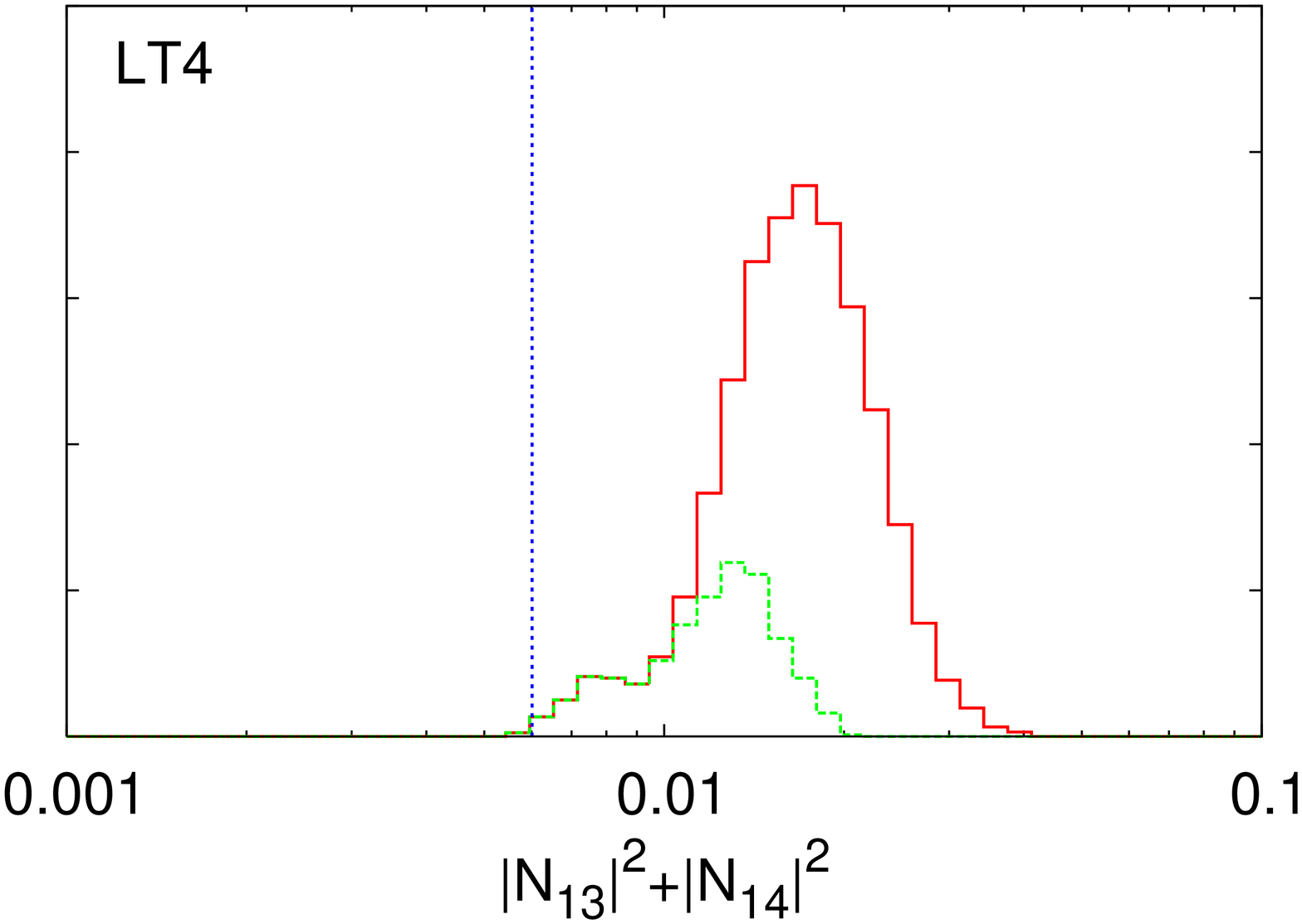}
\\
\hspace{-0.8cm}
\includegraphics[width=2.5in,angle=-90]{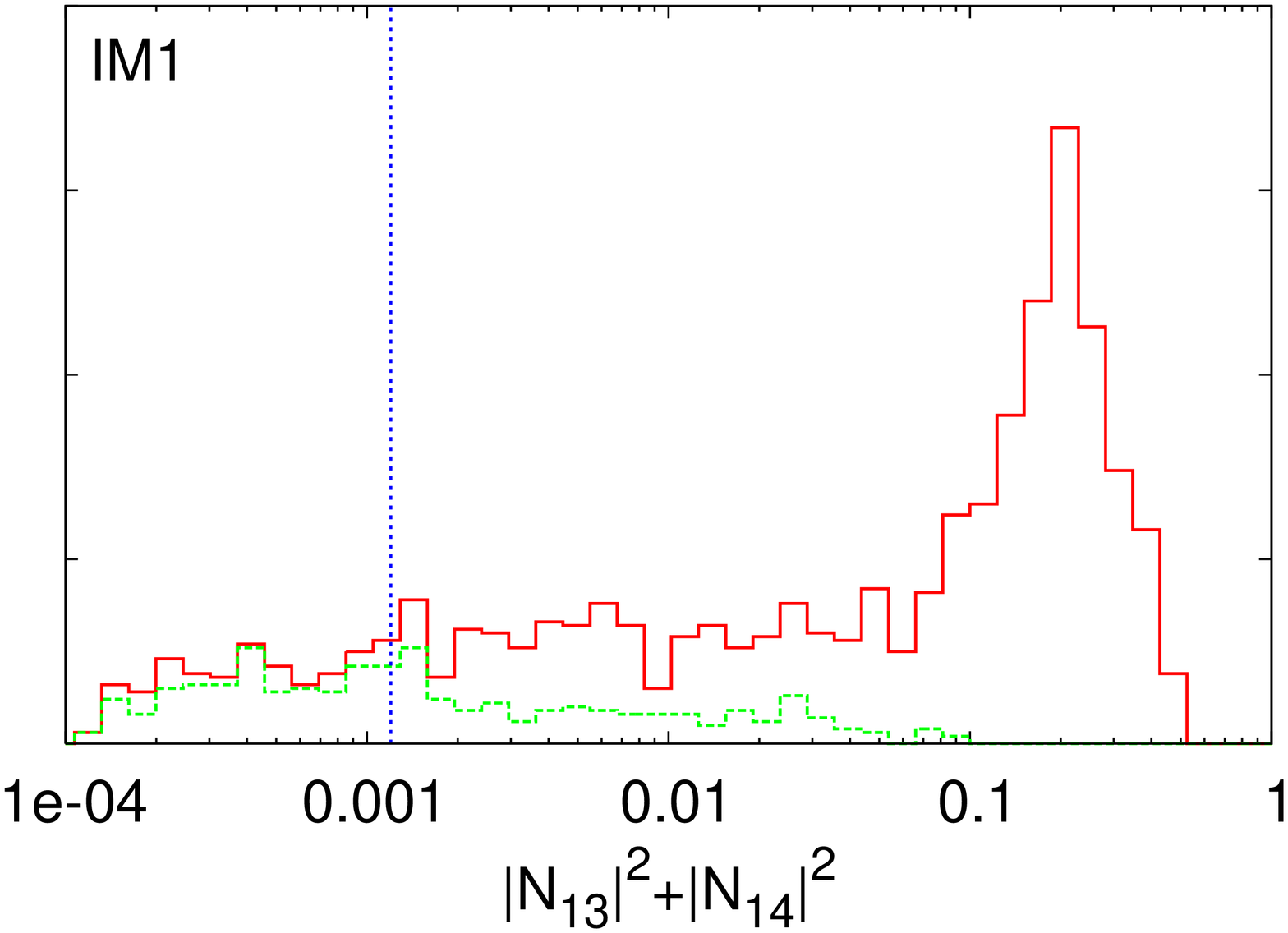}
\hspace{-0.8cm}
\includegraphics[width=2.5in,angle=-90]{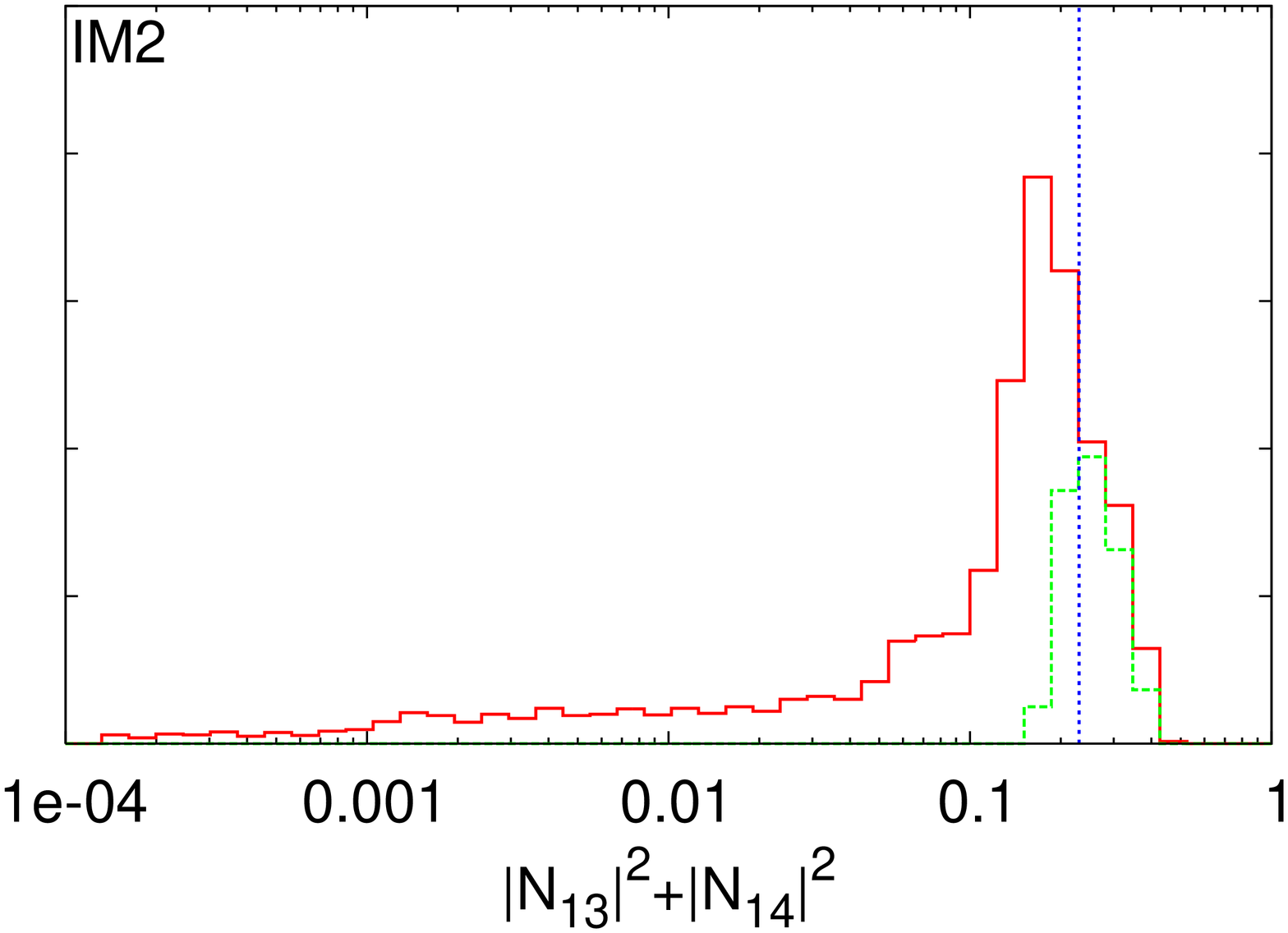}
\\
\caption{The same as Fig.~\ref{muhist1}, but showing the ability of future astrophysical and collider data to constrain the higgsino fraction of the lightest neutralino for the first six of our benchmark models.}
\label{highist1}
\end{figure}

\begin{figure}[!tbp]
\hspace{-0.8cm}
\includegraphics[width=2.5in,angle=-90]{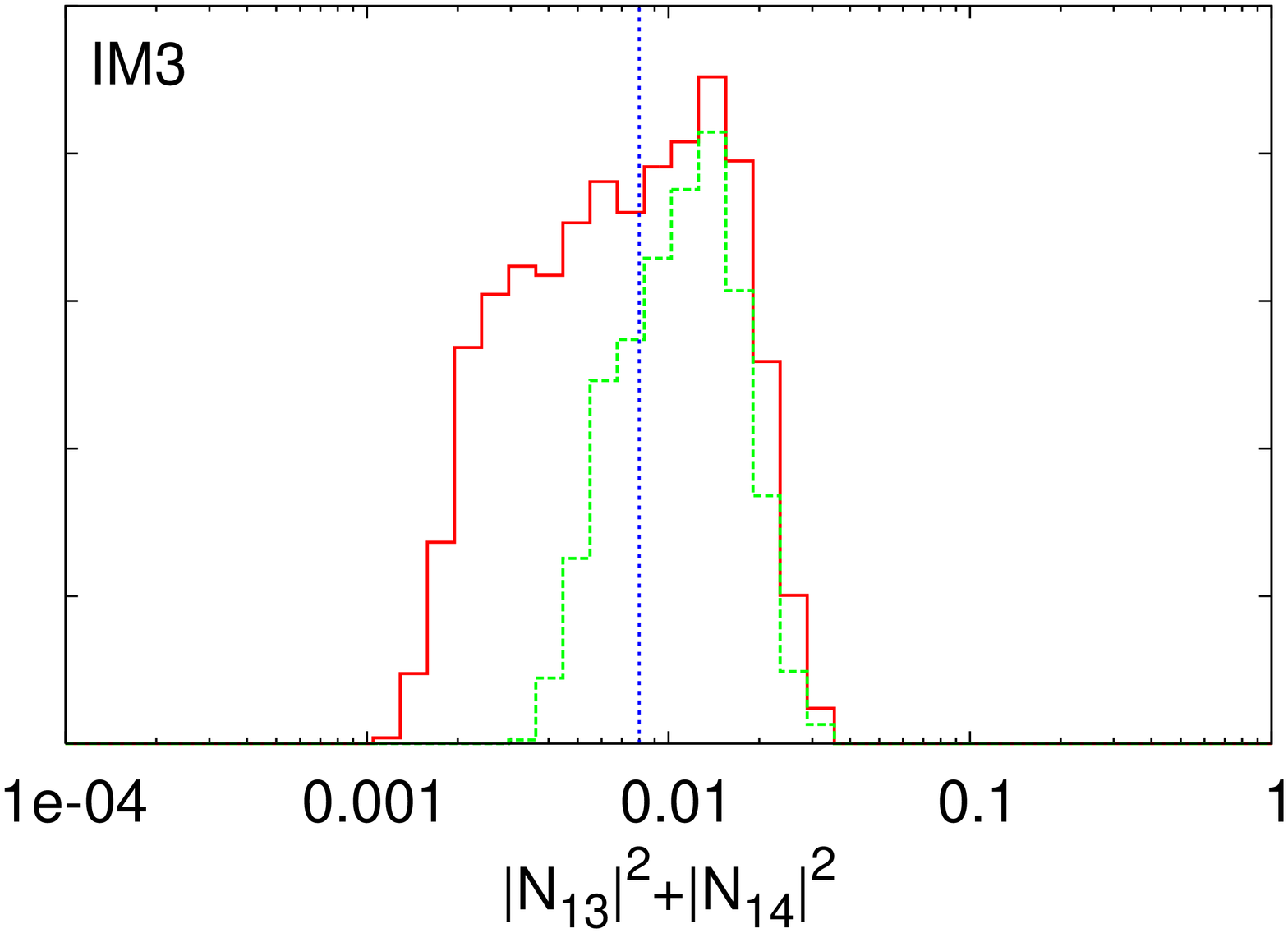}
\hspace{-0.8cm}
\includegraphics[width=2.5in,angle=-90]{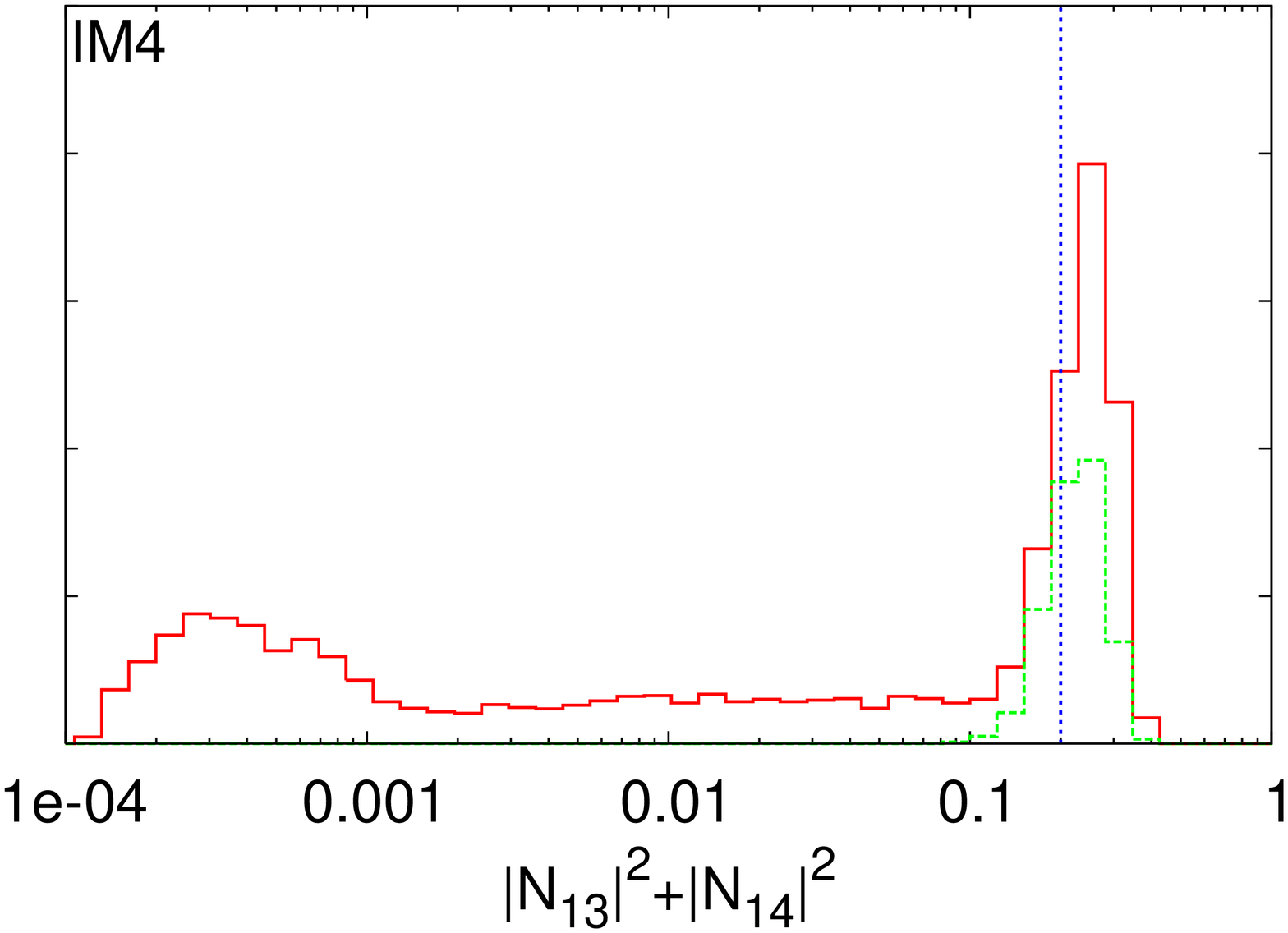}
\\
\hspace{-0.8cm}
\includegraphics[width=2.5in,angle=-90]{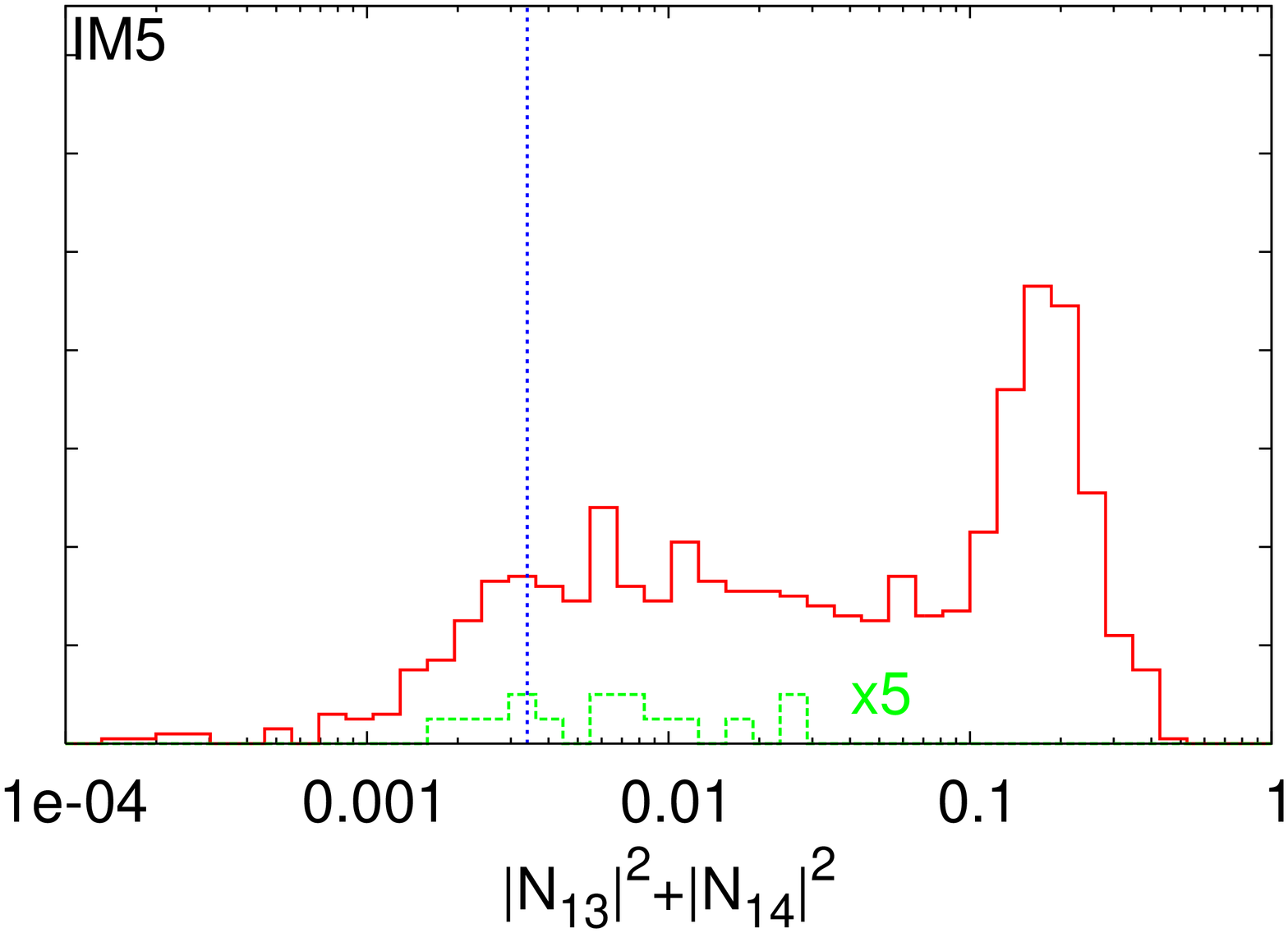}
\hspace{-0.8cm}
\includegraphics[width=2.5in,angle=-90]{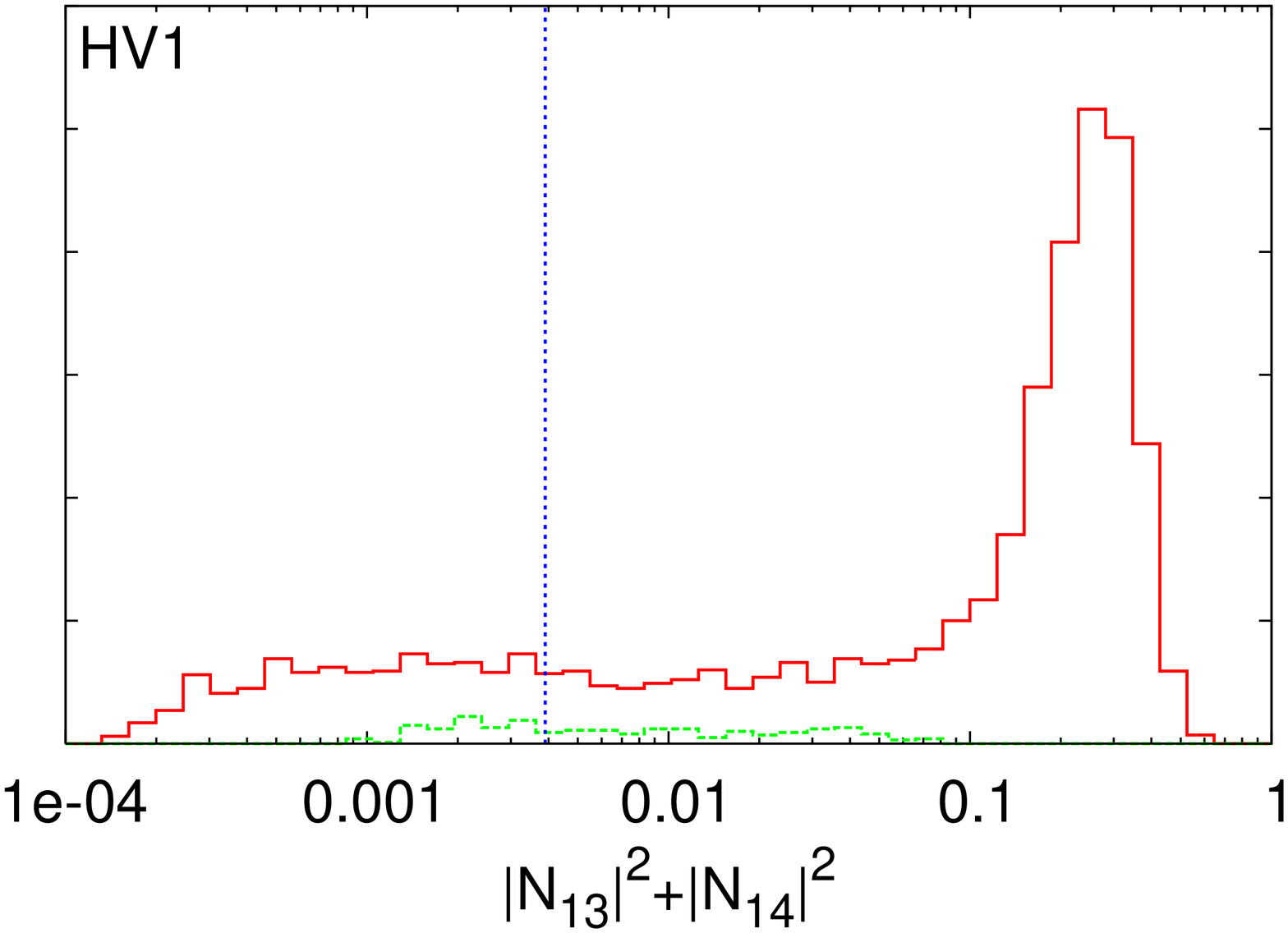}
\\
\hspace{-0.8cm}
\includegraphics[width=2.5in,angle=-90]{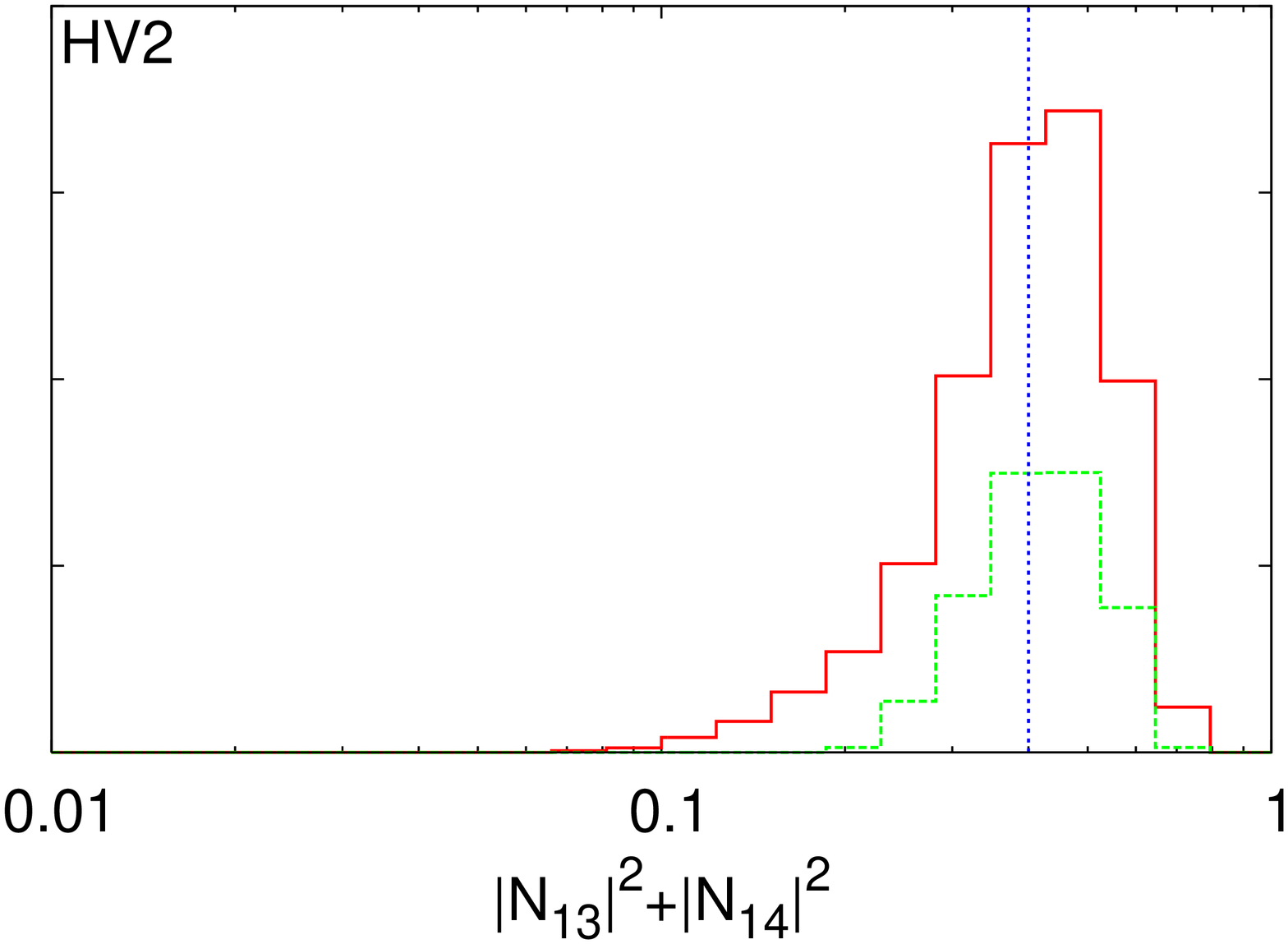}
\hspace{-0.8cm}
\includegraphics[width=2.5in,angle=-90]{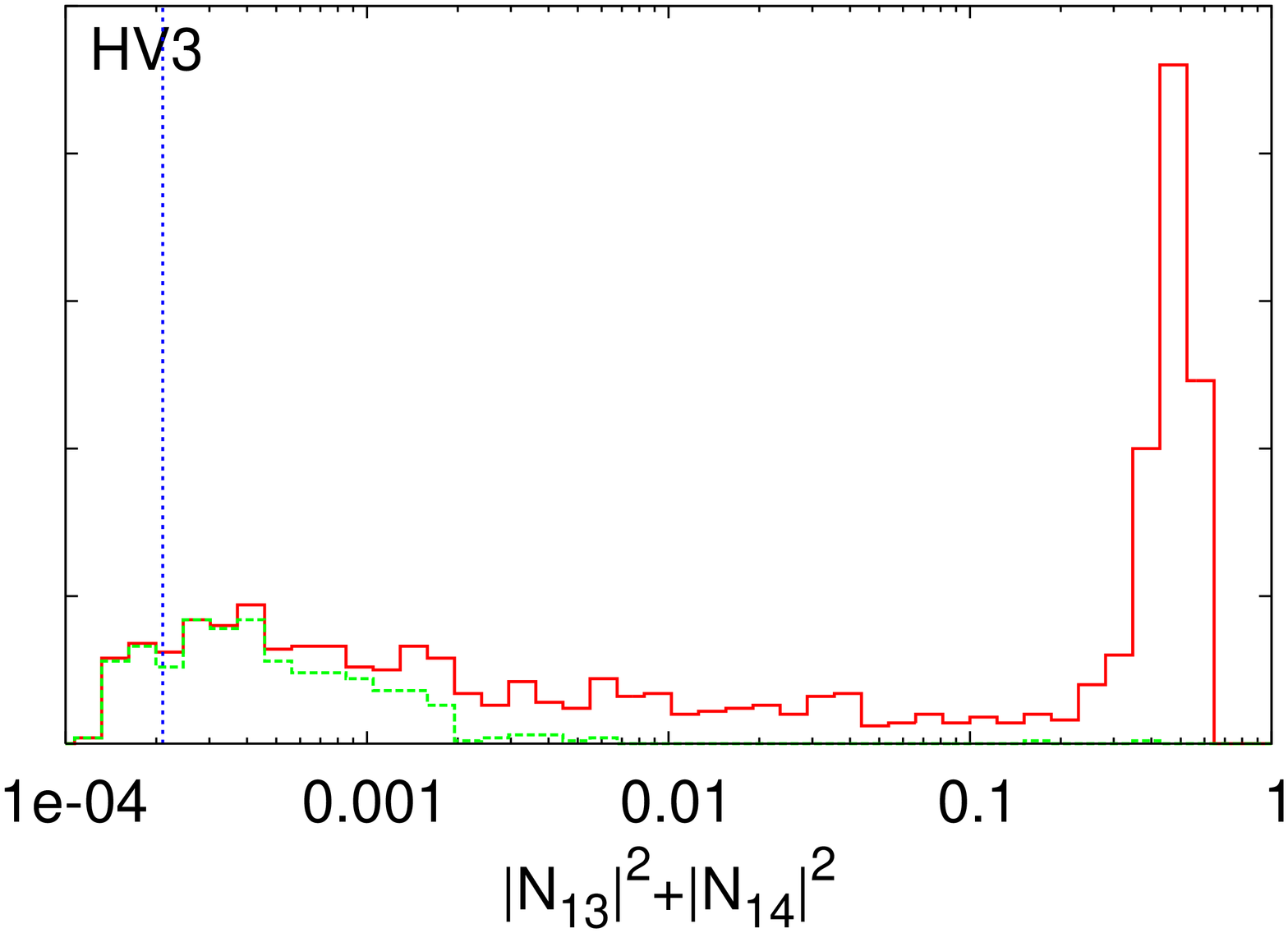}
\\
\caption{The same as Fig.~\ref{highist1}, but for the last six of our benchmark models.  For model IM5, the lower curve has been multipled by 5 for illustration.}
\label{highist2}
\end{figure}

In Figs.~\ref{muhist1} and~\ref{muhist2}, we plot the ability of the set of observations given in table~\ref{t3} to constrain the parameter $\mu$. For each benchmark model, we have randomly scanned over SUSY parameters\footnote{We have scanned linearly over $M_2$, $\mu$, $m_A$, $\tan \beta$, $A_t$, $A_b$, and $m_{\tilde{f}}$. $M_2$, $\mu$ and $m_{\tilde{f}}$ have each been varied up to 4 TeV, while $m_A$ and $\tan \beta$ have been varied up to 1 TeV and 60, respectively. $A_t$ and $A_b$ have been varied up to 3 times $m_{\tilde{f}}$.} for models with observable quantities within the range of the measurements given in the table. For each frame ({\it ie.} for each benchmark model), the upper histogram reflects distribution of those models we found that satisfied the LHC measurements given in table~\ref{t3}, in addition to satisfying the relic density constraint. The lower histogram for each benchmark model are those models which {\it also} satisfy the astrophysical measurements given in table~\ref{t3}. The vertical dotted line in each frame denotes the actual value of $\mu$ for the given benchmark model. Alternatively, instead of the quantity $\mu$, we could consider the ability of these measurements to constrain the lightest neutralino's higgsino fraction. Those results are shown in Figs.~\ref{highist1} and~\ref{highist2}.

For many of the benchmark models, the inclusion of astrophysical data in the analysis allows for $\mu$ (and the higgsino fraction) to be significantly more constrained than by the LHC and relic abundance considerations alone. In a minority of the benchmarks, this additional information is not very useful, however. For model LT2, for example, the two histograms fall nearly on top of each other, indicating that astrophysical data does not allow for $\mu$ to be constrained by astrophysical data much further than with collider data (plus relic density) alone. Even in such a case, astrophysical data may be used to provide an important confirmation on the model assumptions that have been used.

\begin{figure}[!tbp]
\hspace{-0.8cm}
\includegraphics[width=2.5in,angle=-90]{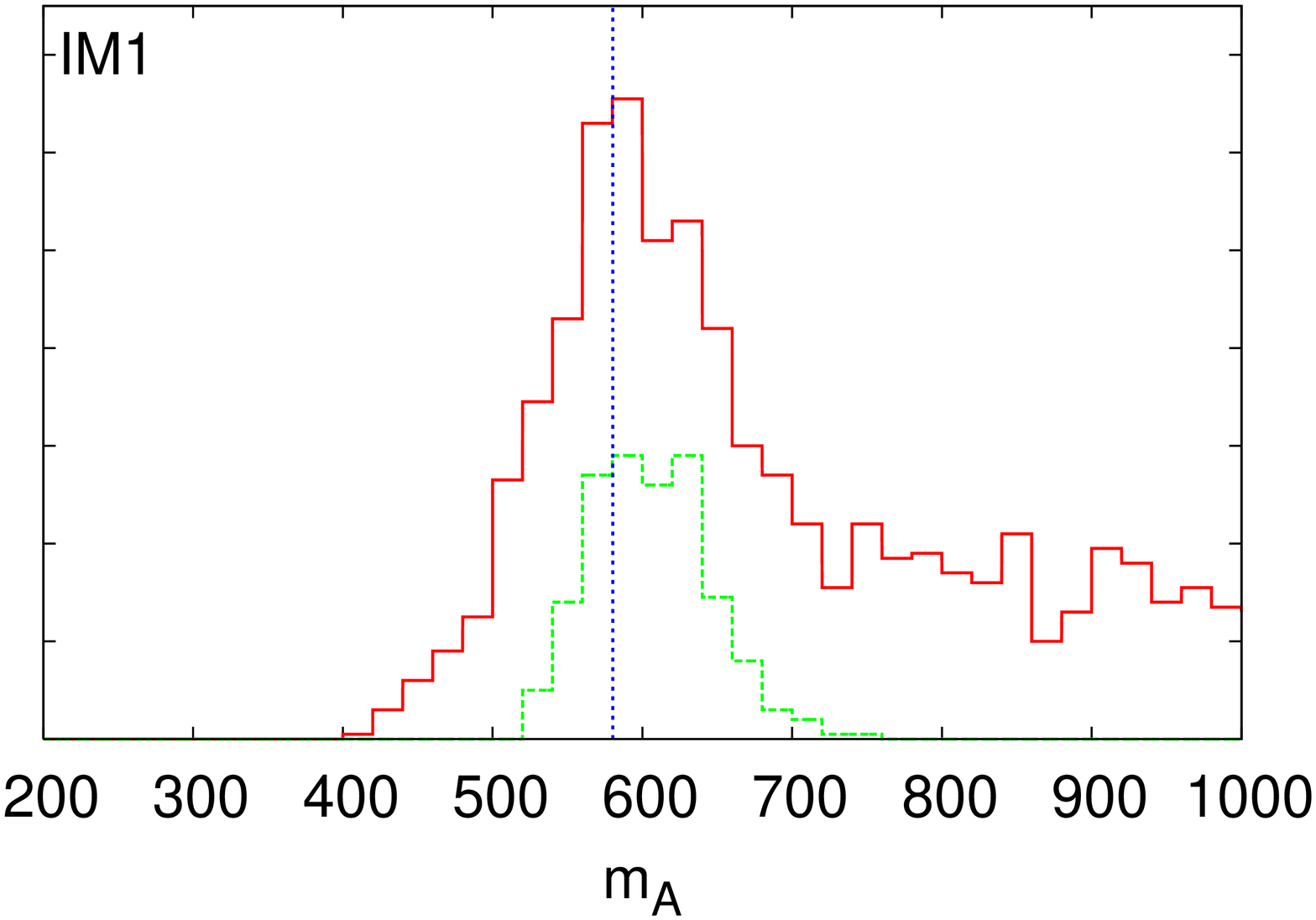}
\\
\hspace{-0.8cm}
\includegraphics[width=2.5in,angle=-90]{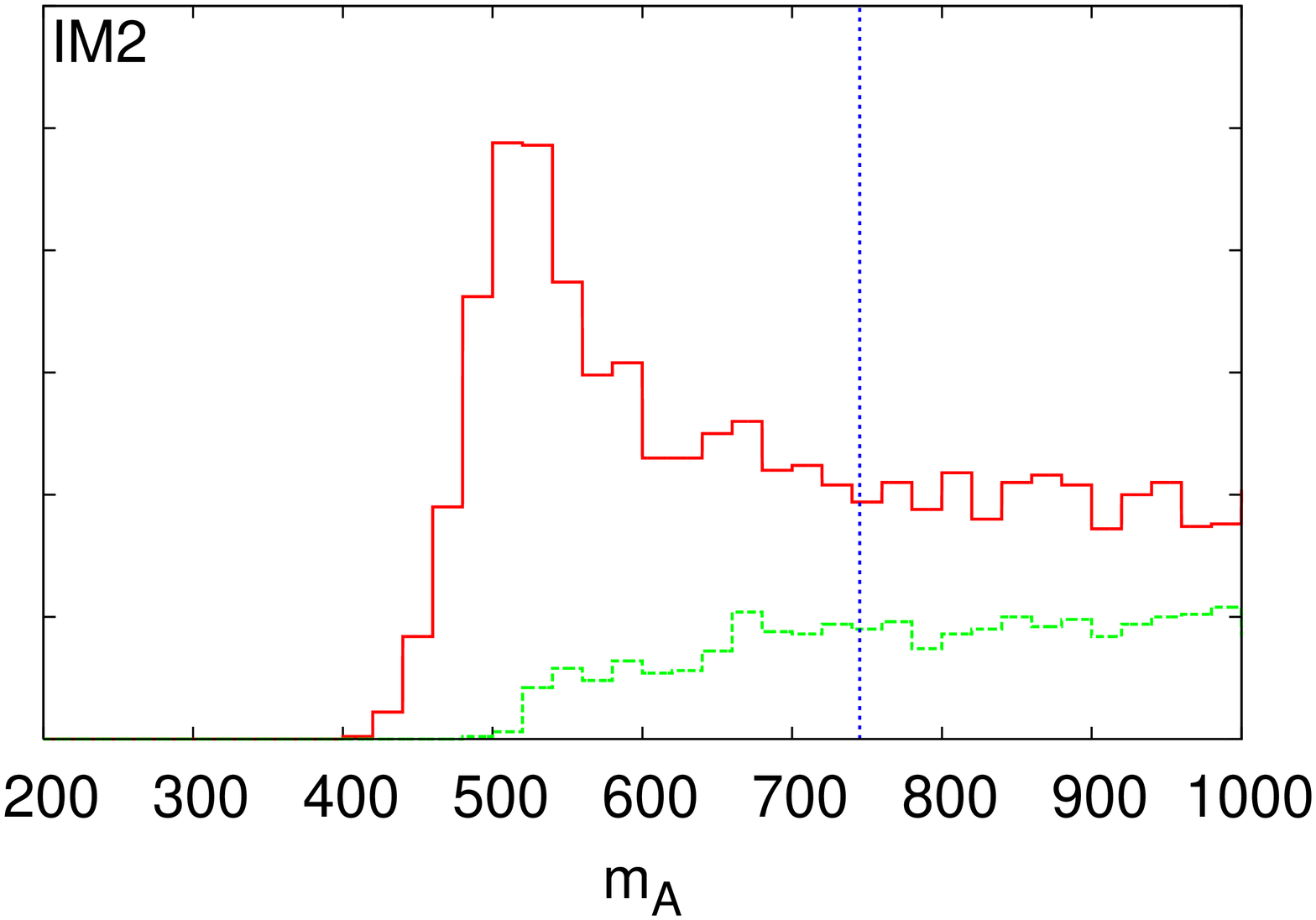}
\hspace{-0.8cm}
\includegraphics[width=2.5in,angle=-90]{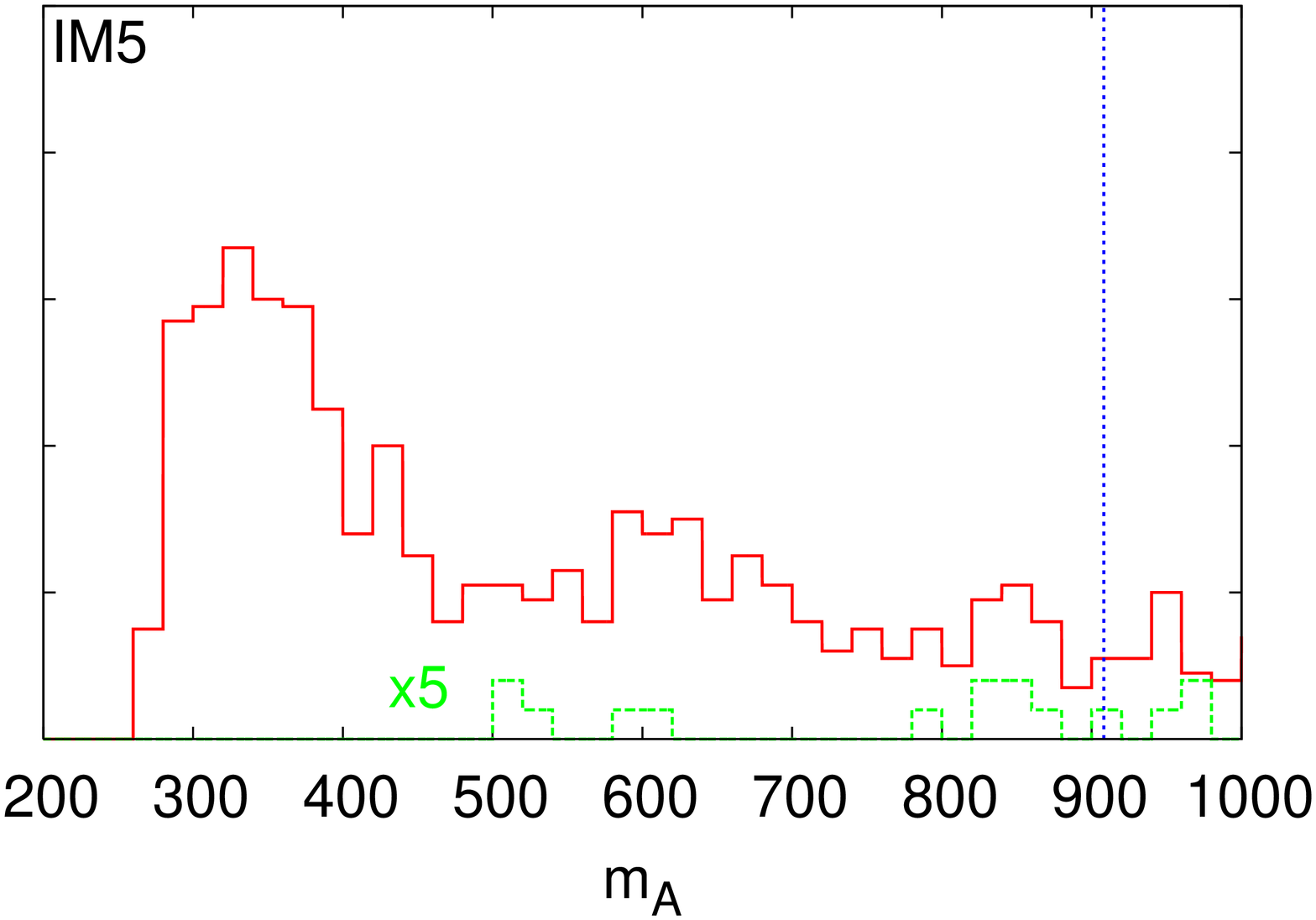}
\\
\hspace{-0.8cm}
\includegraphics[width=2.5in,angle=-90]{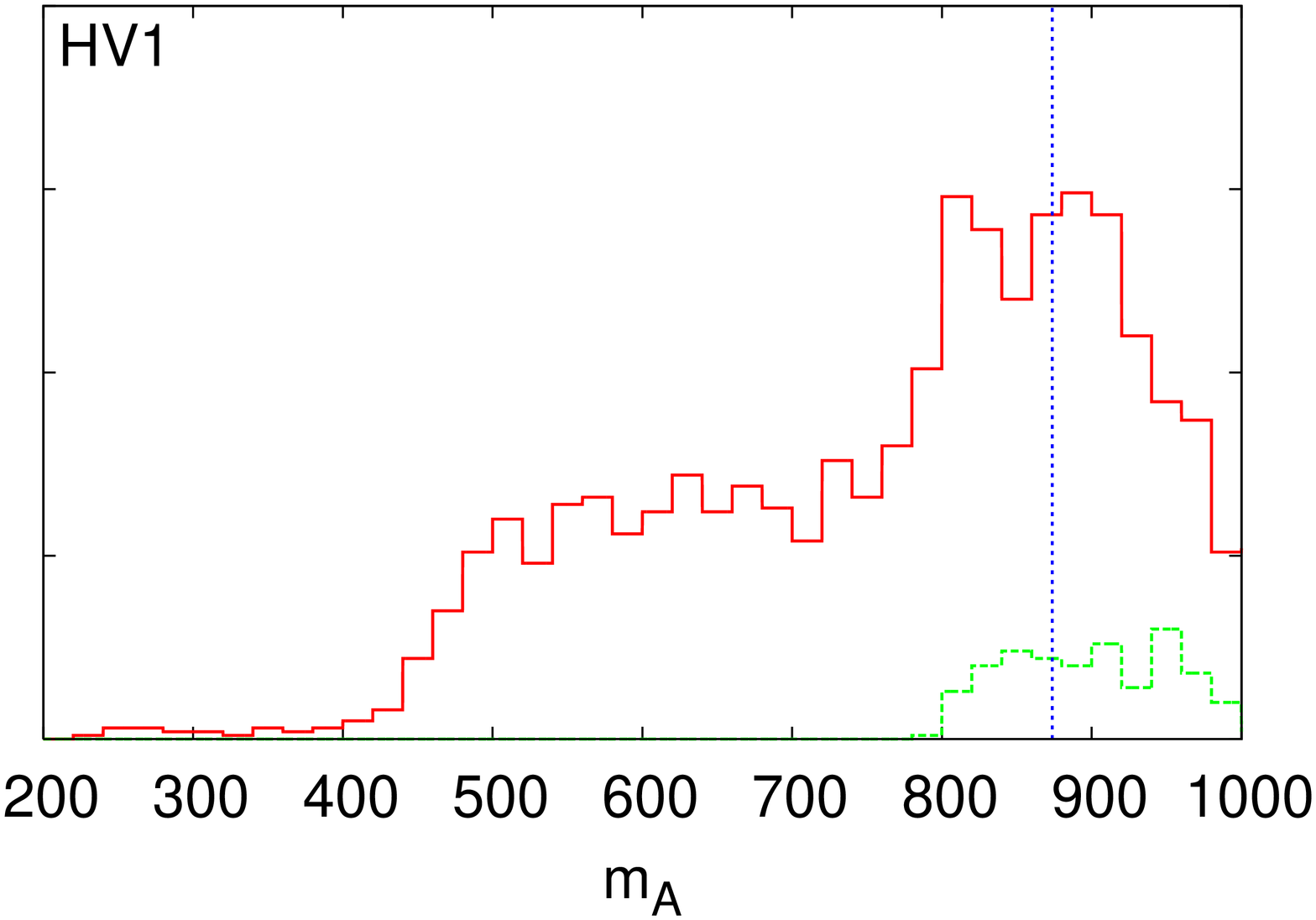}
\hspace{-0.8cm}
\includegraphics[width=2.5in,angle=-90]{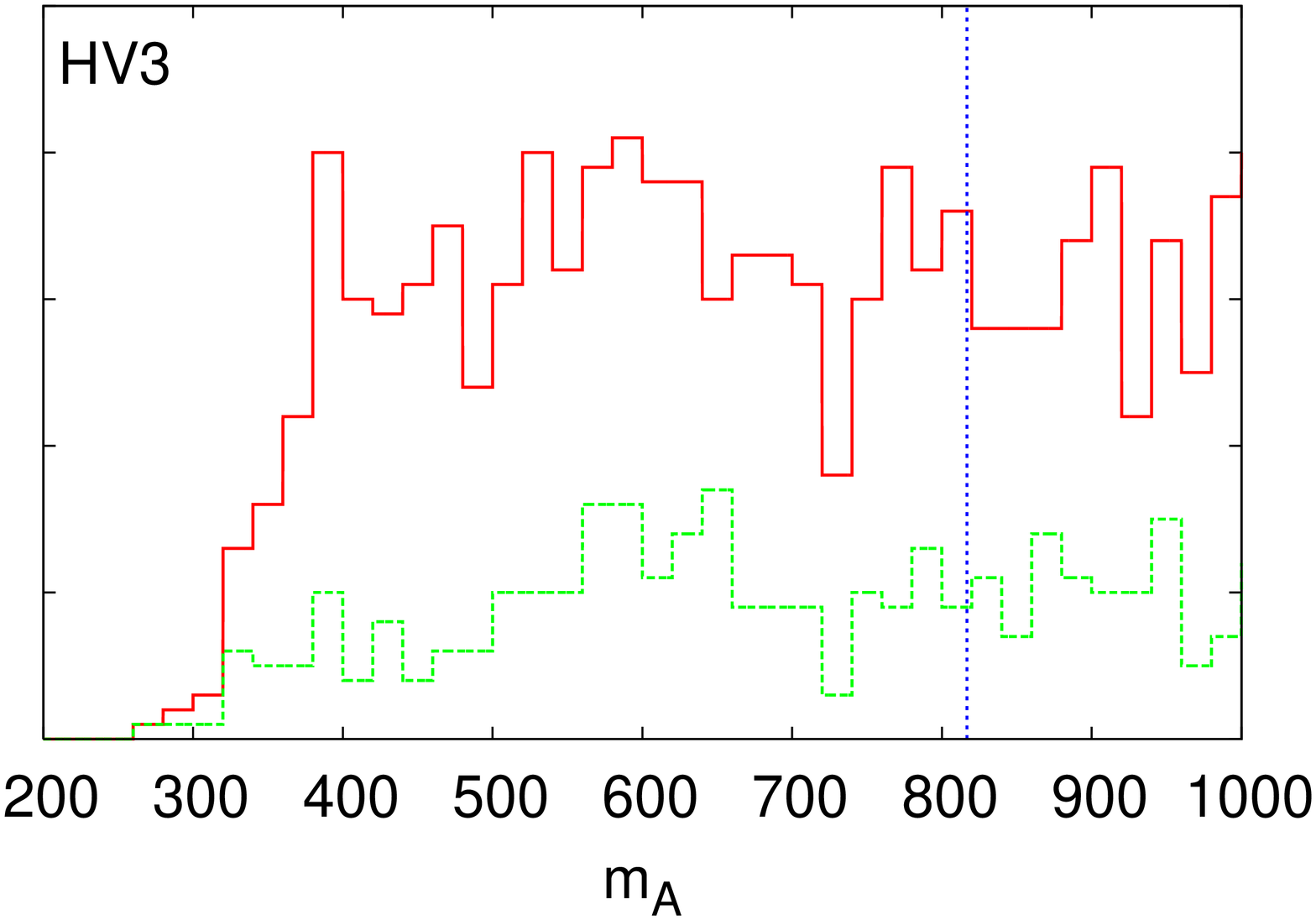}
\\
\caption{The ability of future astrophysical and collider (LHC) data to constrain the mass of the CP-odd Higgs boson, $m_A$, in those of our benchmark models in which it is unlikely to be measured at the LHC. In each frame, the upper histogram describes the distribution of models found in our random parameter scan which satisfy the LHC measurements given in table~\ref{t3}, in addition to the relic density constraint. The lower histogram for each benchmark model contains those models which {\it also} satisfy the astrophysical measurements given in table~\ref{t3}. The vertical dotted line in each frame denotes the actual value of $m_A$ for the given benchmark model. For models in the $A$-funnel region of paramter space (IM1 and HV1) astrophysical data allow for $m_A$ to be quite well constrained. The models IM2, IM5 and especially HV3 benefit far less from astrophysical data. For model IM5, the lower curve has been multipled by 5 for illustration.}
\label{mahist}
\end{figure}

\begin{figure}[!tbp]
\hspace{-0.8cm}
\includegraphics[width=2.5in,angle=-90]{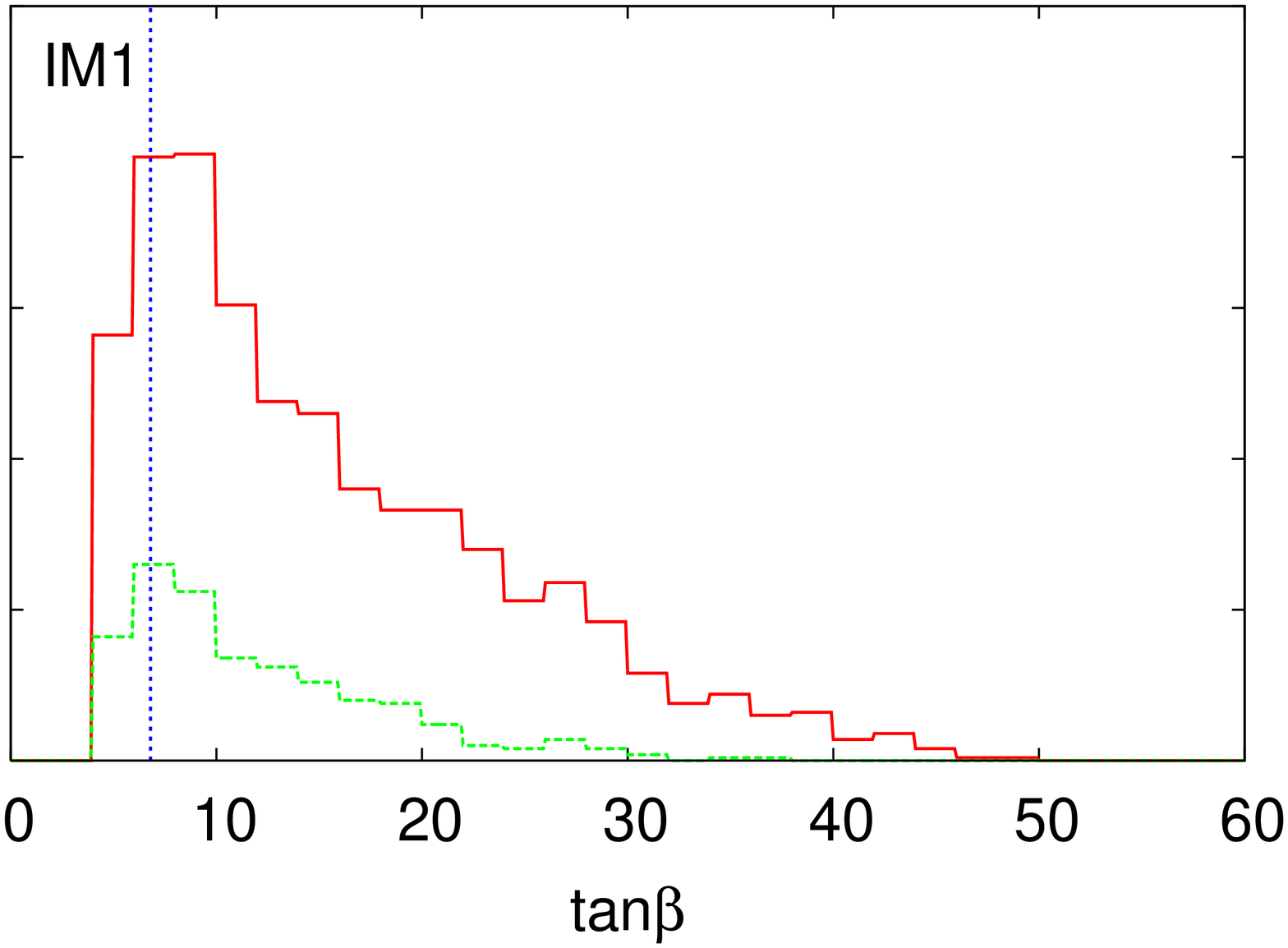}
\\
\hspace{-0.8cm}
\includegraphics[width=2.5in,angle=-90]{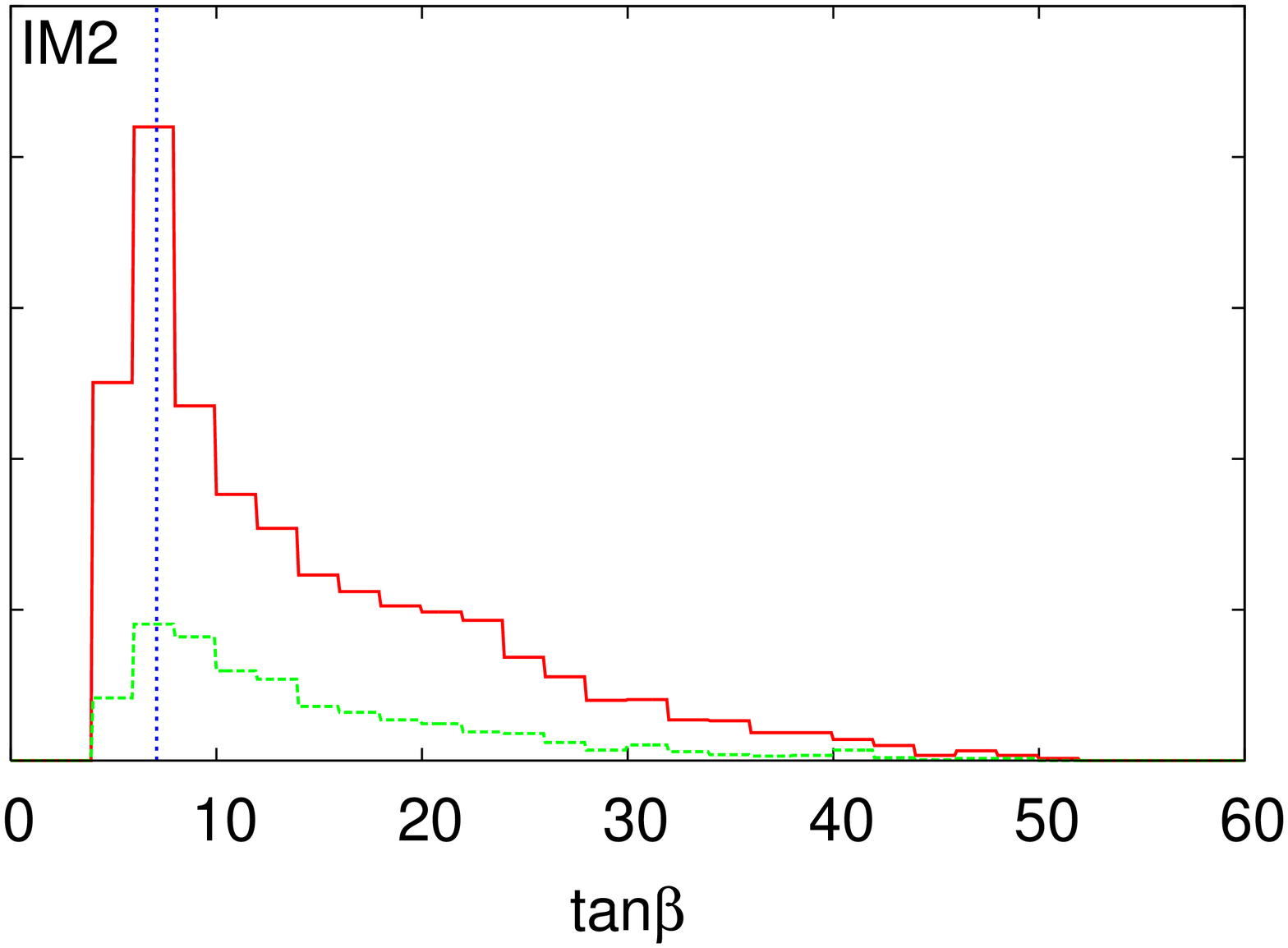}
\hspace{-0.8cm}
\includegraphics[width=2.5in,angle=-90]{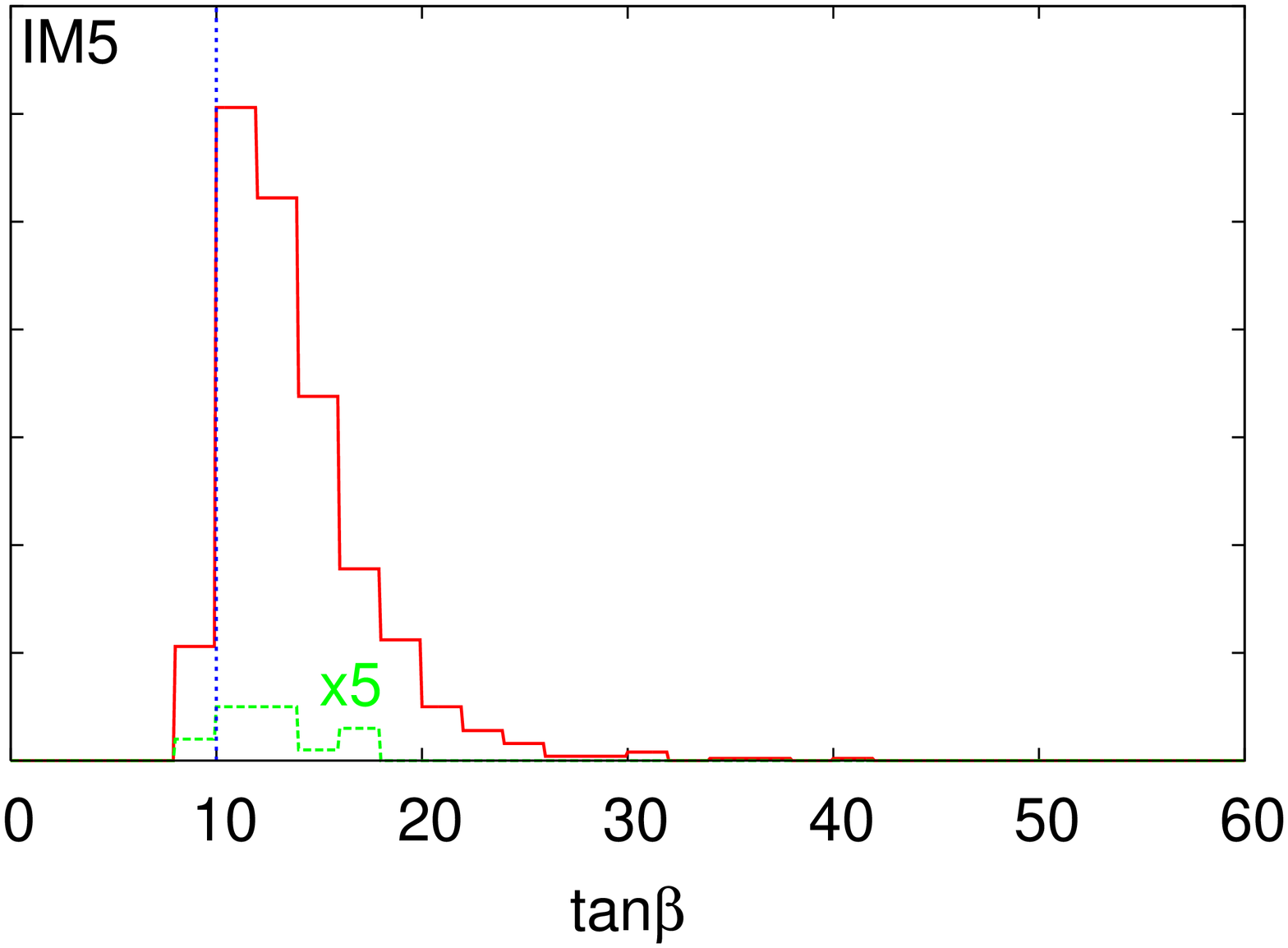}
\\
\hspace{-0.8cm}
\includegraphics[width=2.5in,angle=-90]{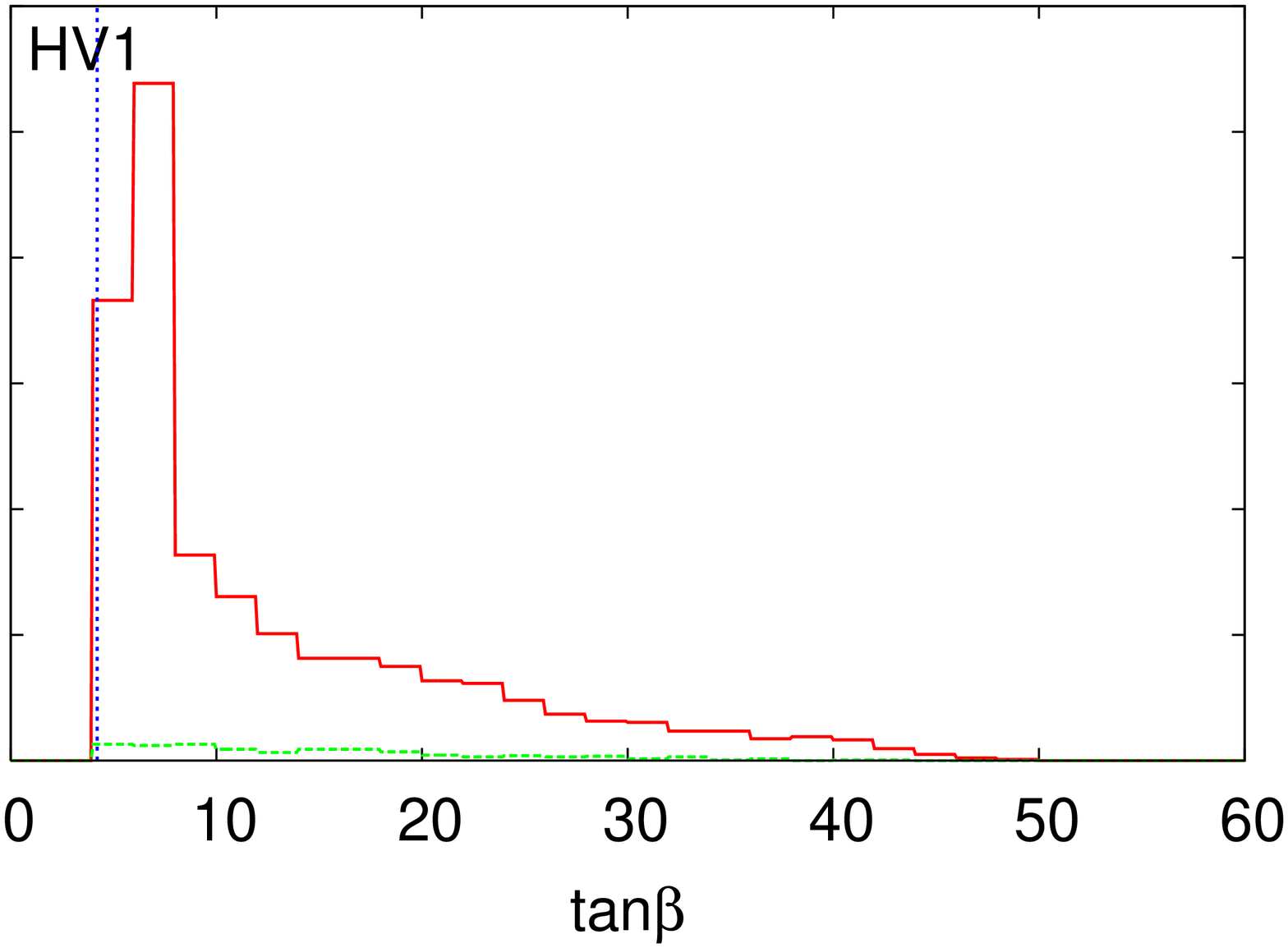}
\hspace{-0.8cm}
\includegraphics[width=2.5in,angle=-90]{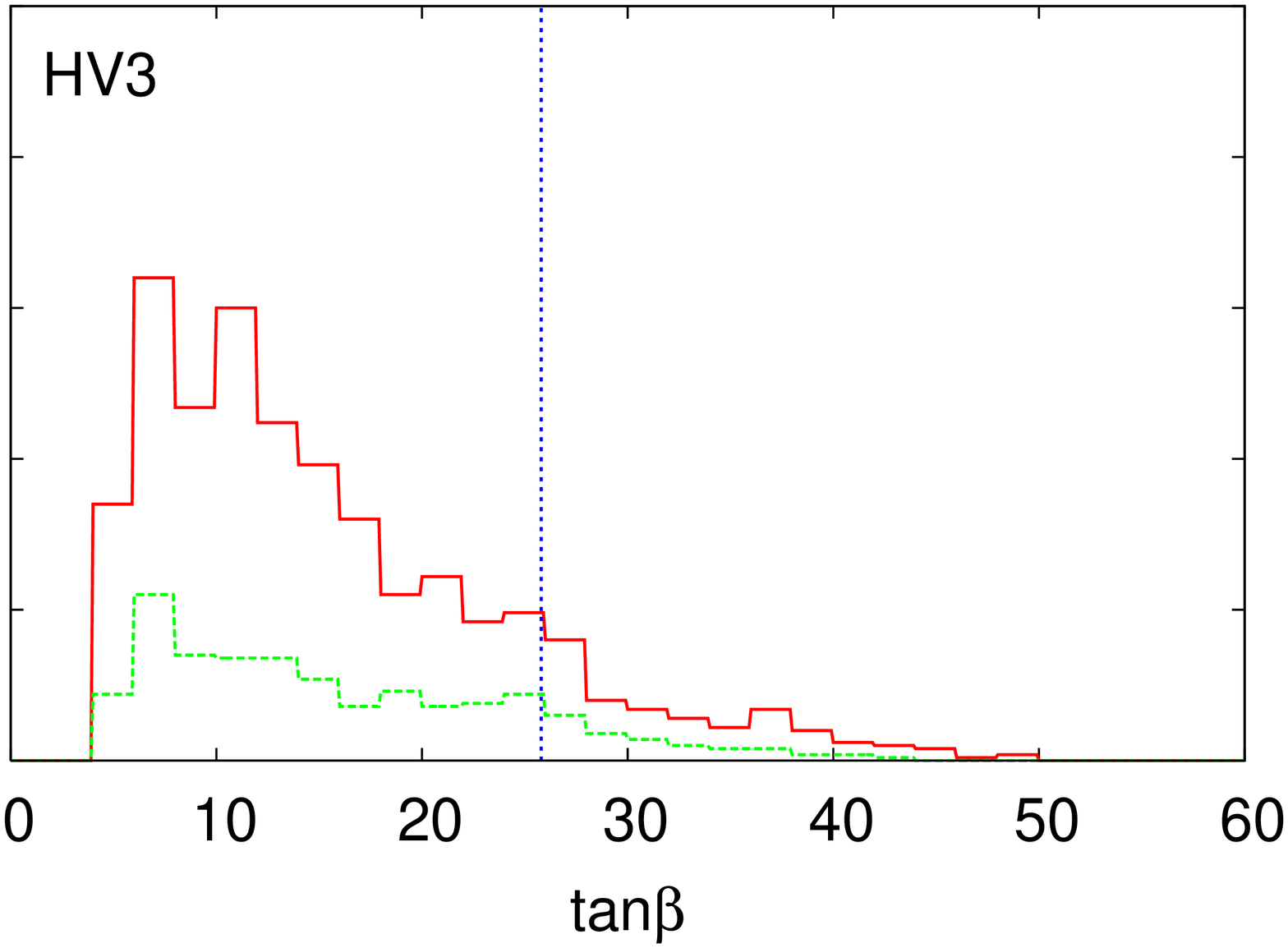}
\\
\caption{The same as in Fig.~\ref{mahist}, but reflecting the ability of future astrophysical and collider data to constrain $\tan \beta$ in those benchmark models that it is not determined at the LHC. In none of these models does astrophysical data substantially assist in determining the value of $\tan\beta$. For model IM5, the lower curve has been multipled by 5 for illustration.}
\label{tanbhist}
\end{figure}

In five of our twelve benchmark models (IM1, IM2, IM5, HV1 and HV3), the quantities $m_A$ and $\tan \beta$ are unlikely to be determined by the LHC. We have studied whether these quantities can be constrained by astrophysical observations and have plotted the histograms of the distribution of values of $m_A$ and $\tan \beta$ in Figs.~\ref{mahist} and \ref{tanbhist} (using the same methodology as Figs.~\ref{muhist1}-\ref{highist2}). In benchmark models, IM1 and HV1, the lightest neutralino annihilates primarily through the $A$-resonance (the so-called ``$A$-funnel'' region of parameter space). In such models, $m_A$ can be rather tightly determined if astrophysical data is taken into account. In a model such as HV3, on the other hand, little is gained by such information. $\tan \beta$ does not appear to be significantly constrained by astrophysical data in the models we have considered.

\newpage 

\newpage

\newpage

\newpage

\newpage

\clearpage

\section{Model Variations and Other Caveats}

Throughout this study, we have made a number of simplifying assumptions regarding the nature of supersymmetry. Our scans have been performed over the following set of parameters: $M_2$, $\mu$, $m_A$, $\tan \beta$, $A_t$, $A_b$, and $m_{\tilde{f}}$. $M_2$, $\mu$ and $m_{\tilde{f}}$ have each been varied up to 4 TeV, while $m_A$ and $\tan \beta$ have been varied up to 1 TeV and 60, respectively. $A_t$ and $A_b$ have been varied up to 3 times $m_{\tilde{f}}$. Positive and negative values of each parameter (except for $\tan \beta$) have been allowed.

These seven parameters obviously do not span the entire 120 dimensional parameter space of the MSSM. We have reduced the number of free parameters considerably by adopting a single universal scalar mass ($m_{\tilde{f}}$), and by fixing $M_1$ and $M_3$ to $M_2$ through the requirement that they unify at the GUT scale. Motivated by a desire to avoid flavor changing neutral currents and large electron and neutron electric dipole moments, we have taken all quantities to be diagonal in flavor space and have not considered the presence of any CP-violating phases.

Of course we expect our results to change somewhat if any or all of these assumptions are relaxed. From the standpoint of the neutralino dark matter sector, the most dramatic variations occur when $M_1$ is not related to $M_2$ by the GUT relationship. If $M_2$ is comparable to or smaller than $M_1$, for example, the lightest neutralino can be primarily wino-like or a mixed bino-wino. This is found in the case of Anomaly Mediated Supersymmetry Breaking (AMSB) models, for example, in which the gaugino masses are related by the ratios: $M_1:M_2:M_3 \approx 2.8:1:1.7$, leading to the lightest neutralino being a nearly pure wino. The wino-like neutralinos found in AMSB models have a distinct phenomenology, annihilating very efficiently, including to $\gamma \gamma$ and $\gamma Z$ final states (with nearly the largest cross section possible for a neutralino to these states)~\cite{ullio,wang}. The large overall annihilation cross section leads to a thermal abundance well below the measured dark matter density, however, and thus AMSB scenarios require a non-thermal mechanism to generate the universe's dark matter~\cite{amsbnonthermal}. Assuming that such a process exists to generate the observed dark matter density in the form of wino-like neutralinos, this also leads to large cosmic positron and anti-proton fluxes~\cite{ullio,wang}, possibly capable of generating the positron excess observed by the HEAT experiment \cite{heat} (and to some degree by AMS-01~\cite{amsheat}) without a large degree of local inhomogeneities ({\it ie.} without a large boost factor)~\cite{wangpositron}. The elastic scattering cross section of the wino-like neutralinos found in these models, in contrast, is somewhat smaller than is found in most other supersymmetric scenarios~\cite{wang}.

If the GUT relationship between $M_1$ and $M_2$ is broken in the opposite way, such that $M_1$ is less than $M_2/2$, then the existence of a very light (10-50 GeV), bino-like neutralino becomes possible without violating the limits on charginos placed by LEP ($m_{\chi^{\pm}} > 104$ GeV) \cite{lightlsp}. If we look beyond the MSSM to models such as the NMSSM (the Next-to-Minimal Supersymmetric Standard Model), even lighter neutralinos are possible \cite{nmssm}.
 
If, instead of adopting a single scale for the sfermion masses, we had selected values of $m^2_Q$, $m^2_{\bar{u}}$, $m^2_{\bar{d}}$, $m^2_L$ and $m^2_{\bar{e}}$ each independently (and independently for each family as well), the spectrum of squarks and sleptons could be considerably different from those considered in our study. In scenarios in which the scalar masses unify at a common scale, such as within the constrained MSSM (the CMSSM, or mSUGRA), $m_Q$, $m_{\bar{u}}$ and $m_{\bar{d}}$ are typically similar is magnitude, where as $m_L$ and $m_{\bar{e}}$ are sometimes smaller. This is a consequence of contributions to  $m^2_Q$, $m^2_{\bar{u}}$ and $m^2_{\bar{d}}$ (but not to $m^2_L$ and $m^2_{\bar{e}}$) from the renormalization group running terms being proportional to the gluino mass. This leads roughly to $m^2_L \approx m^2_{\bar{e}} \sim m^2_Q - 8 M^2_2$. Therefore, within the CMSSM, sleptons lighter than squarks are often predicted, particularly in the case of sfermions lighter than or comparable in mass to $M_2$. The masses of the sleptons are not of particularly great importance for our study, however, as they do not effect the neutralino spin-independent or spin-dependent elastic scattering cross sections with nucleons, and typically contribute little to the neutralino annihilation cross section to lines. The most important effect of the sleptons is in the relic density calculation. If sleptons only slightly heavier than the lightest neutralino are observed at the LHC, for example, some of our conclusions may be modified.

Of course there is another important assumption we have adopted throughout this study: that the spectrum of new particles being observed at colliders and in astrophysical experiments are supersymmetric particles. Even if a number of superpartners are discovered at the LHC, it is not at all clear that those particles will be identified conclusively as such. In particular, it has shown to be a challenge to distinguish supersymmetry from models with Universal Extra Dimensions (UED) at the LHC~\cite{ued}, especially if the Kaluza-Klein states appear with masses of $\sim1$ TeV or above. Fortunately, the Kaluza-Klein dark matter found in these models~\cite{kkdm} possesses features which could potentially be distinguished from neutralinos, such as a high neutrino rate from the Sun~\cite{kribsneutrinos}, a hard positron spectrum~\cite{kribspositrons} and fairly predictable direct detection rates~\cite{kkelastic}. Furthermore, other scenarios, such as T-parity conserving little Higgs models~\cite{littlehiggs}, may be challenging to distinguish from supersymmetry at the LHC, but might be assisted by dark matter observations.

\section{Summary Of Our Results}

In this article, we have studied the ability of astrophysical dark matter experiments to constrain the parameters of supersymmetry. A number of observable quantities can be useful in this respect, including rates at direct detection experiments, the flux of high-energy neutrinos from the Sun, the brightness of gamma-ray lines and the flux of high-energy cosmic positrons.

In exploring ways to use these observables to learn about the properties of supersymmetry, we have exploited a number of correlations:

\begin{itemize}

\item{In the case that the scalar (spin-independent) neutralino-nucleon elastic scattering cross section is relativity large ($\gsim 10^{-7}$ pb), this process is dominated by the exchange of the heavy, CP-even Higgs boson, coupled to $s$ and $b$-quarks. In this case, the cross section scales as $\sigma_{\chi N} \propto \tan^2\beta |N_{11}|^2 |N_{13}|^2/m^4_A$. For smaller cross sections, contributions from light CP-even Higgs exchange and squark exchange can also play dominant roles, making the correlation less powerful, but still potentially useful.}

\item{As constraints on the spin-independent, neutrino-nucleon elastic scattering cross section become stronger, the only way that an observable flux of high-energy neutrinos from neutralino annihilations in the Sun could be generated is if neutralinos are captured in the Sun primarily through spin-dependent (axial-vector) couplings. In observable models, this is dominated by $Z$-exchange diagrams, leading to the strong correlation, $\sigma^{\rm{SD}}_{\chi p} \propto  (|N_{13}|^2- |N_{14}|^2)^2$. The rate in a neutrino telescope scales with this quantity as well (as long as the neutralino mass is well above the energy threshold of the experiment).}

\item{In most models, the neutralino annihilation cross sections to gamma-ray line producing final states ($\gamma \gamma, \gamma Z$) are dominated by diagrams that include a chargino-$W^{\pm}$ loop. In this class of diagrams, the cross section scales as some combination of the higgsino and wino fractions of the neutralino. A measurement of the annihilation cross section to gamma-ray lines (or even the relative fraction of annihilations which produce lines) could be used to constrain the higgsino and wino components of the lightest neutralino.}

\item{The total neutralino annihilation cross section (in the low velocity limit) depends on a large number of supersymmetric parameters. Nevertheless, this quantity would have some discriminating power if it could be determined by measurements of the cosmic positron spectrum, which samples the dark matter annihilation rate in the local halo.}

\end{itemize}

Although the mass of the lightest neutralino will likely be measured to roughly 10\% precision at the LHC, its composition will be considerably more difficult to constrain. To study the ability of astrophysical observations to assist in determining the composition of the lightest neutralino, we have considered a number of benchmark models. Asumming that the mass of the lightest neutralino and the squark masses (and in some cases $\tan \beta$ and $m_A$) can be measured at the LHC, we performed parameter scans looking for models consistent with a given set of LHC observations, and consistent with the measured dark matter density. Comparing the distribution of the value of $\mu$ among these models (or alternatively, the higgsino fraction of the lightest neutralino) to the distribution after constraints from plausible future astrophysical observations were made, we find that in many scenarios, the astrophysical measurement can play an important role in determining the value of $\mu$ (or the higgsino fraction).

In particular, we have found that:

\begin{itemize}
\item{In the majority of our benchmark models (LT1, LT3, LT4, IM2, IM3, IM4 and HV3), we found that the inclusion of information from astrophysical dark matter experiments (most importantly direct detection) allowed for a constraint on the magnitude of $\mu$ (or alternatively, the higgsino fraction of the lightest neutralino) that is considerably more stringent than can be obtained from collider experiments and relic density considerations alone. In addition to the important role of direct detection in constraining $\mu$, we note that in those models which generate an observable neutrino flux from the Sun (IM2, IM4 and HV2) that $|\mu|$ (and the higgsino fraction) are quite tightly constrained by this measurement. This is further enhanced by the very large annihilation cross sections neutralinos have to $\gamma \gamma$ and $\gamma Z$ in these models. None of these measurements assist significantly in determining the sign of $\mu$.}

\item{In a few of the benchmark models, astrophysical dark matter measurements provide far less information regarding the value of $\mu$ (and the higgsino fraction). For the case of model LT2, for example, nearly all of the supersymmetric models we found which were consistent with the projected measurements at the LHC were also found to predict similar elastic scattering cross sections and other astrophysically relevant quantities. For this reason, astrophysical measurements provide little discriminating power, although can still provide a valuable confirmation. Additionally, in the cases of models IM5 and HV1, little is learned about $\mu$ (or the higgsino fraction) from astrophysical measurements. In these cases, the LHC will likely not be capable of measuring $\tan\beta$ or $m_A$, leading to a situation where a number of very different sets of parameters could lead to the measured relic abundance and astrophysical observables. This is further exasperated by the significant role of coannihilations in the freeze-out process of model IM5 and of the resonant annihilations through the CP-odd Higgs in model HV1.}

\item{In two of our benchmark models (IM1 and HV1), the mass of the CP-odd Higgs could be quite constrained by astrophysical dark matter measurements. This is very interesting considering that in these models, $A$ is too heavy and $\tan \beta$ is too small for heavy neutral MSSM Higgs bosons ($A$, $H_1$) to be detected at the LHC. In each of these models, neutralino annihilations in the freeze-out process is dominated by the resonant exchange of a CP-odd Higgs. When direct detection experiments determine that the lightest neutralino's elastic scattering cross section with nucleons is very small ($7\times 10^{-11}$ and $2\times 10^{-10}$ pb, respectively), it demonstrates that these neutralinos are quite bino-like (a small higgsino fraction). This is further supported by the small annihilation cross section to $\gamma \gamma$ and $\gamma Z$ and the small rate in neutrino telescopes found in these models. A very bino-like neutralino requires an annihilation resonance (or possibly coannihilations) to avoid being overproduced in the early universe, thus it can be determined that $m_A \approx 2 m_{\chi^0}$. In such models, $m_A$ can be constrained in this way to roughly $\pm 100$ GeV.}

\item{We have also studied the ability of astrophysical measurements to be used to constrain the quantity $\tan\beta$. We find that the prospects for this determination are far less optimistic.}

\end{itemize}

\section{Conclusions}

If low energy supersymmetry exists in nature, it will likely be discovered in the next few years at the LHC (or possibly at the Tevatron). Over roughly the same period of time, the prospects for direct and indirect searches for neutralino dark matter are also very encouraging. Each of these windows into the characteristics of supersymmetry can provide us with useful and complementary information. In this paper, we have studied the ability to constrain the parameters of supersymmetry, beyond what can be done at the LHC, using direct and indirect dark matter experiments.

Although the presence of a neutralino LSP can in most cases be confirmed at the LHC, the composition and corresponding couplings of such a state will likely go unconstrained by such an experiment. Direct and indirect dark matter experiments, on the other hand, can probe the lightest neutralino's interactions with nucleons and with themselves (annihilations),  potentially allowing for a determination of its composition (and the value of $|\mu|$). We also find that direct and indirect dark matter measurements, combined with relic abundance considerations, can in some cases (the $A$-funnel region) determine the mass of the CP-odd Higgs boson ($A$), even when beyond the reach of the LHC.

\acknowledgements{We would like to thank Tilman Plehn, Ted Baltz, Bob McElrath and Gianfranco Bertone for insightful discussions and helpful comments. DH is supported by the US Department of Energy and by NASA grant NAG5-10842.}

\end{document}